\documentclass[review]{elsarticle}

\usepackage{lineno,hyperref}
\usepackage{amsmath,amssymb,amsfonts}
\usepackage{graphicx}
\usepackage{epsfig}
\usepackage{subfigure}
\usepackage{hhline}
\usepackage{multirow}
\usepackage{xcolor}
\modulolinenumbers[5]

\journal{Computerized Medical Imaging and Graphics}

\begin{document}

\begin{frontmatter}

\title{Deep Multi-Magnification Networks for Multi-Class Breast Cancer Image Segmentation}





\author[memorial]{David Joon Ho\corref{mycorrespondingauthor}}
\cortext[mycorrespondingauthor]{Corresponding author}
\ead{hod@mskcc.org}

\author[memorial]{Dig V. K. Yarlagadda}

\author[memorial]{Timothy M. D'Alfonso}

\author[memorial]{Matthew G. Hanna}

\author[memorial]{Anne Grabenstetter}

\author[memorial]{Peter Ntiamoah}

\author[memorial]{Edi Brogi}

\author[memorial]{Lee K. Tan}

\author[memorial,weill]{Thomas J. Fuchs}

\address[memorial]{Department of Pathology, Memorial Sloan Kettering Cancer Center, New York, NY 10065 USA}
\address[weill]{Weill Cornell Graduate School for Medical Sciences, New York, NY 10065 USA}

\begin{abstract}
Pathologic analysis of surgical excision specimens for breast carcinoma is important to evaluate the completeness of surgical excision and has implications for future treatment. 
This analysis is performed manually by pathologists reviewing histologic slides prepared from formalin-fixed tissue. 
In this paper, we present Deep Multi-Magnification Network trained by partial annotation for automated multi-class tissue segmentation by a set of patches from multiple magnifications in digitized whole slide images. 
Our proposed architecture with multi-encoder, multi-decoder, and multi-concatenation outperforms other single and multi-magnification-based architectures by achieving the highest mean intersection-over-union, and can be used to facilitate pathologists' assessments of breast cancer.
\end{abstract}


\end{frontmatter}


\section{Introduction}
\label{sec:introduction}
Breast carcinoma is the most common cancer to be diagnosed for women \cite{bray2018}.
Approximately 12$\%$ of women in the United States will be diagnosed with breast cancer during their lifetime \cite{desantis2019}. 
Pathologists diagnose breast carcinoma based on a variety of morphologic features including tumor growth pattern and nuclear cytologic features.
Pathologic assessment of breast tissue dictates the clinical management of the patient and provides prognostic information.
Breast tissue from a variety of biopsies and surgical specimens is evaluated by pathologists. 
For example, patients with early-stage breast cancer often undergo breast-conserving surgery, or lumpectomy, which removes a portion of breast tissue containing the cancer \cite{moo2014}. 
To determine the completeness of the surgical excision, the edges of the lumpectomy specimen, or margins, are evaluated microscopically by a pathologist. 
Achieving negative margins (no cancer found touching the margins) is important to minimize the risk of local recurrence of the cancer \cite{gage1996}. 
Accurate analysis of margins by the pathologist is critical for determining the need for additional surgery. 
Pathologic analysis of margin specimens involves the pathologist reviewing roughly 20-40 histologic slides per case, and this process can be time-consuming and tedious. 
With the increasing capabilities of digitally scanning histologic glass slides, computational pathology approaches could potentially improve the efficiency and accuracy of this process by evaluating whole slide images (WSIs) of specimens \cite{fuchs2011}. 

Various approaches have been used to analyze WSIs. 
Most models include localization, detection, classification, and segmentation of objects (i.e. histologic features) in digital slides. 
Histopathologic features include pattern-based identification, such as nuclear features, cellular/stromal architecture, or texture. 
Computational pathology has been used in nuclei segmentation to extract nuclear features such as size, shape, and relationship between them \cite{gurcan2009,veta2014}.
Nuclei segmentation is done by adaptive thresholding and morphological operations to find regions where nuclei density is high \cite{petushi2006}.
A breast cancer grading method can be developed by gland and nuclei segmentation using a Bayesian classifier and structural constraints from domain knowledge \cite{naik2008}.
To segment overlapping nuclei and lymphocytes, an integrated active contour based on region, boundary, and shape is presented in \cite{ali2012}.
A gland segmentation and classification method in prostate tissue is introduced where structural and contextual features from nuclei, cytoplasm, and lumen are used to classify artifact, normal gland, and cancer gland \cite{nguyen2012}.
These nuclei-segmentation-based approaches are challenging because shapes of nuclei and structures of cancer regions may have large variations in the tissue samples captured in the WSIs.

Recently, deep learning, a type of machine learning, has been widely used for automatic image analysis due to the availability of large training datasets and the advancement of graphics processing units (GPUs) \cite{lecun2015}.
Deep learning models composed of deep layers with non-linear activation functions enable to learn more sophisticated features.
Especially, convolutional neural networks (CNNs) learning spatial features in images have shown outstanding achievements in image classification \cite{krizhevsky2012}, object detection \cite{girshick2014}, and semantic segmentation \cite{long2015}.
Fully Convolutional Network (FCN) in \cite{long2015} developed for semantic segmentation, also known as pixel-wise classification, can understand location, size, and shape of objects in images.
FCN is composed of an encoder and a decoder, where the encoder extracts low-dimensional features of an input image and the decoder utilizes the low-dimensional features to produce segmentation predictions.
To improve segmentation predictions, SegNet introduces max-unpooling layers where max-pooling indices in an encoder are stored and used at the corresponding upsampling layers in a decoder \cite{badrinarayanan2017}.
Semantic segmentation has been used on medical images to automatically segment biological structures.
For example, U-Net \cite{ronneberger2015} is used to segment cells in microscopy images. 
U-Net architecture has concatenations transferring feature maps from an encoder to a decoder to preserve spatial information.
This architecture has shown more precise segmentation predictions on biomedical images. 

Deep learning has recently received high attention in the computational pathology community \cite{janowczyk2016,litjens2017,robertson2018}.
Investigators have shown automated identification of invasive breast cancer detection in WSIs by using a simple 3-layer CNN \cite{cruz2017}.
A method of classifying breast tissue slides to invasive cancer or benign by analyzing stroma regions using CNNs is described in \cite{bejnordi2018a}.
More recently, a multiple-instance-learning-based CNN achieves 100$\%$ sensitivity where the CNN is trained by 44,732 WSIs from 15,187 patients \cite{campanella2019}.
The availability of public pathology datasets contributes to develop many deep learning approaches for computational pathology.
For example, a breast cancer dataset to detect lymph node metastases was released for the CAMELYON challenges \cite{bejnordi2018b,bandi2019} and several deep learning techniques to analyze breast cancer datasets are developed \cite{wang2016,liu2017,lee2018}.

One challenge of using deep learning on WSIs is that the size of a single, entire WSI is too large to be processed into GPUs.
Images can be downsampled to be processed by pretrained CNNs \cite{kohl2018,kone2018} but critical details needed for clinical diagnosis in WSIs would be lost.
To solve this, patch-based approaches are generally used instead of slide-level approaches. 
Here, patches are extracted from WSIs to be processed by CNNs.
A patch-based process followed by a multi-class logistic regression to classify in slide-level is described in \cite{hou2016}.
The winner of the CAMELYON16 challenge uses the Otsu thresholding technique \cite{otsu1979} to extract tissue regions and trains a patch-based model to classify tumor and non-tumor patches \cite{wang2016}.
To increase the performance, class balancing between tumor and non-tumor patches and data augmentation techniques such as rotation, flip, and color jittering are used in \cite{liu2017}.
The winner of the CAMELYON17 challenge additionally develops patch-overlapping strategy for more accurate predictions \cite{lee2018}.
In \cite{mehta2017}, a patch is processed with an additional larger patch including border regions in the same magnification to segment subtypes in breast WSIs.
Alternatively, Representation-Aggregation CNNs to aggregate features generated from patches in WSIs are developed to share representations between patches \cite{agarwalla2017,shaban2019}.
The main limitations of these patch-based approaches extracted from a single magnification are (1) the field-of-view becomes narrow and (2) morphological features from a lower magnification are not used.

\begin{figure}[t]
\centering
\subfigure[Deep Single-Magnification Network]{\epsfig{figure=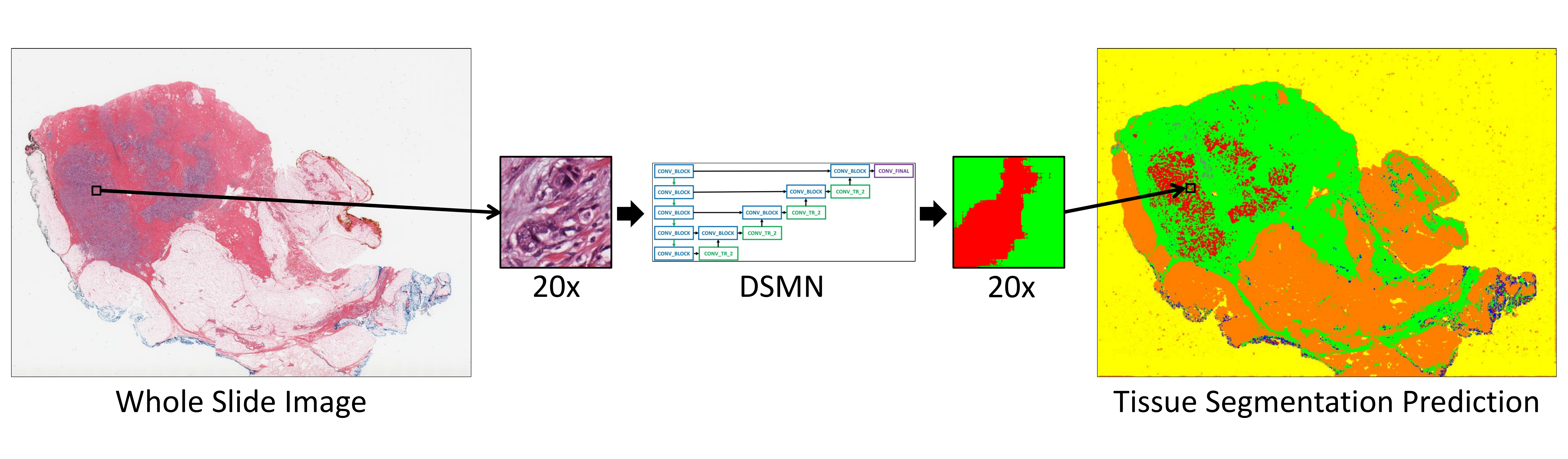,width = 0.95\textwidth}}
\subfigure[Deep Multi-Magnification Network]{\epsfig{figure=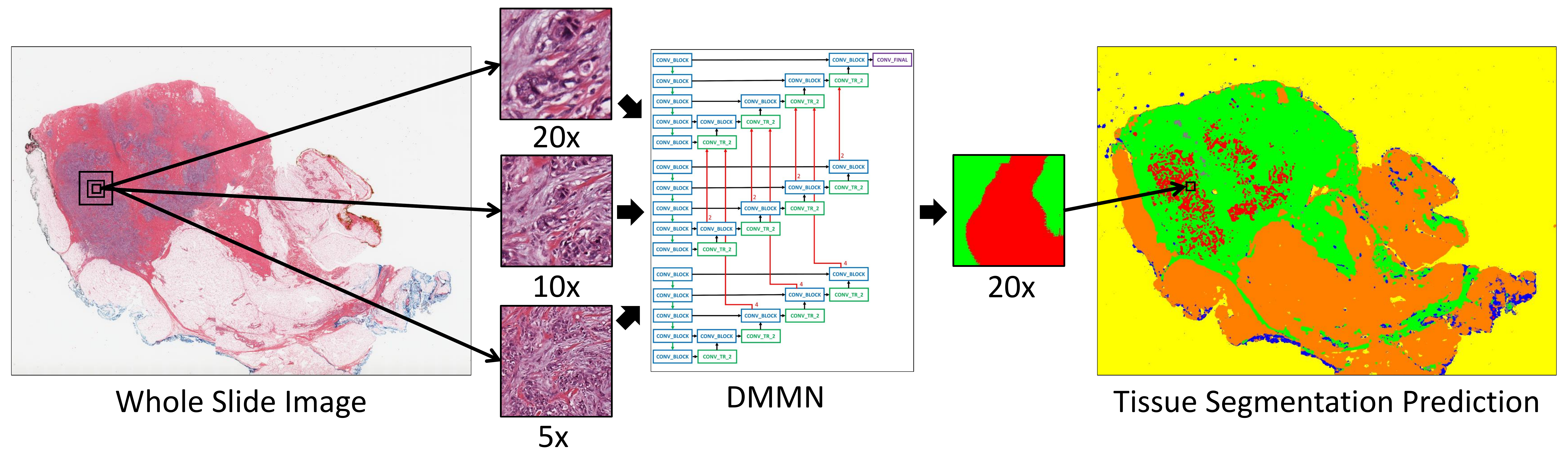,width = 0.95\textwidth}}
\caption{Introduction of a Deep Single-Magnification Network (DSMN) and a Deep Multi-Magnification Network (DMMN) for tissue segmentation of whole slide images. (a) A DSMN looks at a patch from a single magnification from a whole slide image with limited field-of-view to generate the corresponding multi-class tissue segmentation prediction. (b) A DMMN looks at a set of patches from multiple magnifications from a whole slide image to have wider field-of-view to generate the corresponding multi-class tissue segmentation prediction. The DMMN can learn both cellular features from a higher magnification and architectural growth patterns from a lower magnification. Here, carcinoma is predicted in red, benign epithelial in blue, background in yellow, stroma in green, necrotic in gray, and adipose in orange.}
\label{fig:introduction}
\end{figure}

Therefore, we develop segmentation CNNs by inputting a set of patches from multiple magnifications to increase the field-of-view and to provide more information from other magnifications.
Figure \ref{fig:introduction} introduces the main difference between a Deep Single-Magnification Network (DSMN) and a Deep Multi-Magnification Network (DMMN) for tissue segmentation of whole slide images.
An input to a DSMN in Figure \ref{fig:introduction}(a) is a patch from a single magnification which limits a field-of-view.
An input to a DMMN in Figure \ref{fig:introduction}(b) is a set of patches from multiple magnifications allowing a wider field-of-view.
High magnification patches provide details at the cellular level, such as nuclear features, whereas low magnification patches demonstrate distribution of tissue types and architectural growth patterns of benign and malignant processes.

There are several works using multiple magnifications to analyze images from tissue samples.
A multi-input multi-output CNN is presented by analyzing an input image in multiple resolutions to segment cells in fluorescence microscopy images \cite{raza2017}. 
Similarly, a stain-aware multi-scale CNN is further designed for instance cell segmentation in histology images \cite{graham2018}.
To segment tumor regions in the CAMELYON dataset \cite{bejnordi2018b}, a binary segmentation CNN is described in \cite{gu2018}.
In this work, four encoders for different magnifications are implemented but only one decoder is used to generate the final segmentation predictions.
More recently, a CNN architecture composed of three expert networks for different magnifications, a weighting network to automatically select weights to emphasize specific magnifications based on input patches, and an aggregating network to produce final segmentation predictions is developed in \cite{tokunaga2019}.
Here, feature maps are not shared between the three expert networks until the last layer which can limit utilizing feature maps from multiple magnifications.
Architectures designed in \cite{gu2018} and \cite{tokunaga2019} center-crop feature maps in lower magnifications and then upsample the cropped feature maps to match the size and magnification during concatenations which can also limit the usage of feature maps on cropped boundary regions in lower magnifications.

In this paper, we present a Deep Multi-Magnification Network (DMMN) to accurately segment multiple subtypes in images of breast tissue.
Our DMMN architecture has multiple encoders, multiple decoders, and multiple concatenations between decoders to have richer feature maps in intermediate layers.
To fully utilize feature maps in lower magnifications, we center-crops intermediate feature maps during concatenations.
By concatenating intermediate feature maps in each layer, feature maps from multiple magnifications can be used to produce accurate segmentation predictions.
To train our DMMN, we partially annotate WSIs, similarly done as \cite{bokhorst2019}, to reduce the burden of annotations.
Our DMMN model trained by our partial annotations can learn not only features of each subtype, but also morphological relationship between subtypes especially transitions from one subtype to another subtype on boundary regions, which leads to outstanding segmentation performance.
We test our multi-magnification model on two breast datasets and observe that our model consistently outperforms other architectures.
Our method can be used to automatically segment cancer regions on breast images to assist in diagnosis of patients' status and to decide future treatments.

\section{Proposed Method}
\label{sec:method}
\begin{figure*}[t!]
\centerline{\epsfig{figure=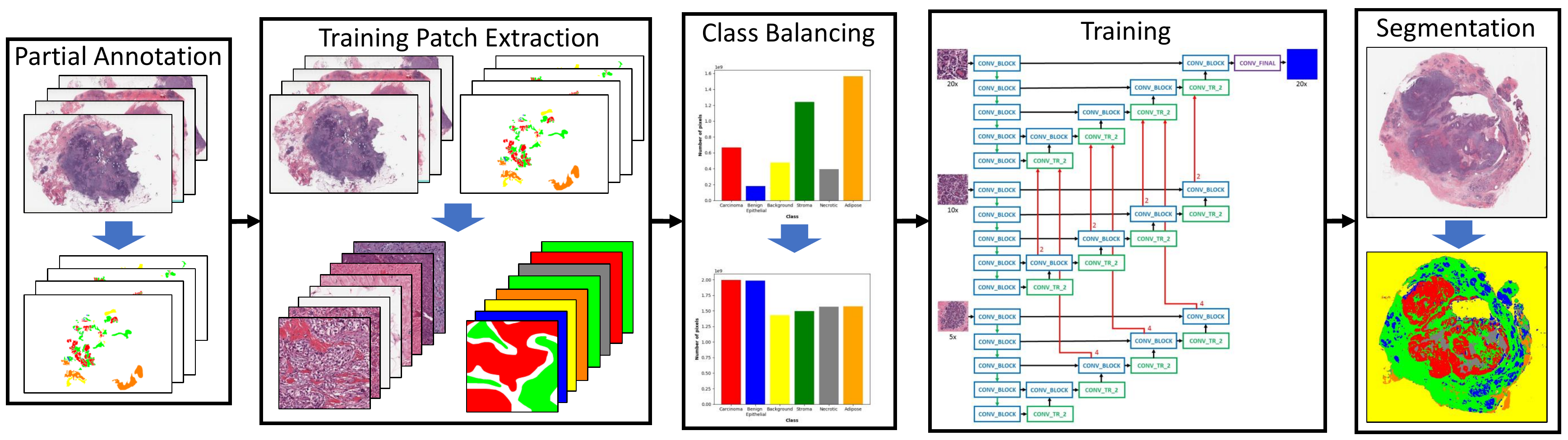,width=\textwidth}}
\caption{Block diagram of the proposed method with our Deep Multi-Magnification Network. The first step of our method is to partially annotate training whole slide images. After extracting training patches from the partial annotations and balancing the number of pixels between classes, our Deep Multi-Magnification Network is trained. The trained network is used for multi-class tissue segmentation of whole slide images.}
\label{fig:block}
\end{figure*}
Figure \ref{fig:block} shows the block diagram of our proposed method.
Our goal is to segment multiple subtypes on breast images using our Deep Multi-Magnification Network (DMMN).
First of all, manual annotation is done on the training dataset with $C$ classes.
Here, this annotation is done partially for an efficient and fast process.
To train our multi-class segmentation DMMN, patches are extracted from whole slide images (WSIs) and the corresponding annotations.
Before training our DMMN with the extracted patches, we use elastic deformation \cite{ronneberger2015,fu2017} to multiply patches belonging to rare classes to balance the number of annotated pixels between classes.
After the training step is done, the model can be used for multi-class segmentation of breast cancer images.
We have implemented our system in PyTorch \cite{paszke2017}. 

\subsection{Partial Annotation}
A large set of annotations is needed for supervised learning, but this is generally an expensive step requiring pathologists' time and effort.
Especially, due to giga-pixel scale of image size, exhaustive annotation to label all pixels in WSIs is not practical.
Many works are done using public datasets such as CAMELYON datasets \cite{bejnordi2018b,bandi2019} but public datasets are designed for specific applications and may not be generalized to others.
To segment multiple tissue subtypes on our breast training dataset, we partially annotate images.
 
\begin{figure}[t]
\centering
\subfigure[Whole Slide Image]{\frame{\epsfig{figure=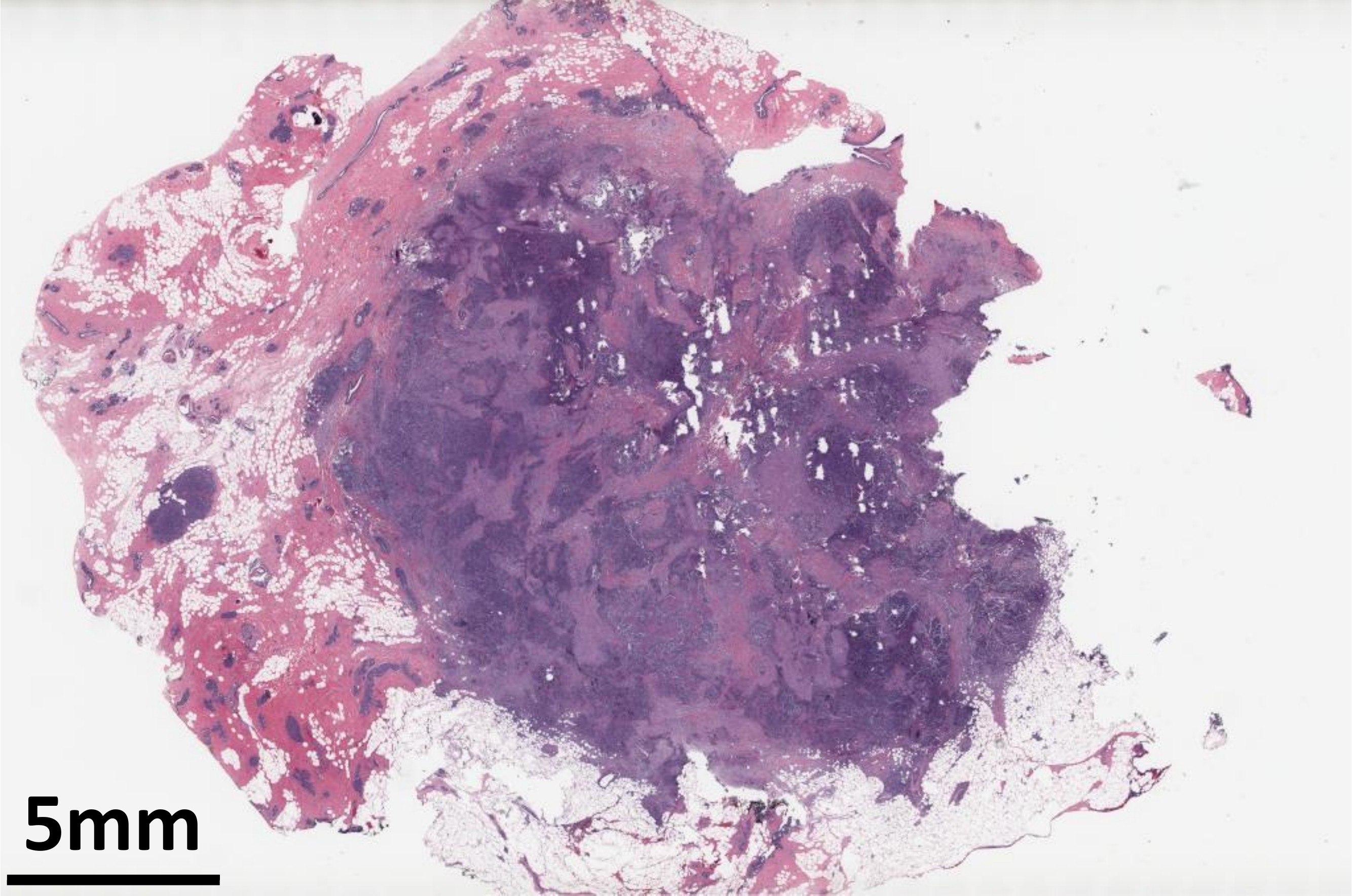,width = 0.4\textwidth}}}
\subfigure[Partial Annotation]{\frame{\epsfig{figure=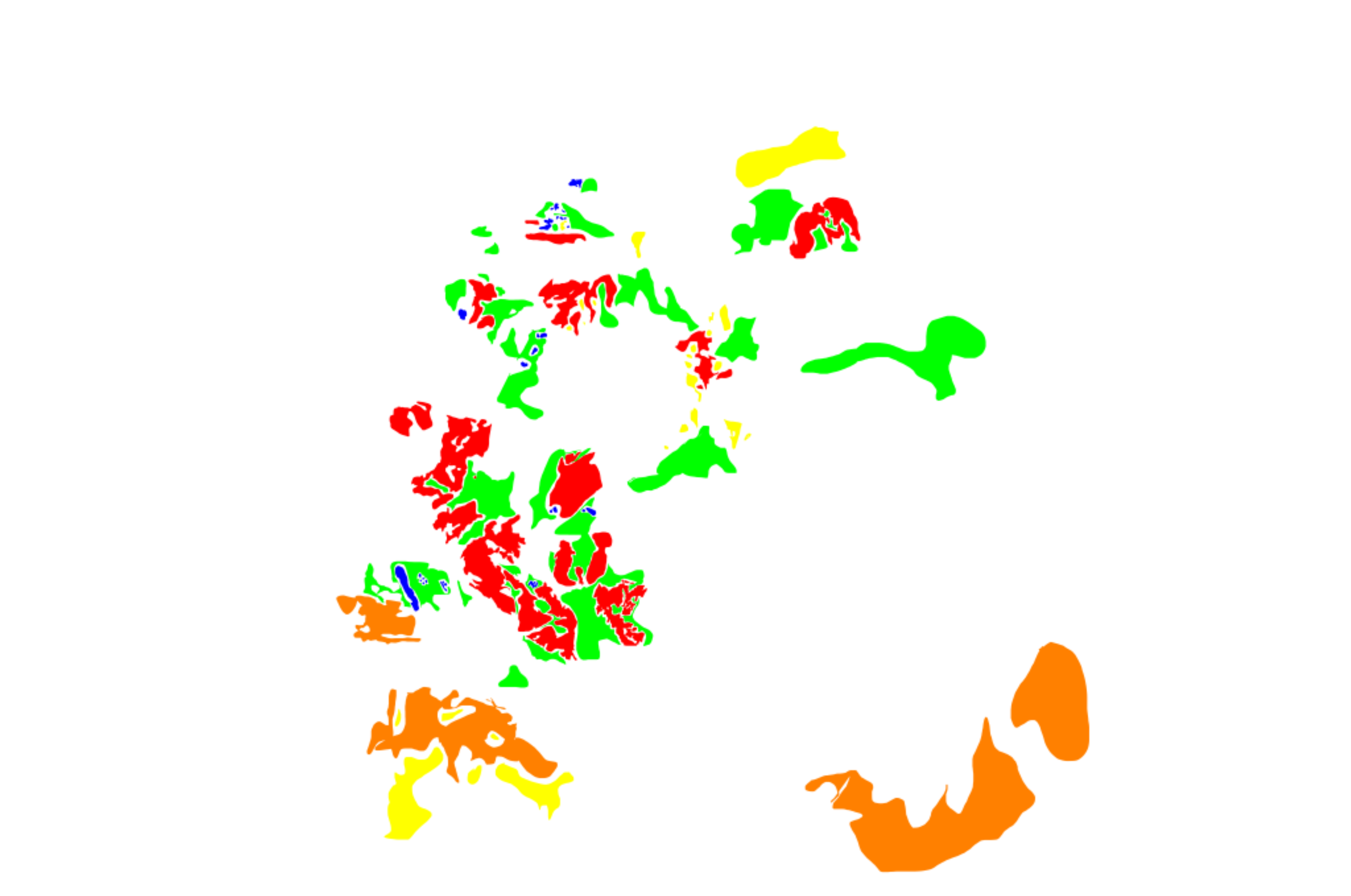,width = 0.4\textwidth}}}
\subfigure[The Partial Annotation Overlaid on the Whole Slide Image]{\frame{\epsfig{figure=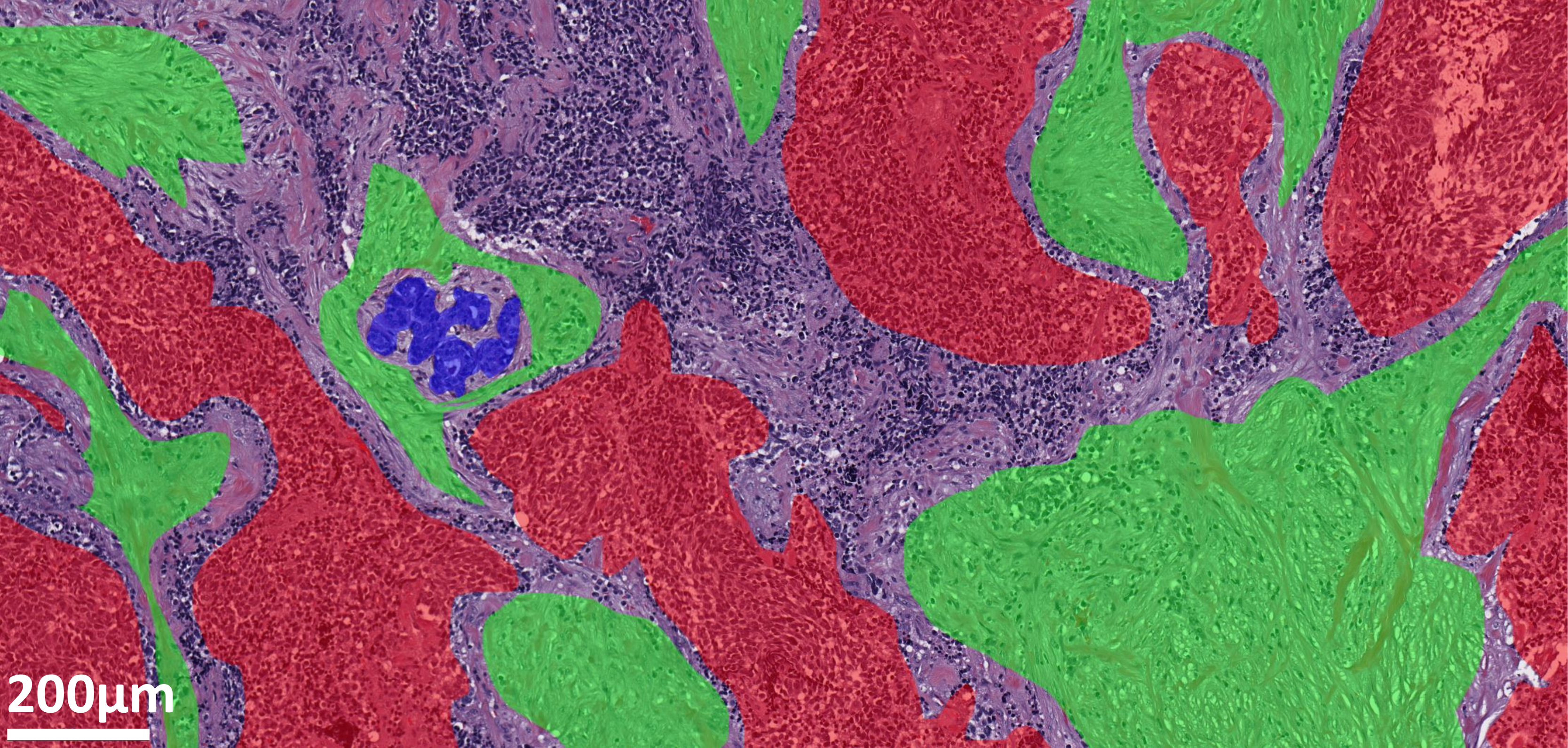,width = 0.75\textwidth}}}
\caption{An example of partial annotation. (a) A whole slide image from breast tissue. (b) A partially annotated image where multiple tissue subtypes are annotated in distinct colors and white regions are unlabeled. (c) The partial annotation overlaid on the whole slide image. Subtype components are annotated without cropping while reducing the thickness of unlabeled regions between the subtype components. Here, carcinoma is annotated in red, benign epithelial in blue, background in yellow, stroma in green, necrotic in gray, and adipose in orange.}
\label{fig:annotations}
\end{figure}

For partial annotations, we (1) annotated the entire subtype components without cropping and (2) reduced the thickness of unlabeled regions between the subtype component.
An example of our proposed partial annotation is shown in Figure \ref{fig:annotations}, where a partially annotated image overlaid on a whole slide image is shown in Figure \ref{fig:annotations}(c).
Note white regions in Figure \ref{fig:annotations}(b) are unlabeled.
Exhaustive annotations, especially on boundary regions, without any overlapping portions and subsequent inaccurate labeling can be challenging given the regions merge into each other seamlessly. 
Additionally, the time required for complete, exhaustive labeling is immense. 
By reducing the thickness of these unlabeled boundary regions, our CNN models trained by our partial annotation can learn the spatial relationships between subtypes such as transitions from one subtype to another subtype and generate precise segmentation boundaries. 
A partially annotated image in Figure \ref{fig:annotations}(c) shows unlabeled regions between carcinoma in red and stroma in green are thinned.
This is different from the partial annotation done in \cite{bokhorst2019} where annotated regions of different subtypes were too widely spaced and thus unsuitable for training spatial relationships between them. 
The work in \cite{bokhorst2019} also suggests exhaustive annotation in subregions of WSIs to reduce annotation efforts, but if the subtype components are cropped the CNN model cannot learn the growth pattern of the different subtypes. 
In this work, we annotated each subtype component entirely to let our CNN model learn the growth pattern of all subtypes. 
According to our proposed partial annotation, an experienced pathologist can spend approximately 30 minutes to annotate one WSI.

\subsection{Training Patch Extraction}
\label{sec:patch_extraction}
Whole slide images are generally too large to process in slide-level using convolutional neural networks.
To analyze WSIs, patch-based methods are used where patches extracted from an image is processed by a CNN and then the outputs are combined for slide-level analysis.
One limitation of the patch-based methods is that they only look at patches in a single magnification with a limited field-of-view.

To have a wider field-of-view, a set of multi-magnification patches is extracted to train our DMMN.
In this work, we set the size of a target patch to be analyzed in a WSI be 256 $\times$ 256 pixels in 20$\times$ magnification.
To analyze the target patch, an input patch with size of 1024 $\times$ 1024 pixels in 20$\times$ is extracted from the image where the target patch is located at the center of the input patch.
From this input patch, a set of three multi-magnification patches is extracted.
The first patch is extracted from the center of the input patch with size of 256 $\times$ 256 pixels in 20$\times$, which is the same location and magnification with the target patch.
The second patch is extracted from the center of the input patch with size of 512 $\times$ 512 pixels and downsampled by a factor of 2 to become size of 256 $\times$ 256 pixels in 10$\times$.
Lastly, the third patch is generated by downsampling the input patch by a factor of 4 to become size of 256 $\times$ 256 pixels in 5$\times$.
The set of three patches in different magnifications becomes the input to our DMMN to segment cancer in the target patch with size of 256 $\times$ 256 pixels.
Input patches are extracted from training images if more than 1$\%$ of pixels in the corresponding target patches are annotated.
The stride to $x$ and $y$-directions is 256 pixels to avoid overlapping target patches.
Note target patches may have multiple class labels.

\subsection{Class Balancing}
\label{sec:class_balancing}
Class balancing is a prerequisite step for training CNNs for accurate performance \cite{buda2018}.
When the number of training patches in one class dominates the number of training patches in another class, CNNs cannot properly learn features from the minor class.
In this work, class imbalance is observed in our annotations.
For example, the number of annotated pixels in carcinoma regions dominates the number of annotated pixels in benign epithelial regions.
To balance between classes, elastic deformation \cite{ronneberger2015,fu2017} is used to multiply training patches belonging to minor classes.

Elastic deformation is widely used as a data augmentation technique in biomedical images due to the squiggling shape of biological structures.
To perform elastic deformation on a patch, a set of grid points in the patch is selected and displaced randomly by a normal distribution with a standard deviation of $\sigma$.
According to the displacements of the grid points, all pixels in the patch are displaced by bicubic interpolation.
In this work, we empirically set the grid points by $17 \times 17$ and $\sigma = 4$ to avoid excessive distortions of nuclei to lose their features.

The number of patches to be multiplied needs to be carefully selected to balance the number of pixels between classes.
Here, we define a rate of elastic deformation for a class $c$, denoted as $r_c$, to be the number of patches to be multiplied for the class $c$ and a class order to decide the order of classes when multiplying patches.
The rate can be selected based on the number of pixels in each class.
The rate is a non-negative integer and elastic deformation is not performed if the rate is 0.
The class order can be decided based on applications.
For example, if one desires an accurate segmentation on carcinoma regions, then a class of carcinoma would have a higher order than other classes.
To multiply patches, each patch needs to be classified to a class $c$ if the patch contains a pixel label classified to $c$.
If a patch contains pixels in multiple classes, a class with a higher class order becomes the class of the patch.
After patches are classified, $r_c$ number of patches will be multiplied for each patch in class $c$ using elastic deformation.
Once class balancing is done, all patches are used to train CNNs.

\subsection{CNN Architectures}
\begin{figure*}[t!]
\centering
\begin{minipage}[b]{0.45\textwidth}
   \subfigure[U-Net]{\epsfig{figure=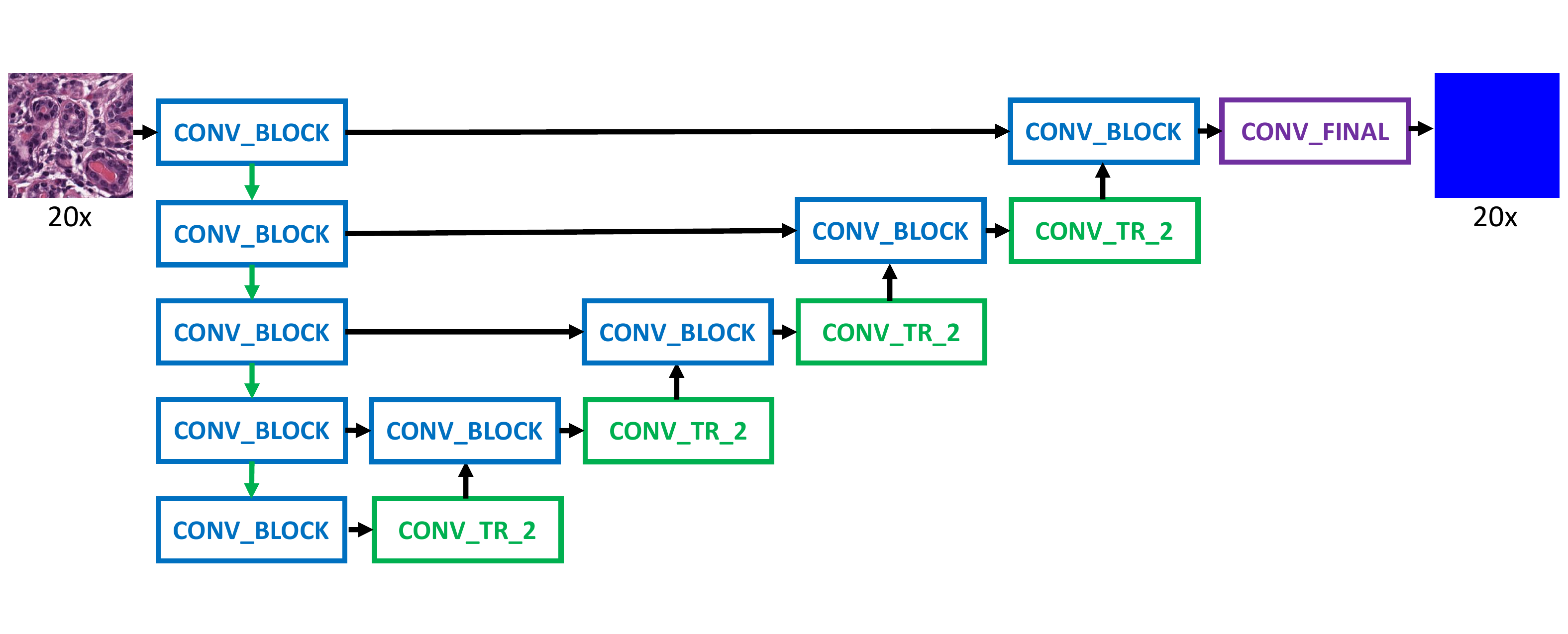,width = \textwidth}}
   \subfigure[DMMN-S2]{\epsfig{figure=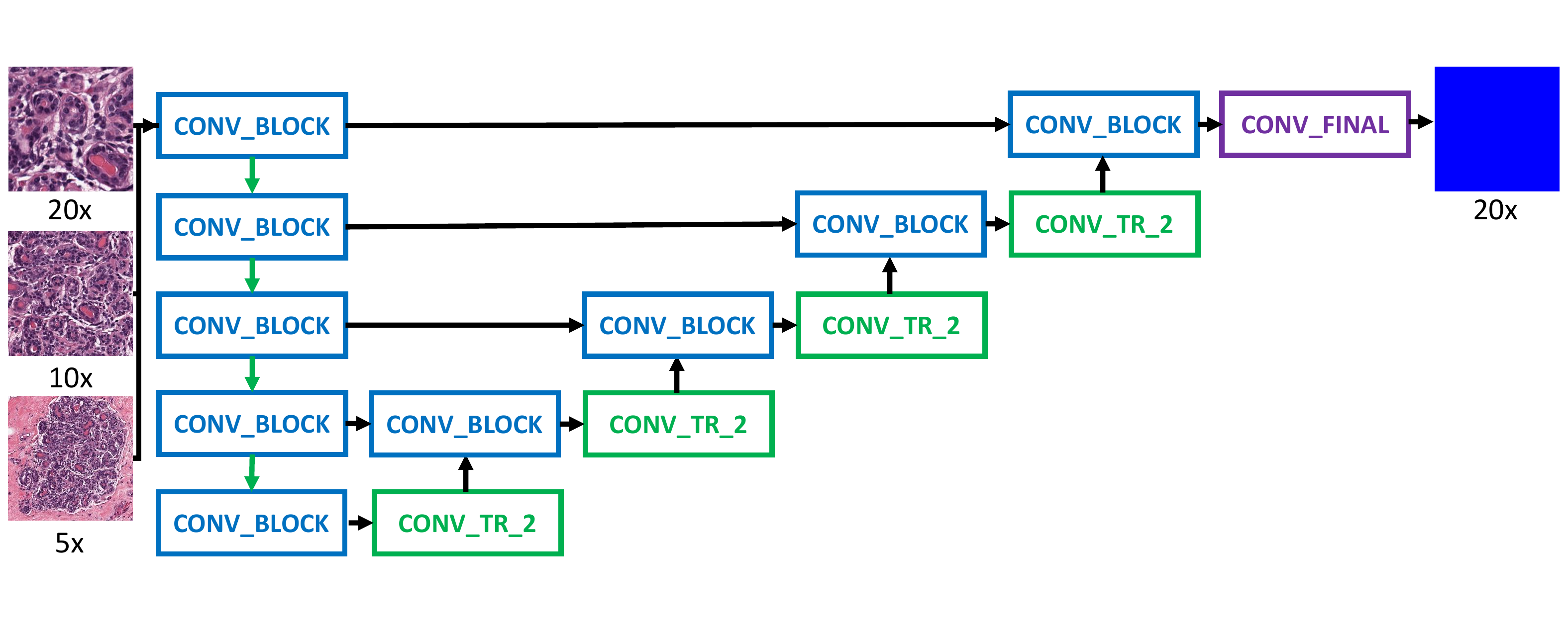,width = \textwidth}}
 \end{minipage}
\subfigure[DMMN-MS]{\epsfig{figure=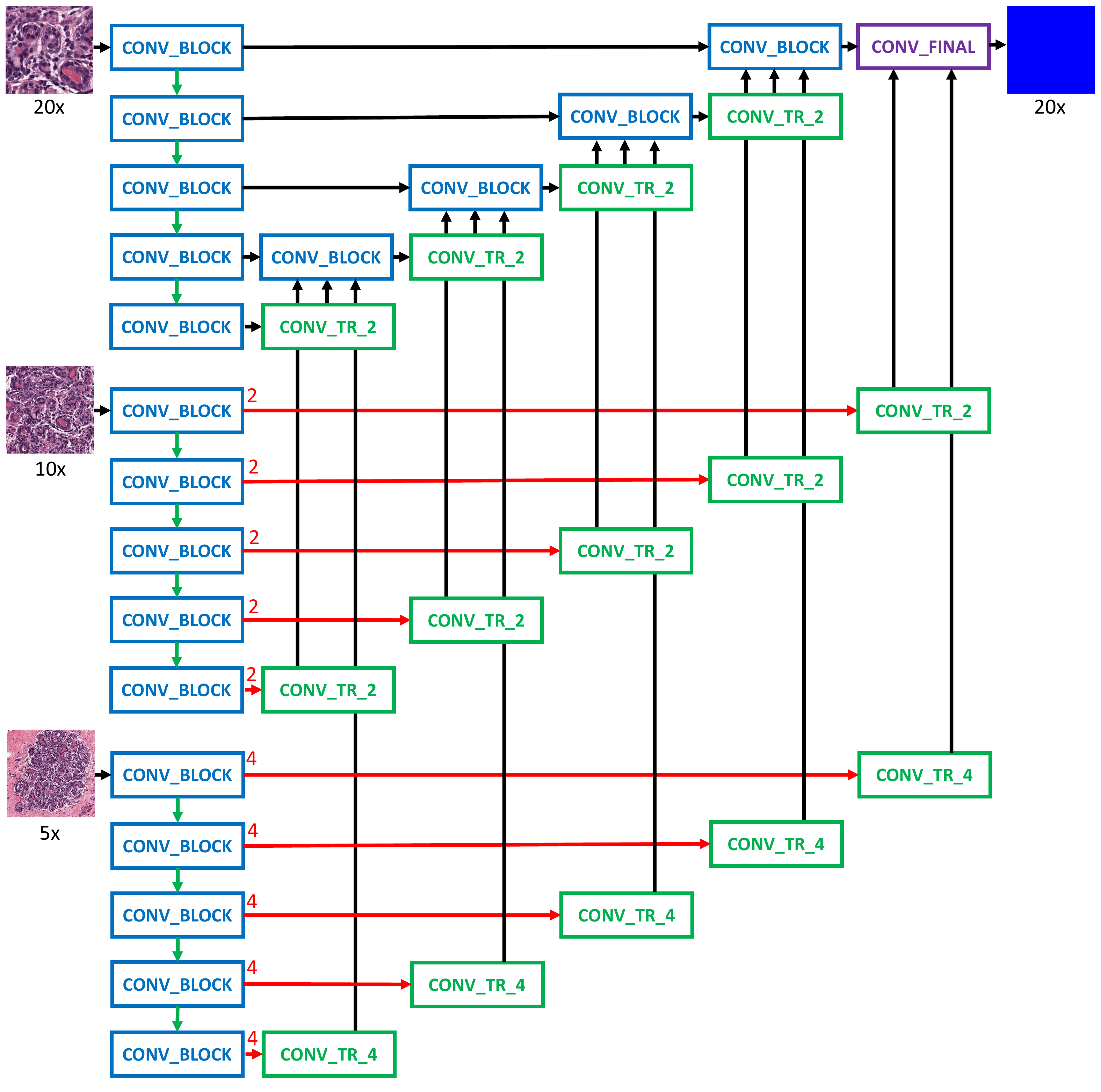,width = 0.45\textwidth}}
\subfigure[DMMN-M2S]{\epsfig{figure=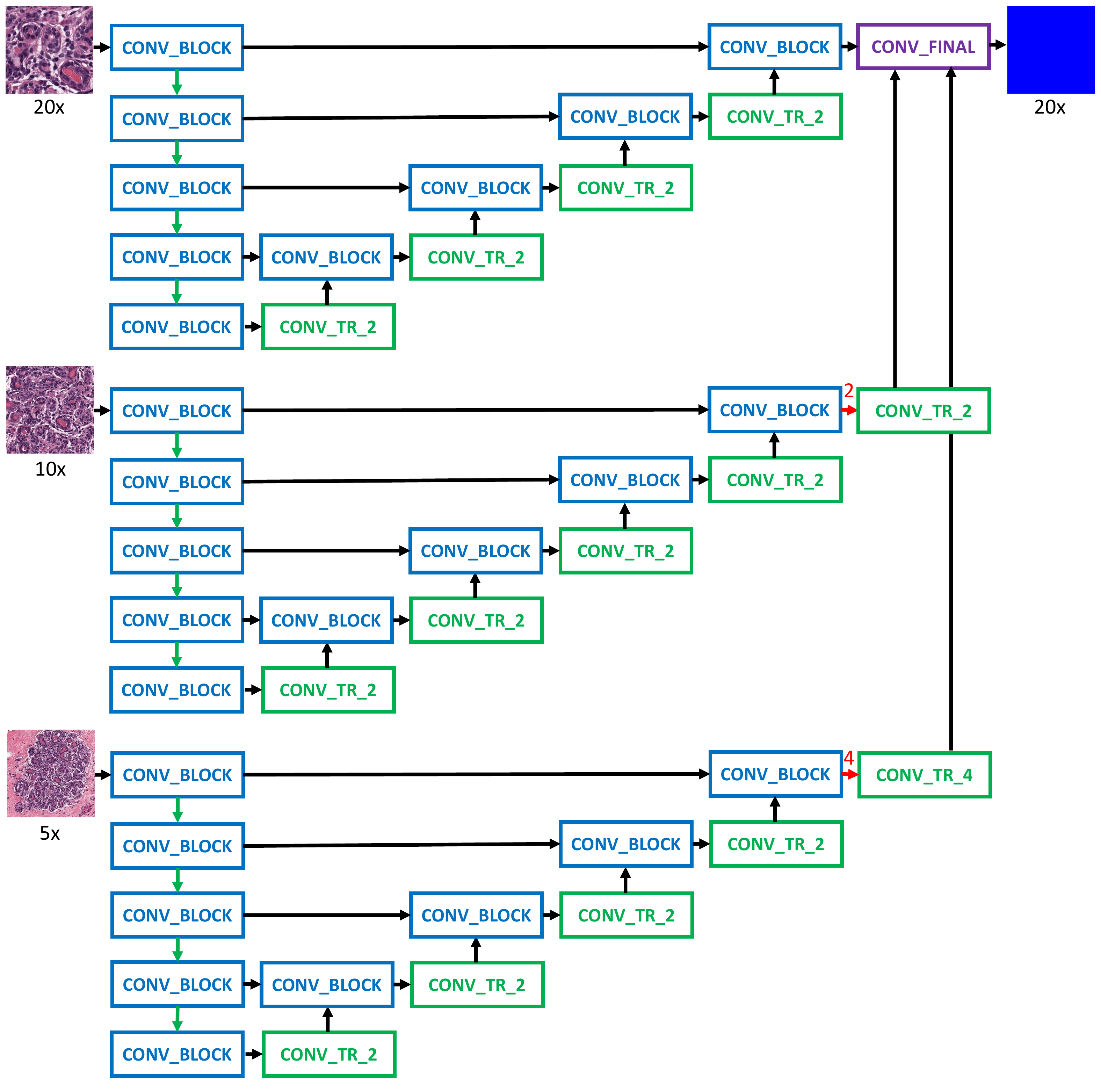,width = 0.45\textwidth}}
\subfigure[DMMN-M3]{\epsfig{figure=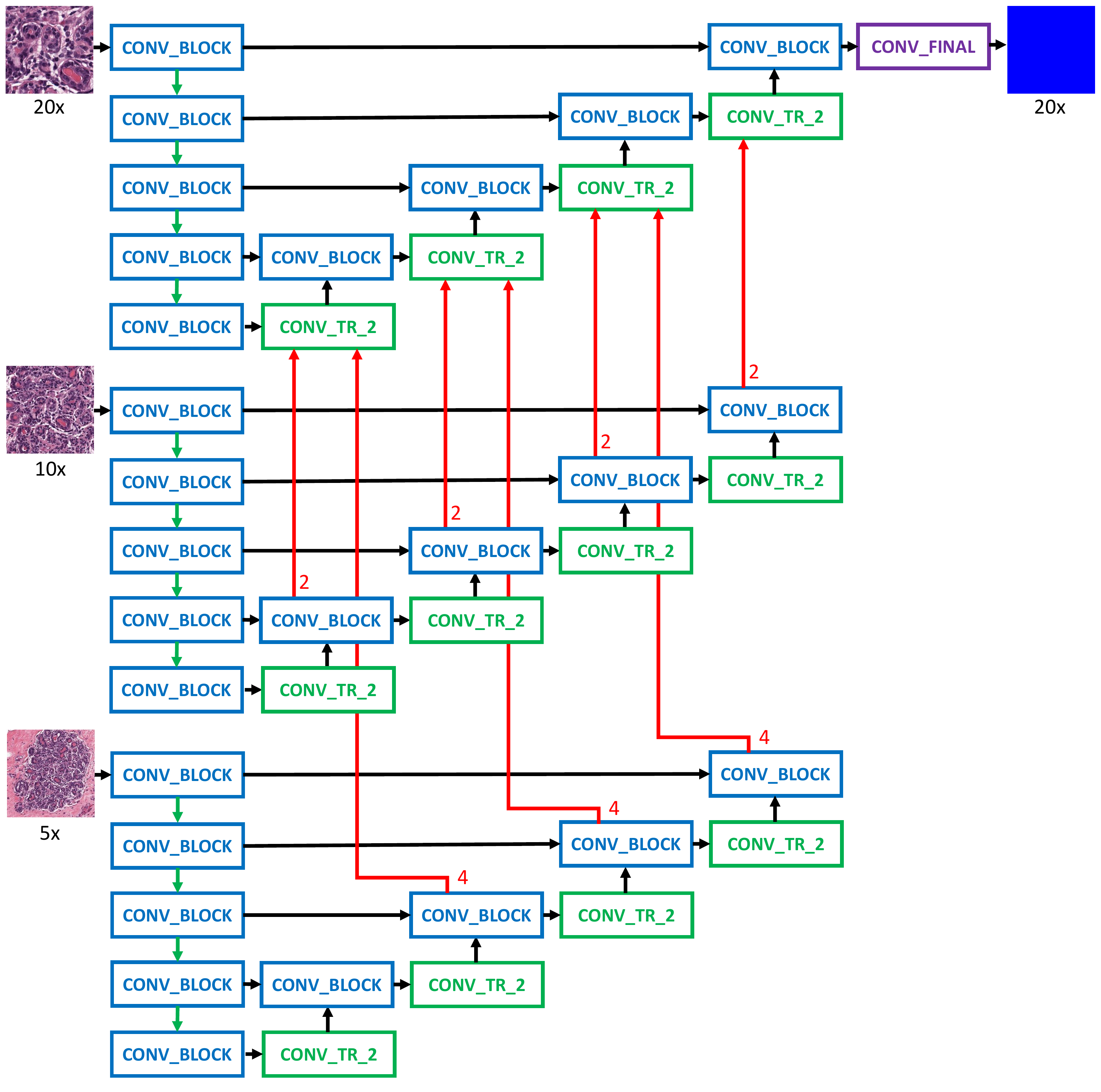,width = 0.45\textwidth}}
\caption{CNN architectures for multi-class tissue segmentation of a Deep Single-Magnification Network (DSMN) in (a) utilizing a patch from a single magnification and Deep Multi-Magnification Networks (DMMNs) in (b-e) utilizing multiple patches in various magnifications. (a) U-Net \cite{ronneberger2015} is used as our DSMN architecture. (b) Single-Encoder Single-Decoder (DMMN-S2) is a DMMN architecture where multiple patches are concatenated and used as an input to the U-Net architecture. (c) Multi-Encoder Single-Decoder (DMMN-MS) is a DMMN architecture having only one decoder. (d) Multi-Encoder Multi-Decoder Single-Concatenation (DMMN-M2S) is a DMMN architecture where feature maps from multiple magnifications are only concatenated at the final layer. (e) Our proposed Multi-Encoder Multi-Decoder Multi-Concatenation (DMMN-M3) is a DMMN architecture where feature maps are concatenated during intermediate layers to enrich feature maps in the decoder of the highest magnification.}
\label{fig:architectures}
\end{figure*}
Figure \ref{fig:architectures} shows architectures of a Deep Single-Magnification Network (DSMN) and Deep Multi-Magnification Networks (DMMNs) for multi-class tissue segmentation.
The size of input patches is 256 $\times$ 256 pixels and the size of an output prediction is 256 $\times$ 256 pixels.
CONV{\_}BLOCK contains two sets of a convolutional layer with kernel size of $3 \times 3$ with padding of 1 followed by a rectified linear unit (ReLU) activation function in series.
CONV{\_}TR{\_}$u$ contains a transposed convolutional layer followed by the ReLU activation function where $u$ is an upsampling rate.
Note CONV{\_}TR{\_}4 is composed of two CONV{\_}TR{\_}2 in series.
CONV{\_}FINAL contains a convolutional layer with kernel size of $3 \times 3$ with padding of 1, the ReLU activation function, and a convolutional layer with kernel size of $1 \times 1$ to output $C$ channels.
The final segmentation predictions are produced using the softmax operation.
Green arrows are max-pooling operations by a factor of 2 and red arrows are center-crop operations where cropping rates are written in red.
The center-crop operations crop the center regions of feature maps in all channels by the cropping rate to fit the size and magnification of feature maps for the next operation.
During the center-crop operations, the width and height of the cropped feature maps become a half and a quarter of the width and height of the input feature maps if the cropping rate is 2 and 4, respectively.

An original U-Net \cite{ronneberger2015} architecture in Figure \ref{fig:architectures}(a) uses a single magnification patch in 20$\times$ to produce the corresponding segmentation predictions.
A Single-Encoder Single-Decoder (DMMN-S2) architecture in Figure \ref{fig:architectures}(b) uses multiple patches in 20$\times$, 10$\times$, and 5$\times$ magnifications, but they are concatenated and used as an input to the U-Net architecture \cite{ronneberger2015}.
A Multi-Encoder Single-Decoder (DMMN-MS) architecture in Figure \ref{fig:architectures}(c), motivated by the work in \cite{gu2018}, uses multiple encoders in 20$\times$, 10$\times$, and 5$\times$ magnifications, but only uses a single decoder in 20$\times$ by transferring feature maps from encoders in 10$\times$ and 5$\times$.
A Multi-Encoder Multi-Decoder Single-Concatenation (DMMN-M2S) architecture in Figure \ref{fig:architectures}(d), motivated by the work in \cite{tokunaga2019}, has multiple encoders and the corresponding decoders in 20$\times$, 10$\times$, and 5$\times$ magnifications, but the concatenation is done only at the end of the encoder-decoder pairs.
Here, the weighting CNN in \cite{tokunaga2019} is excluded for a fair comparison with other architectures.
Lastly, our proposed Multi-Encoder Multi-Decoder Multi-Concatenation (DMMN-M3) architecture in Figure \ref{fig:architectures}(e) has multiple encoders and decoders and has concatenations between the decoders in multiple layers to enrich feature maps from the decoders in 10$\times$ and 5$\times$ to the decoder in 20$\times$.
Additionally, we use center-crop operations while transferring feature maps from the decoders in 10$\times$ and 5$\times$ to the decoder in 20$\times$ to extract features in 10$\times$ and 5$\times$ as much as possible.
Note DMMN-MS and DMMN-M2S use center-crop operations in 10$\times$ and 5$\times$ levels where cropped regions before concatenation can limit feature extraction processes in lower magnifications.

\subsection{CNN Training}
The balanced set of patches from Section \ref{sec:class_balancing} is used to train our multi-class segmentation CNNs.
We used a weighted cross entropy as our training loss function with $N$ pixels in a patch and $C$ classes:
\begin{equation}
    L(t^{gt},t^{pred}) = -\frac{1}{N}\sum_{p=1}^{N} \sum_{c=1}^{C} w_c t^{gt}_c(p) \log t^{pred}_c(p)
\label{eq:loss}
\end{equation}
where $t_c^{gt}$ and $t_c^{pred}$ are two-dimensional ground truth and segmentation predictions for a class $c$, respectively.
$t^{gt}_c(p)$ is a binary ground truth value for a class $c$ at a pixel location $p$, either 0 or 1, and $t^{pred}_c(p)$ is a segmentation prediction value for a class $c$ at a pixel location $p$, between 0 and 1.
In Equation \ref{eq:loss}, a weight for class $c$, $w_c$ is defined as
\begin{equation}
    w_c = 1-\frac{N_c}{\sum_{c}{N_c}}
\end{equation}
where $N_c$ is the number of pixels for class $c$ in a training set.
Unlabeled pixels do not contribute to the training loss function.
We use stochastic gradient descent (SGD) with a learning rate of $5 \times 10^{-5}$, a momentum of 0.99, and a weight decay of $10^{-4}$ for 20 epochs for optimization.
A CNN model with the highest mean intersection-over-union (mIOU) on validation images is selected as the final model.
During training, data augmentation using random rotation, vertical and horizontal flip, brightness, contrast, and color jittering is used.

\subsection{Multi-Class Segmentation}
Multi-class tissue segmentation on breast images can be done using the trained CNN.
The final label in each pixel is selected as a class which has the largest prediction value among the $C$ classes.
An input patch with size of 1024 $\times$ 1024 pixels is extracted from a WSI to generate a set of three patches with size of 256 $\times$ 256 pixels in 20$\times$, 10$\times$, and 5$\times$ magnifications by the process described in Section \ref{sec:patch_extraction}.
The set of three patches is processed by our trained CNN.
The segmentation predictions with size of 256 $\times$ 256 pixels are located at the center location of the input patch.
Input patches are extracted from the top-left corner of the WSI with a stride of 256 pixels in $x$ and $y$ directions to process the entire WSI.
Zero-padding is done to extract input patches on the boundary of WSIs.
The Otsu thresholding technique \cite{otsu1979} can be used before extracting patches as optional to remove background regions to speed up the segmentation process.
No pre-processing step is used during segmentation.

\section{Experimental Results}
\label{sec:results}
Two breast datasets, Dataset-I and Dataset-II, were used to train and evaluate various multi-class tissue segmentation methods.
Dataset-I is composed of whole slide images (WSIs) with Triple-Negative Breast Cancer (TNBC) containing high grade invasive ductal carcinoma (IDC).
Dataset-II is composed of WSIs from lumpectomy and breast margins containing IDC and ductal carcinoma in situ (DCIS) of various histologic grades.
All WSIs in Dataset-I and Dataset-II were from different patients, were hematoxylin and eosin (H$\&$E) stained, and were digitized at Memorial Sloan Kettering Cancer Center.
Dataset-I was digitized by Aperio XT where microns per pixel (MPP) in 20$\times$ is 0.4979 and Dataset-II was digitized by Aperio AT2 where MPP in 20$\times$ is 0.5021.
WSIs in Dataset-I were partially annotated by two pathologists and WSIs in Dataset-II were partially annotated by another pathologist.

\begin{figure}[ht]
\centering
\epsfig{figure=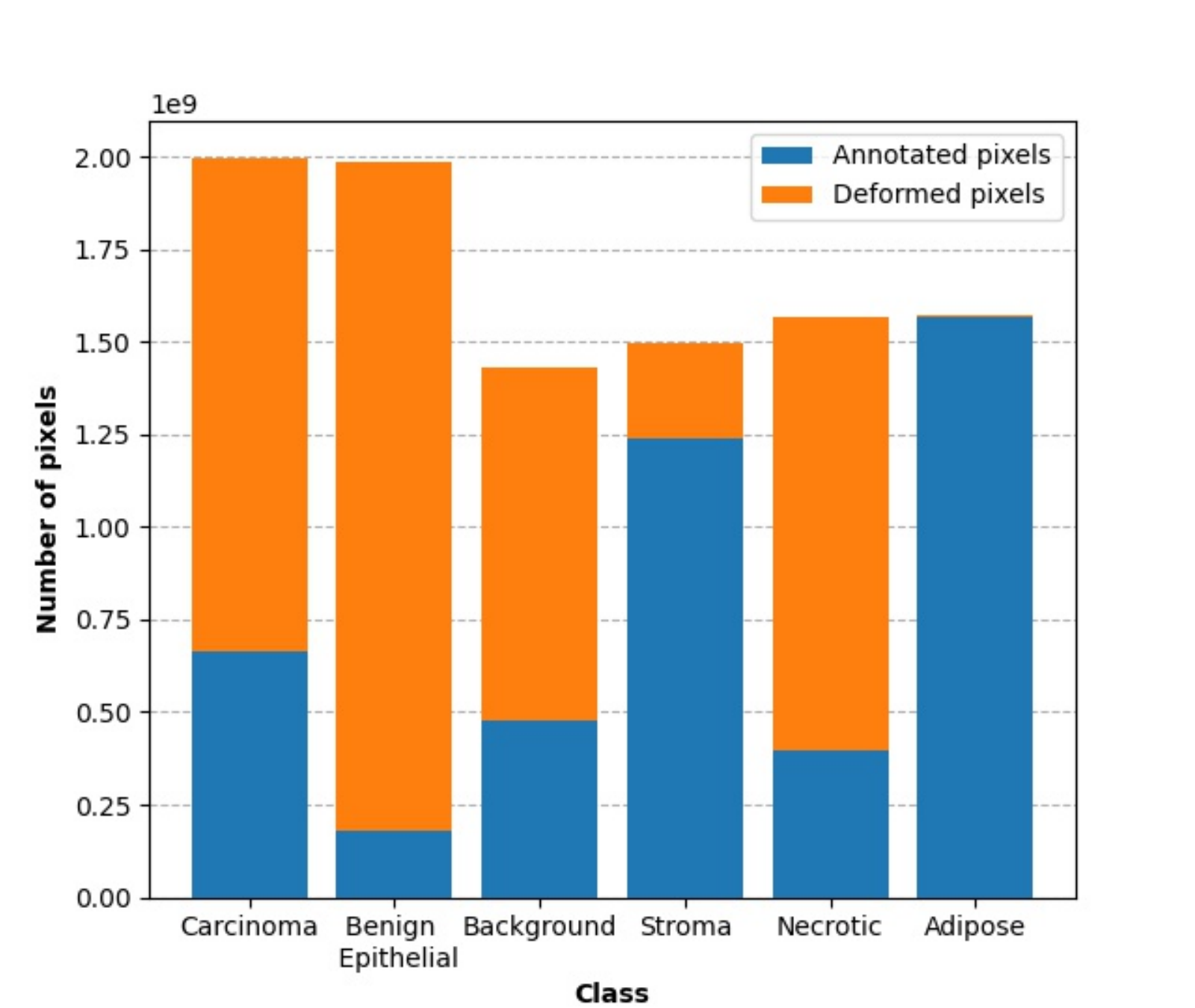,width = 0.8\textwidth}
\caption{Class balancing using elastic deformation in the training breast dataset.}
\label{fig:pixel}
\end{figure}

\begin{figure*}[ht!]
\centering
\subfigure[Image]{\frame{\epsfig{figure=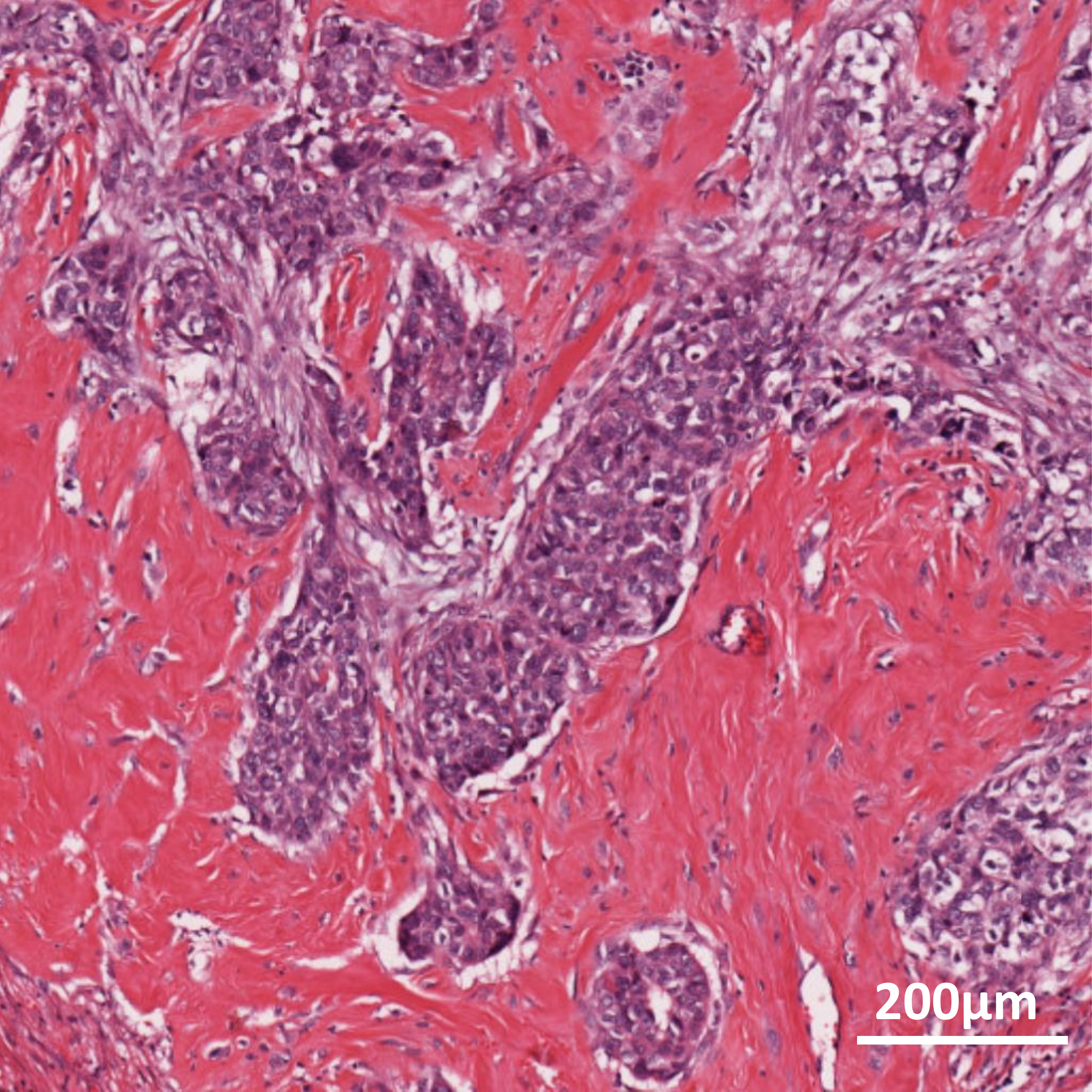,width = 0.32\textwidth}}}
\subfigure[Ground Truth]{\frame{\epsfig{figure=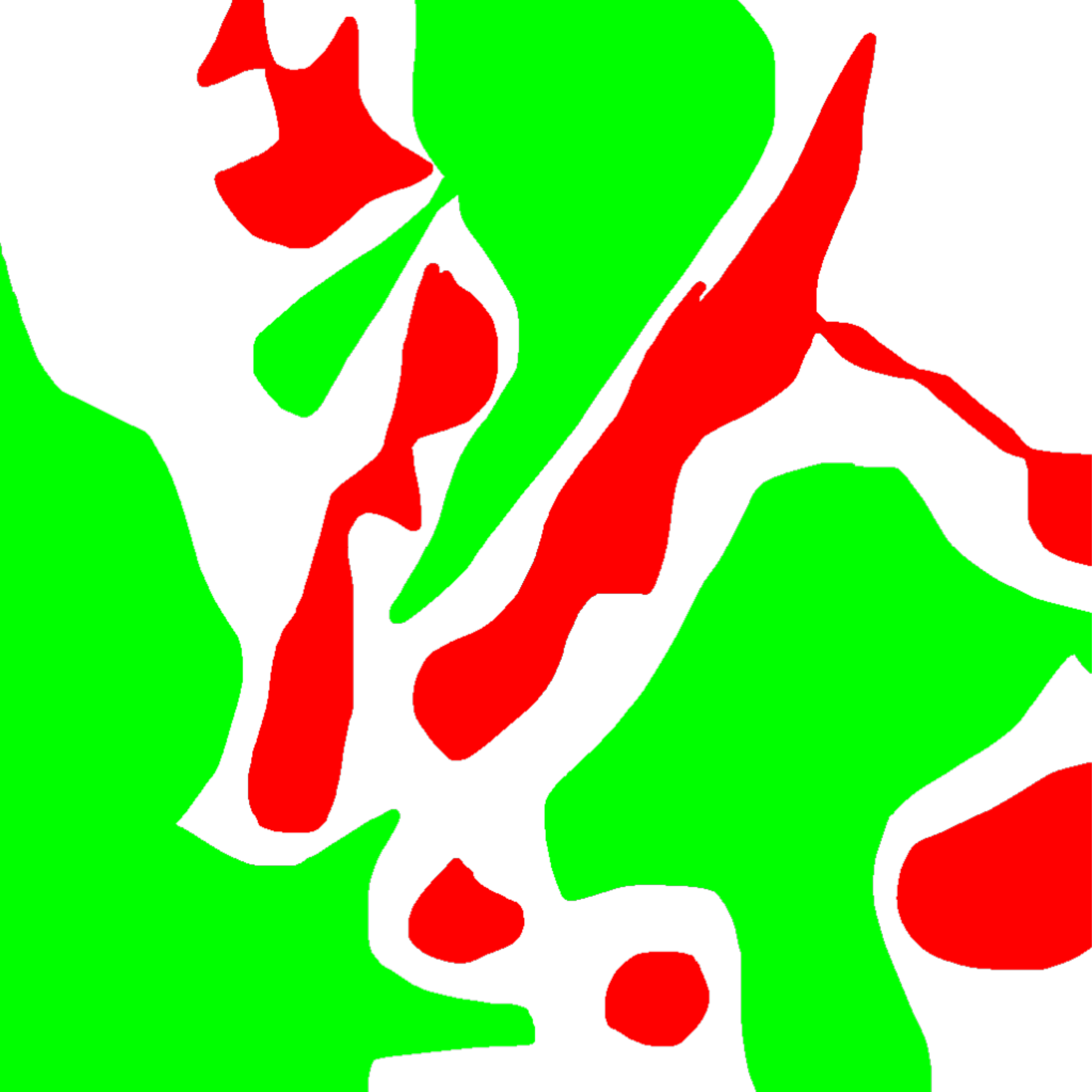,width = 0.32\textwidth}}}
\subfigure[SegNet]{\frame{\epsfig{figure=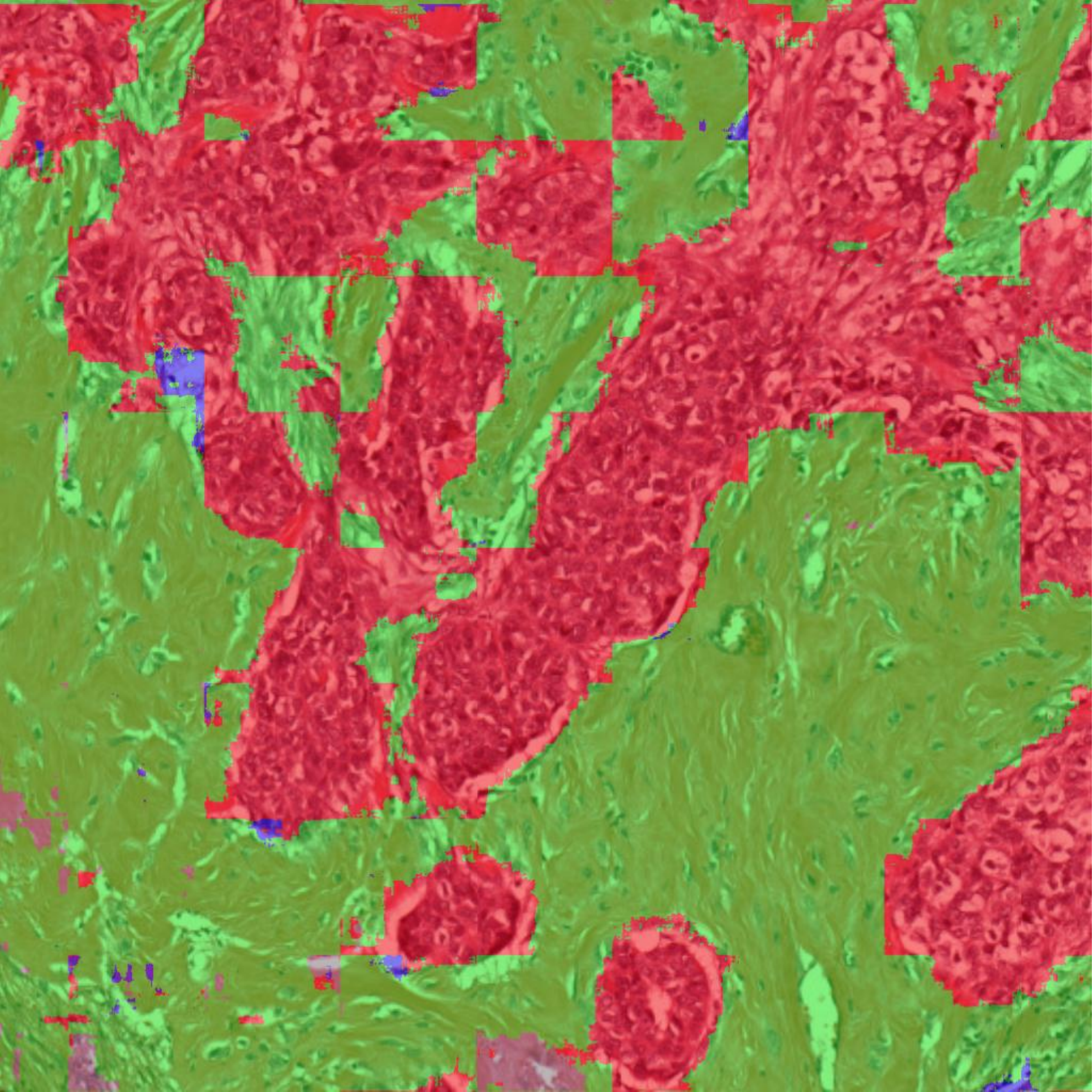,width = 0.32\textwidth}}}

\subfigure[U-Net]{\frame{\epsfig{figure=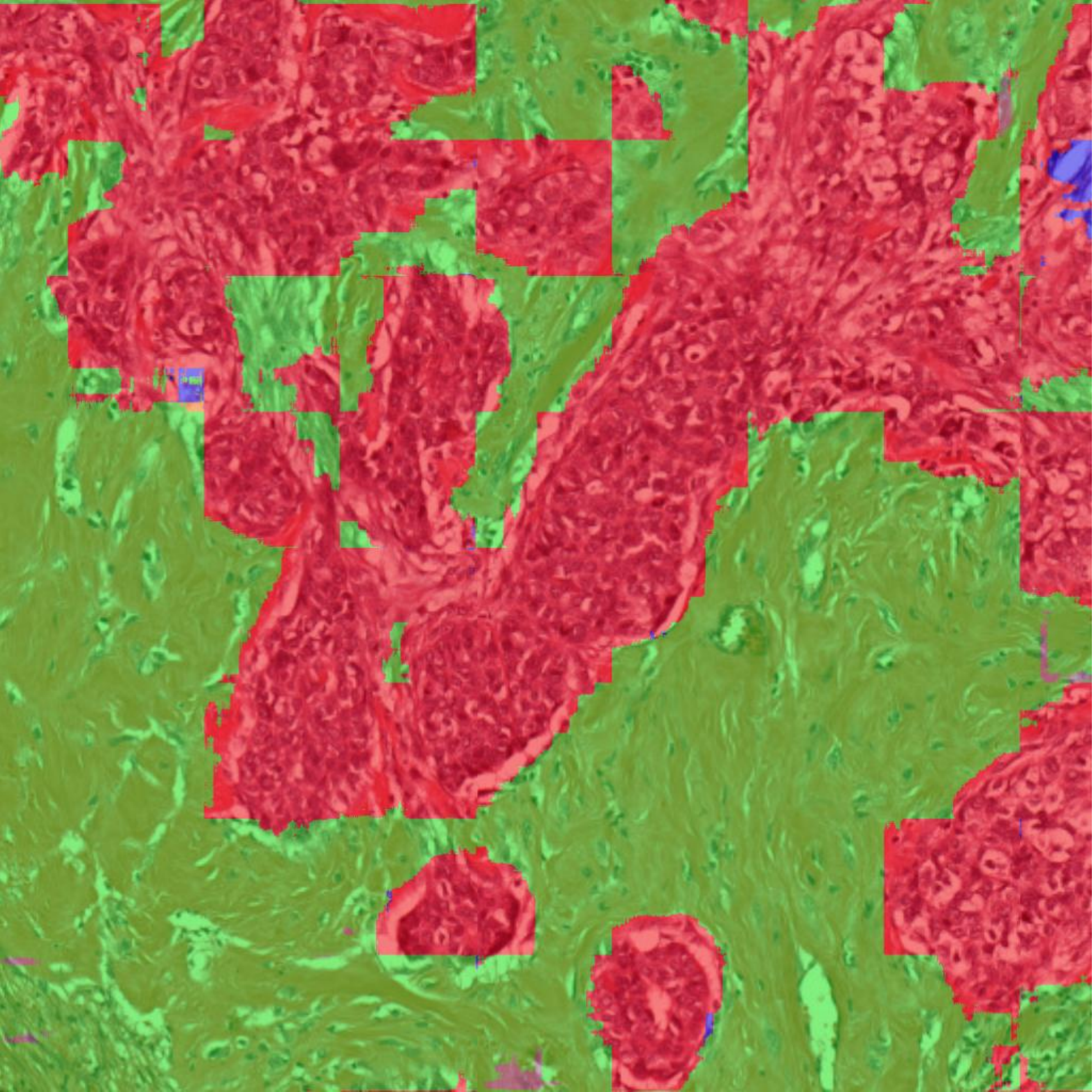,width = 0.32\textwidth}}}
\subfigure[DMMN-S2]{\frame{\epsfig{figure=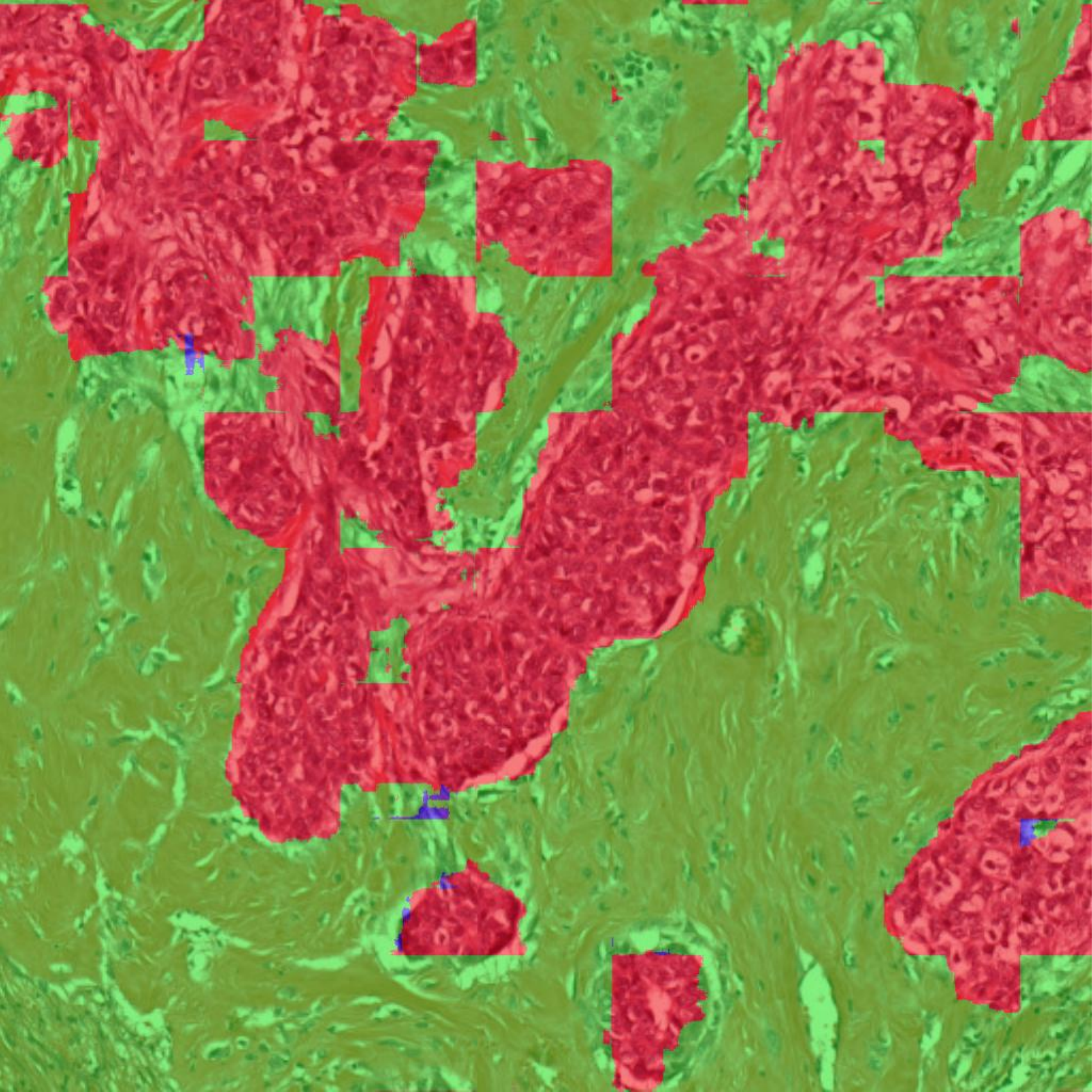,width = 0.32\textwidth}}}
\subfigure[DMMN-MS]{\frame{\epsfig{figure=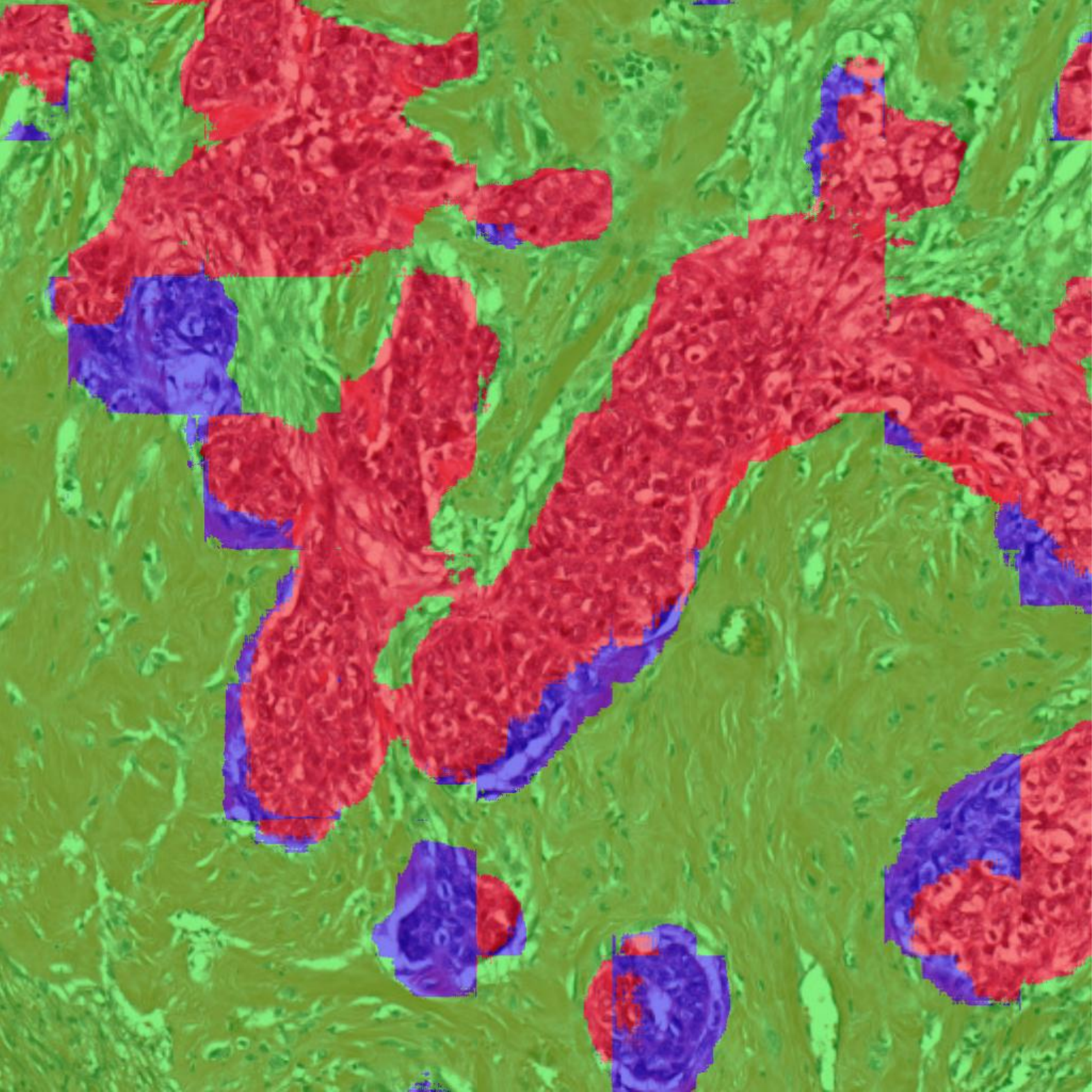,width = 0.32\textwidth}}}

\subfigure[DMMN-M2S]{\frame{\epsfig{figure=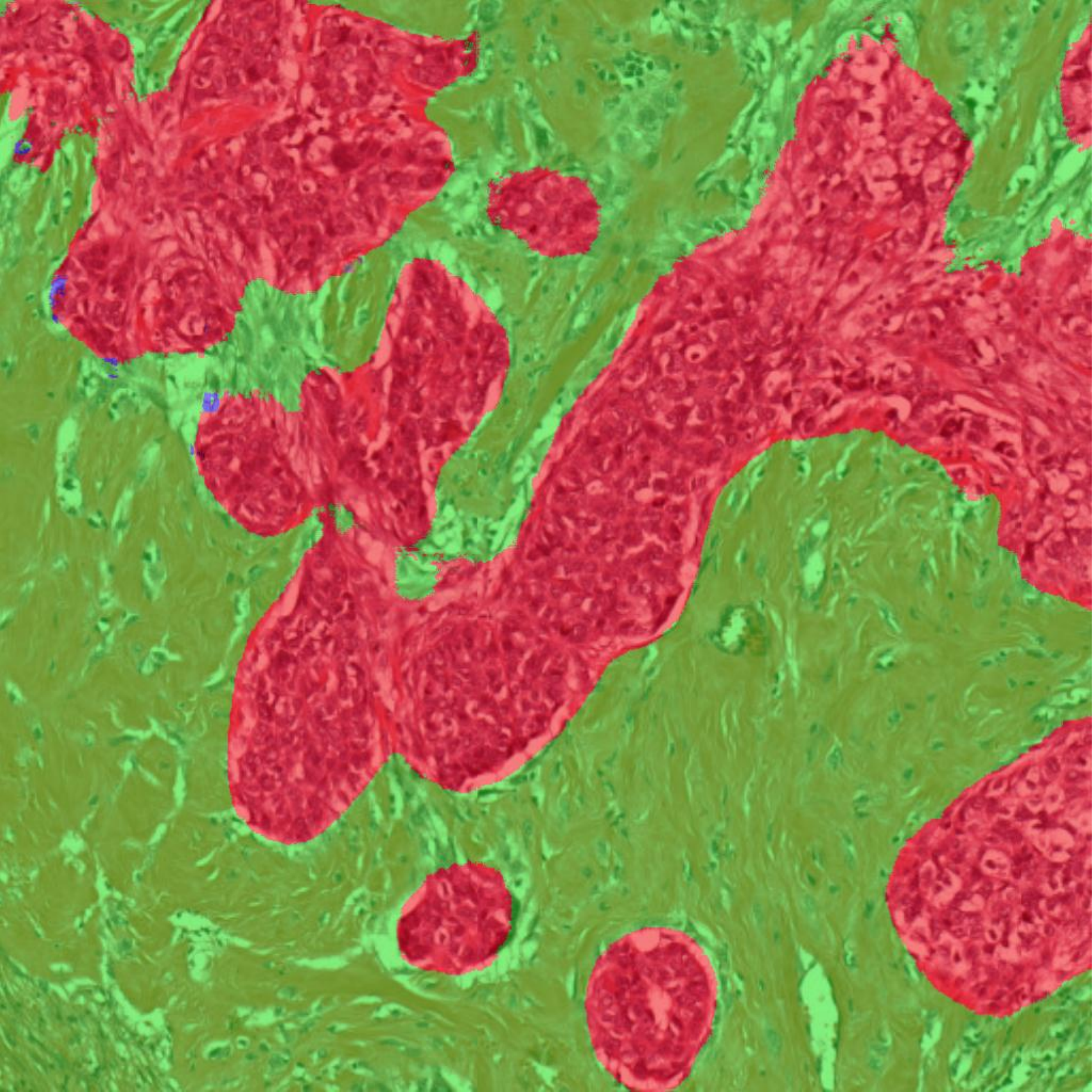,width = 0.32\textwidth}}}
\subfigure[DMMN-M3]{\frame{\epsfig{figure=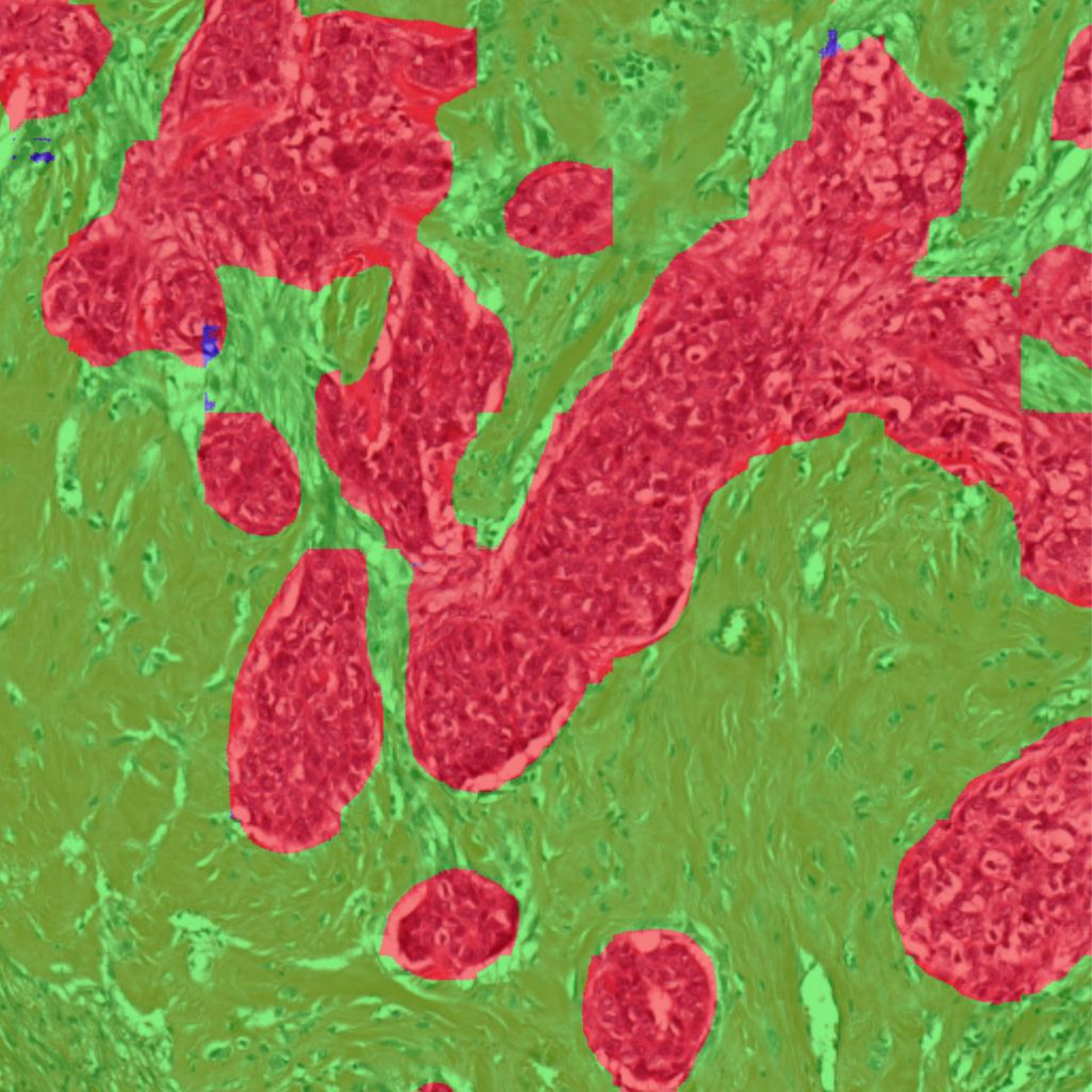,width = 0.32\textwidth}}}
\caption{Multi-class tissue segmentation predictions of invasive ductal carcinoma (IDC) in red from Dataset-I using two Deep Single-Magnification Networks (DSMNs), SegNet \cite{badrinarayanan2017} and U-Net \cite{ronneberger2015}, and four Deep Multi-Magnification Networks (DMMNs), Single-Encoder Single-Decoder (DMMN-S2), Multi-Encoder Single-Decoder (DMMN-MS), Multi-Encoder Multi-Decoder Single-Concatenation (DMMN-M2S), and our proposed Multi-Encoder Multi-Decoder Multi-Concatenation (DMMN-M3).}
\label{fig:393382}
\end{figure*}

Thirty two WSIs from Dataset-I were used to train and validate segmentation models.
The number of whole slide images and the number of patches before and after class balancing to train and validate the models are shown in Table \ref{tab:numbers}.
No images from Dataset-II were used during training.
In our work, only 5.34$\%$ of pixels of training WSIs were annotated.
Our models can predict 6 classes ($C = 6$) which are carcinoma, benign epithelial, background, stroma, necrotic, and adipose.
Note that background is defined as regions which are not tissue.
To balance the number of annotated pixels between classes, we empirically set $r_2 = 10$, $r_1 = 2$, $r_5 = 3$, $r_3 = 1$, $r_4 = 0$, and $r_6 = 0$ where $r_1$, $r_2$, $r_3$, $r_4$, $r_5$, and $r_6$ are rates of elastic deformation of carcinoma, benign epithelial, background, stroma, necrotic, and adipose, respectively.
Benign epithelial was selected as the highest class order followed by carcinoma, necrotic, and background, because we want to accurately segment carcinoma regions and separate benign epithelial to reduce false segmentation.
Figure \ref{fig:pixel} shows the number of annotated pixels between classes are balanced using elastic deformation.
We trained two Deep Single-Magnification Networks (DSMNs), SegNet \cite{badrinarayanan2017} architecture and U-Net \cite{ronneberger2015} architecture, and four Deep Multi-Magnification Networks (DMMNs), Single-Encoder Single-Decoder (DMMN-S2) architecture, Multi-Encoder Single-Decoder (DMMN-MS) architecture, Multi-Encoder Multi-Decoder Single-Concatenation (DMMN-M2S) architecture, and our proposed Multi-Encoder Multi-Decoder Multi-Concatenation (DMMN-M3) architecture.
The number of convolutional layers, the number of downsampling and upsampling layers, and the number of channels are kept the same between the SegNet architecture used in this experiment and the original U-Net architecture.
Also, the number of channels on DMMN-MS, DMMN-M2S, and DMMN-M3 are reduced by a factor of 2 from an original U-Net architecture.
Table \ref{tab:models} lists the models we compared, the number of trainable parameters, and segmentation time, where the segmentation time was measured on a whole slide image in Figure \ref{fig:393382} whose size is 53,711 $\times$ 38,380 pixels with 31,500 patches using a single NVIDIA GeForce GTX TITAN X GPU.

\begin{table}[ht]
\centering
{
\caption{The number of whole slide images and the number of patches before and after class balancing from Dataset-I used to train and validate segmentation models}
\begin{tabular}{| c | c | c | c |}
	\hline
	 & Training & Validation \\
	\hline
	Whole slide images  & 26 & 6 \\
    \hline
    Patches before class balancing & 52,769 & 9,506 \\
    \hline
    Patches after class balancing & 115,844 & 24,119 \\
    \hline
\end{tabular}
\label{tab:numbers}
}
\end{table}
\begin{table}[ht]
\centering
{
\caption{The number of trainable parameters and computational time for multi-class segmentation models}
\begin{tabular}{| c | c | c | c |}
	\hline
	Model & Trainable Parameters & Segmentation Time \\
	\hline
	SegNet \cite{badrinarayanan2017} & 18,881,543 & 7 min 48 sec \\
    \hline
    U-Net \cite{ronneberger2015} & 34,550,663 & 12 min 50 sec \\
    \hline
    DMMN-S2 & 34,554,119 & 13 min 16 sec \\
    \hline
    DMMN-MS & 30,647,207 & 13 min 6 sec \\
    \hline
    DMMN-M2S & 25,947,047 & 16 min 21 sec \\
    \hline
    DMMN-M3 & 27,071,303 & 14 min 52 sec \\
    \hline
\end{tabular}
\label{tab:models}
}
\end{table}

\begin{figure*}[ht!]
\centering
\subfigure[Image]{\frame{\epsfig{figure=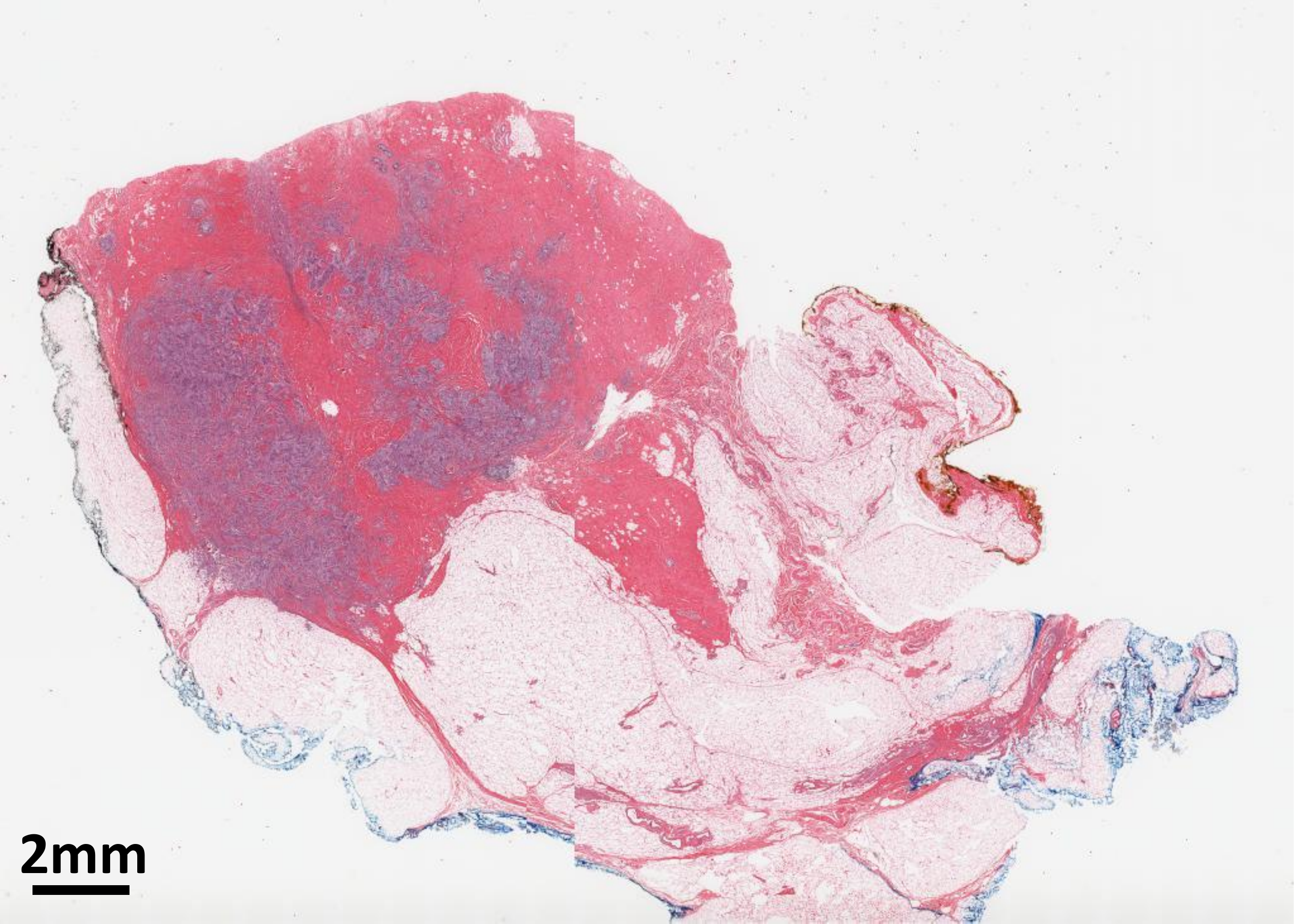,width = 0.32\textwidth}}}
\subfigure[Ground Truth]{\frame{\epsfig{figure=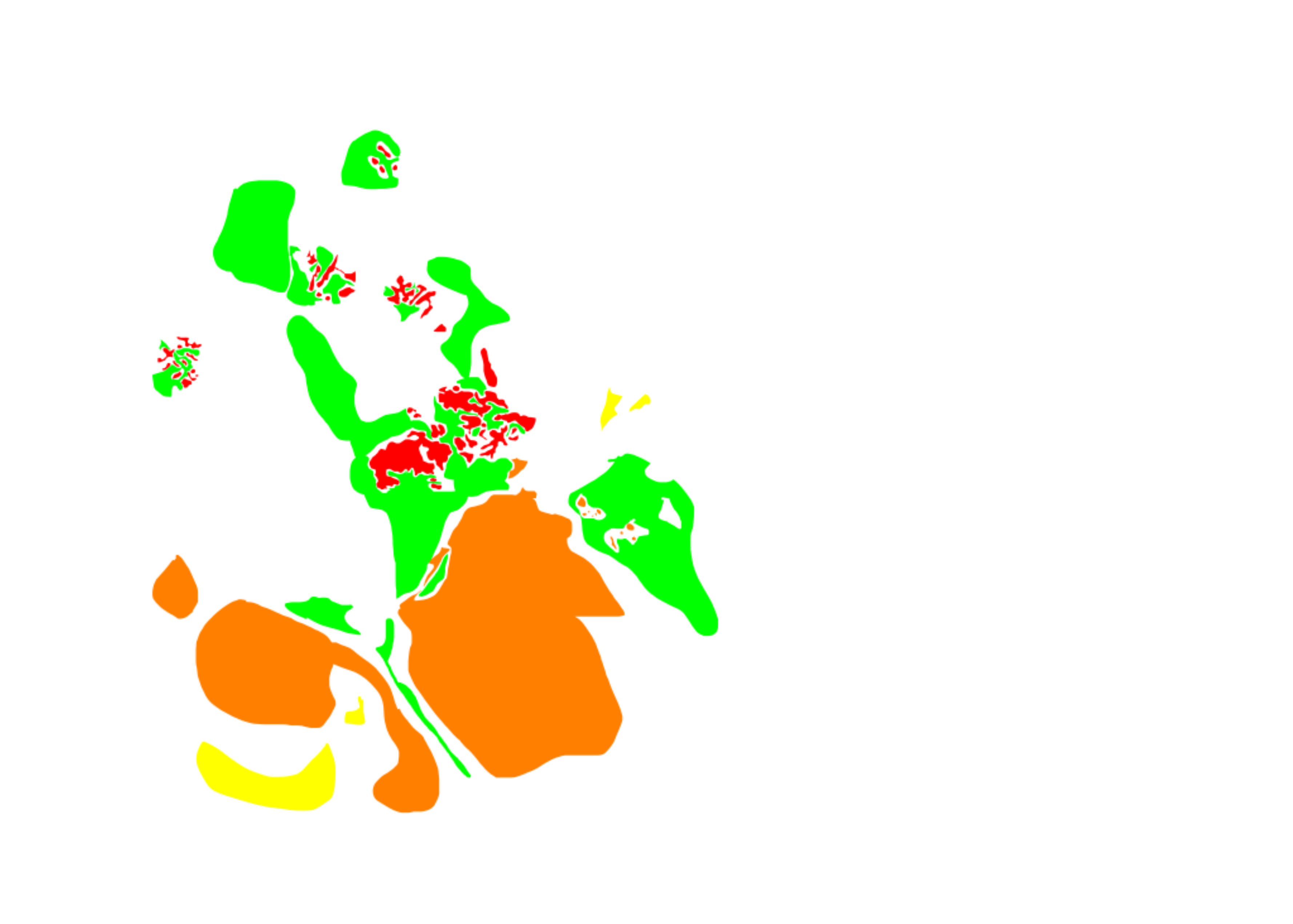,width = 0.32\textwidth}}}
\subfigure[SegNet]{\frame{\epsfig{figure=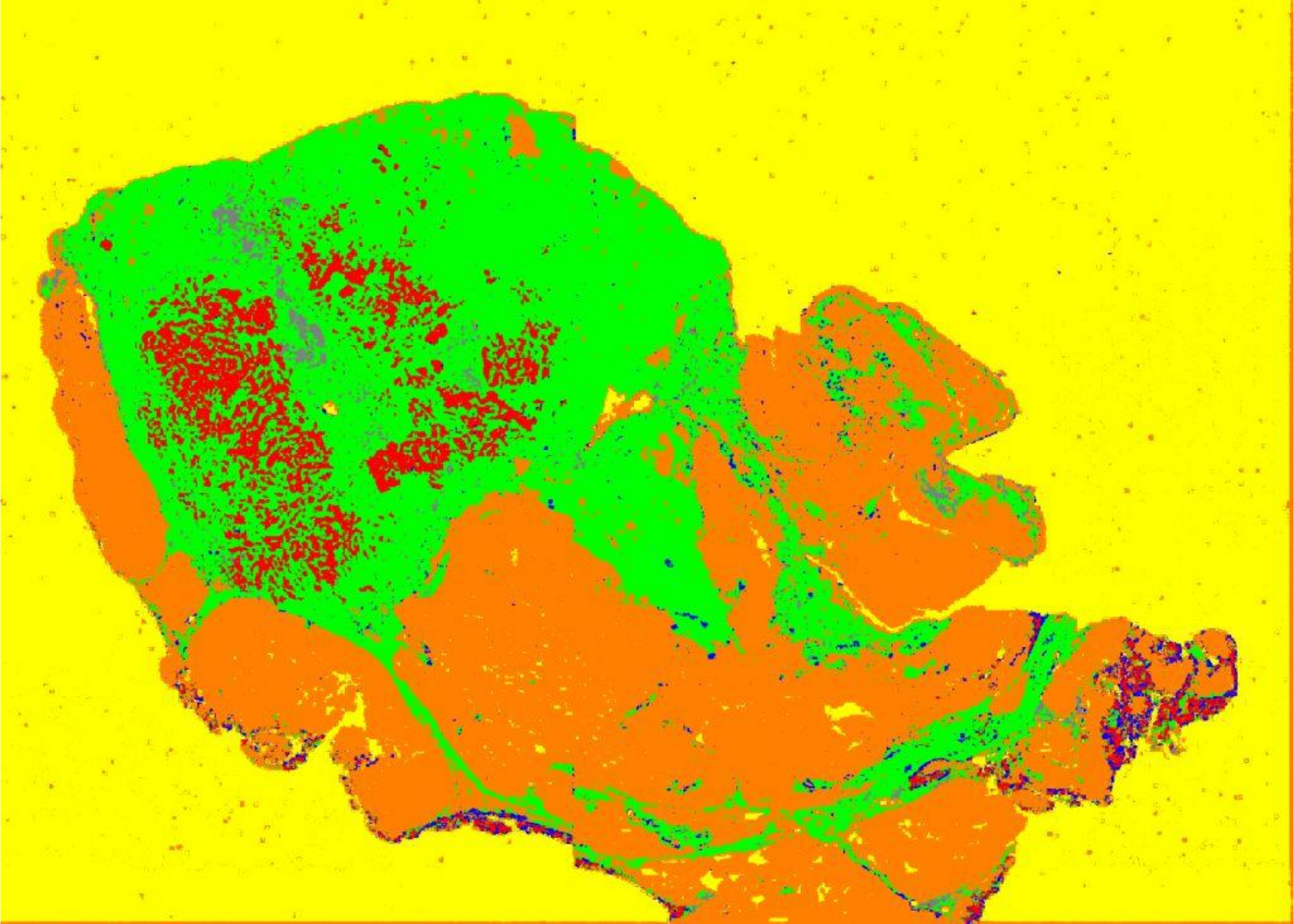,width = 0.32\textwidth}}}

\subfigure[U-Net]{\frame{\epsfig{figure=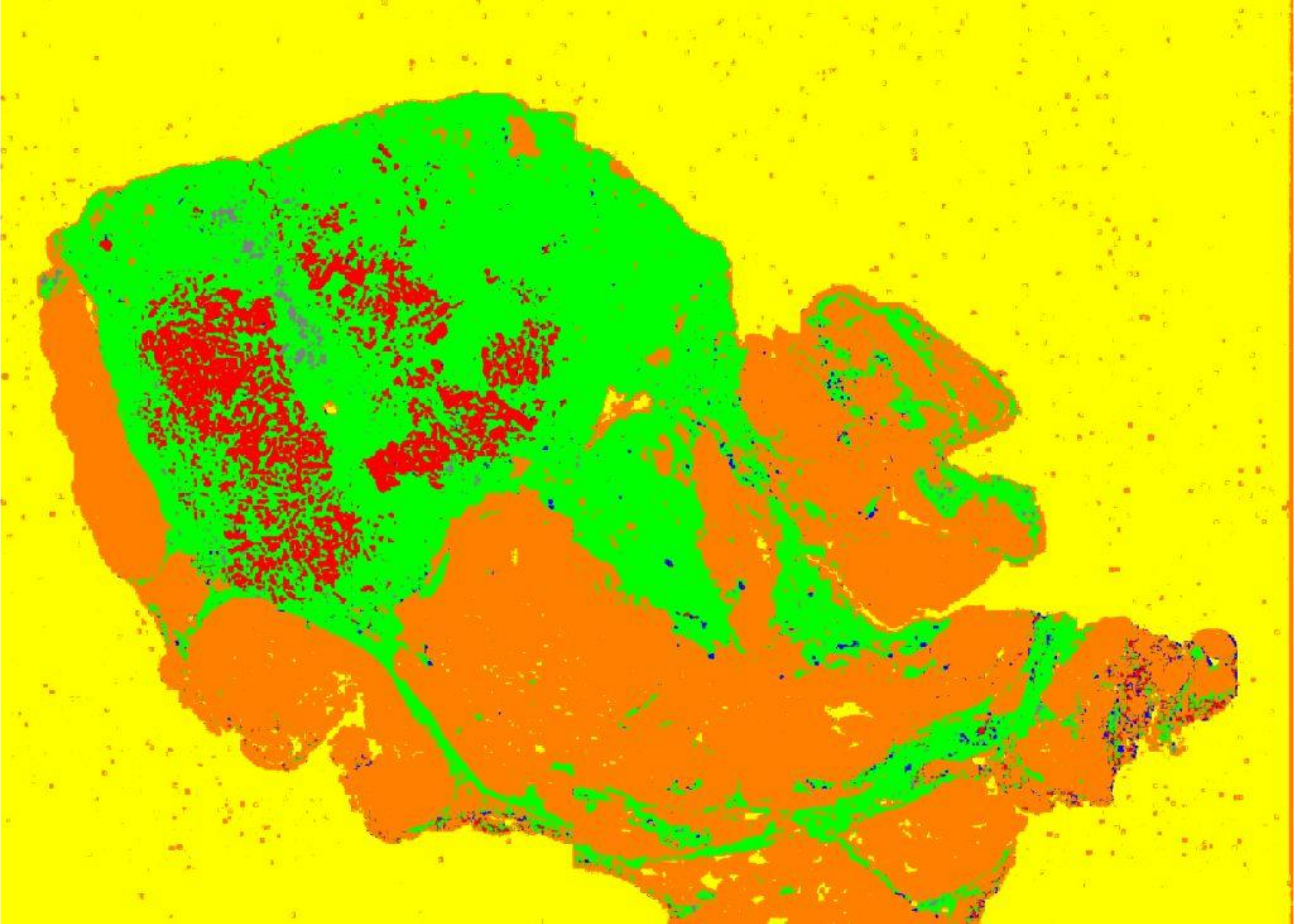,width = 0.32\textwidth}}}
\subfigure[DMMN-S2]{\frame{\epsfig{figure=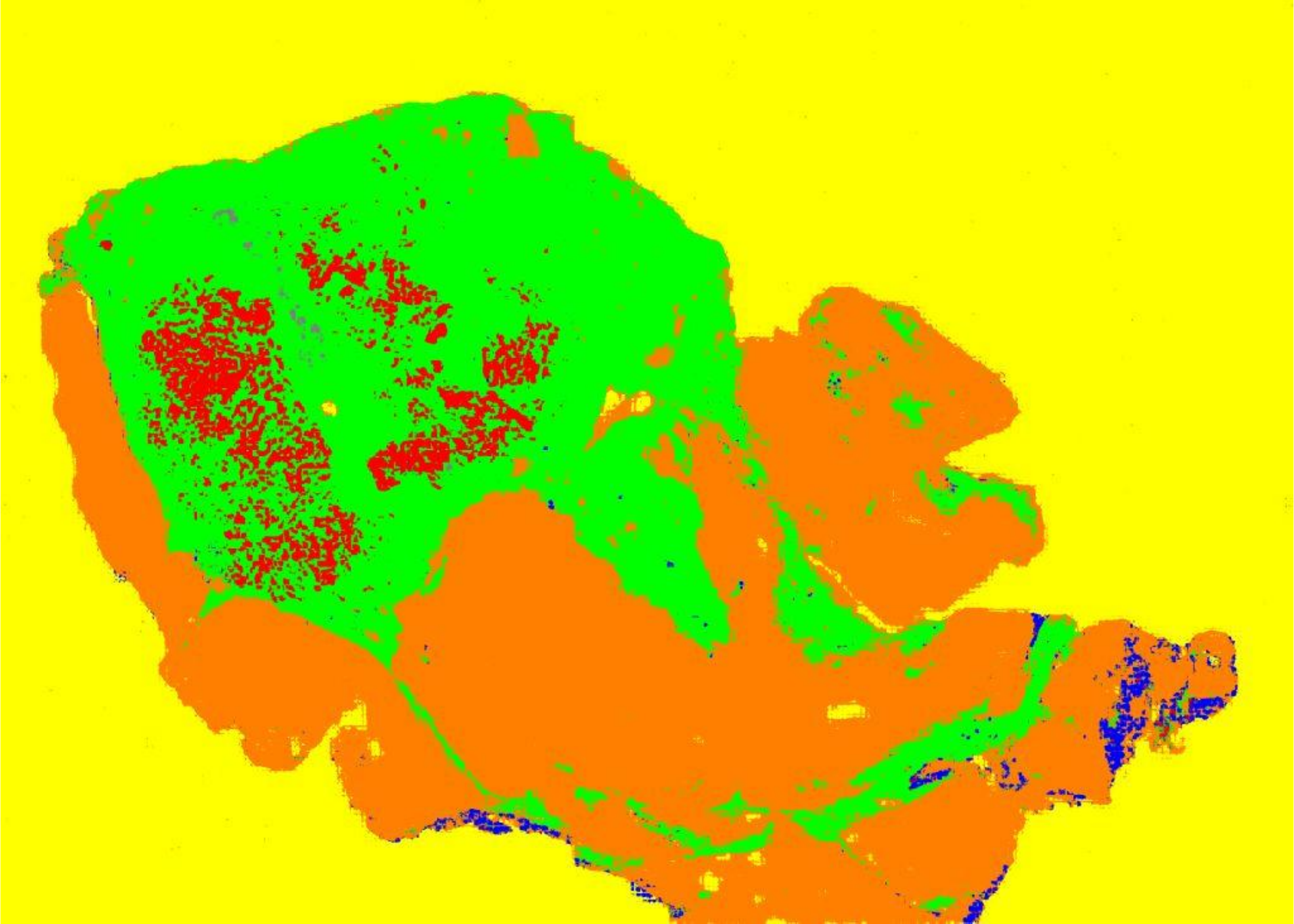,width = 0.32\textwidth}}}
\subfigure[DMMN-MS]{\frame{\epsfig{figure=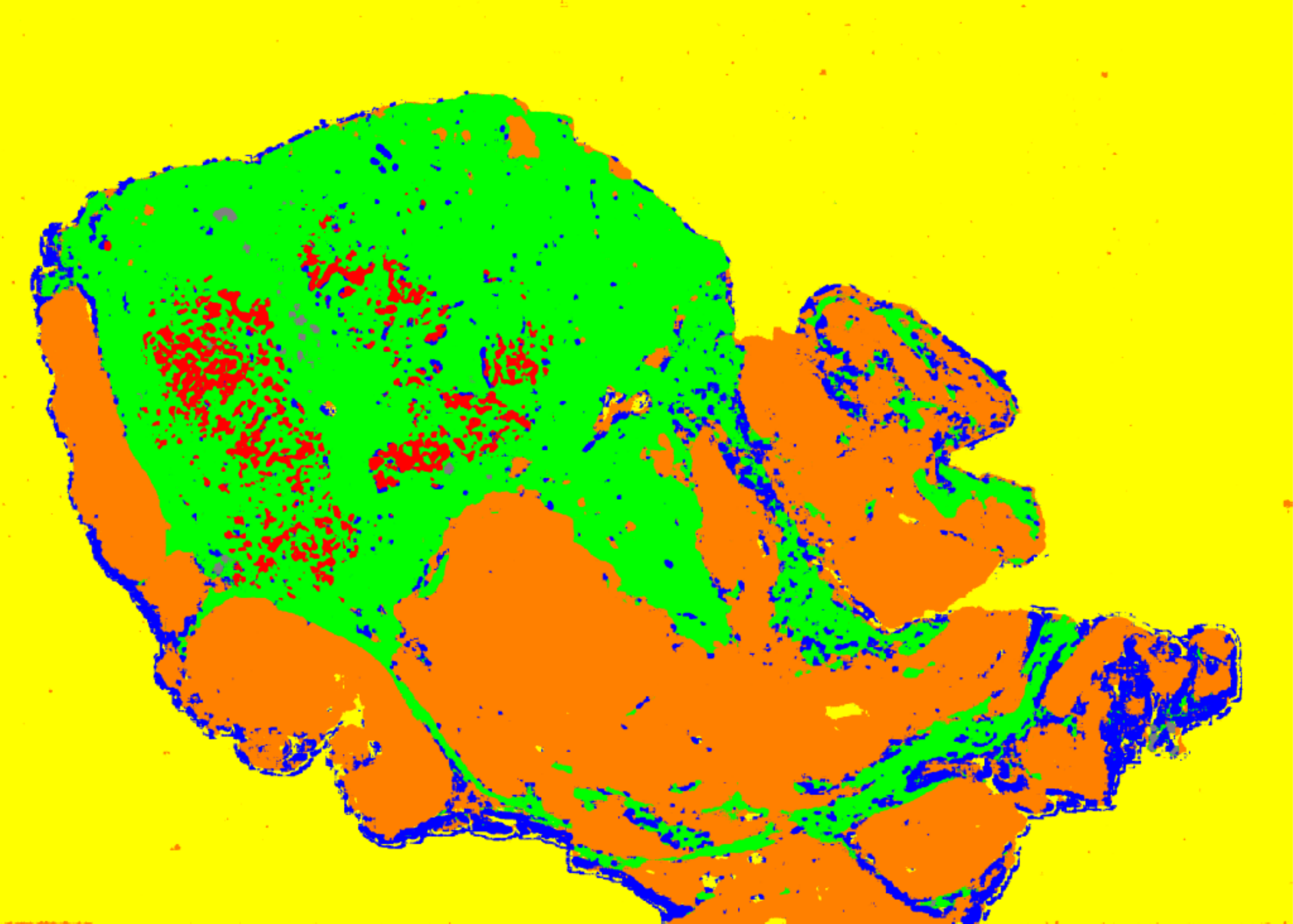,width = 0.32\textwidth}}}

\subfigure[DMMN-M2S]{\frame{\epsfig{figure=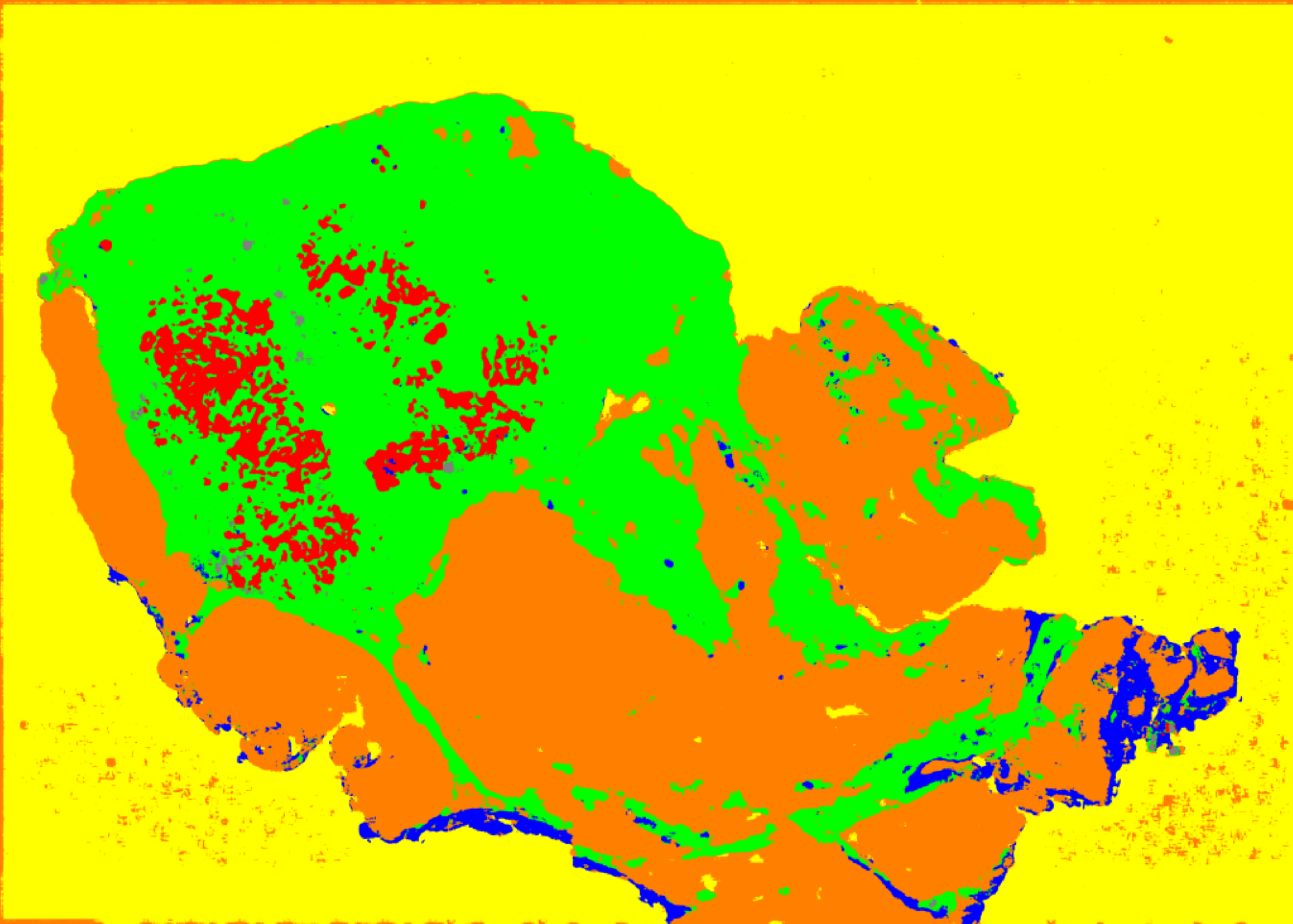,width = 0.32\textwidth}}}
\subfigure[DMMN-M3]{\frame{\epsfig{figure=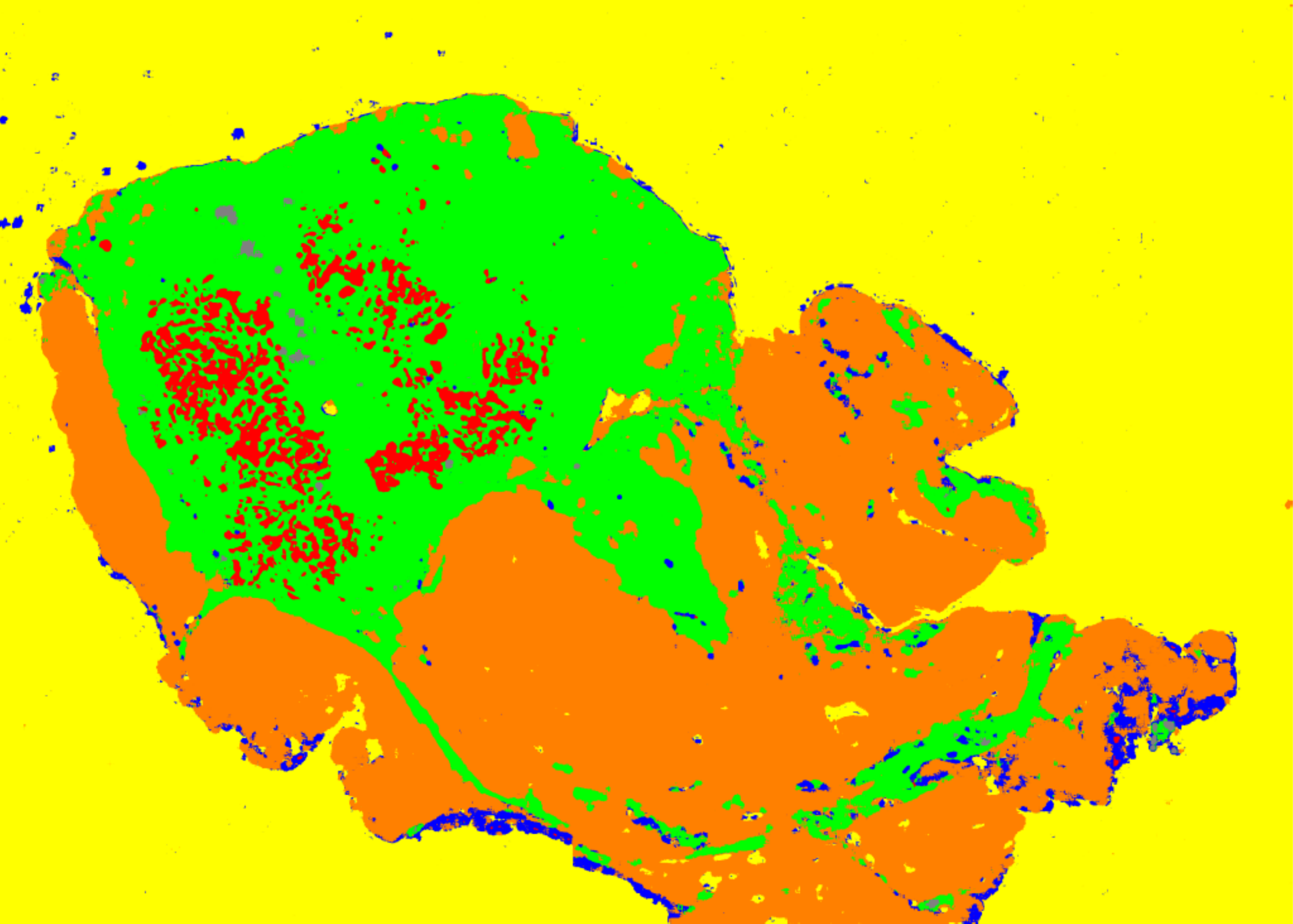,width = 0.32\textwidth}}}
\caption{Multi-class tissue segmentation predictions of a whole slide image from Dataset-I using two Deep Single-Magnification Networks (DSMNs), SegNet \cite{badrinarayanan2017} and U-Net \cite{ronneberger2015}, and four Deep Multi-Magnification Networks (DMMNs), Single-Encoder Single-Decoder (DMMN-S2), Multi-Encoder Single-Decoder (DMMN-MS), Multi-Encoder Multi-Decoder Single-Concatenation (DMMN-M2S), and our proposed Multi-Encoder Multi-Decoder Multi-Concatenation (DMMN-M3).}
\label{fig:393382_WSI}
\end{figure*}

\begin{figure*}[ht!]
\centering
\subfigure[Image]{\frame{\epsfig{figure=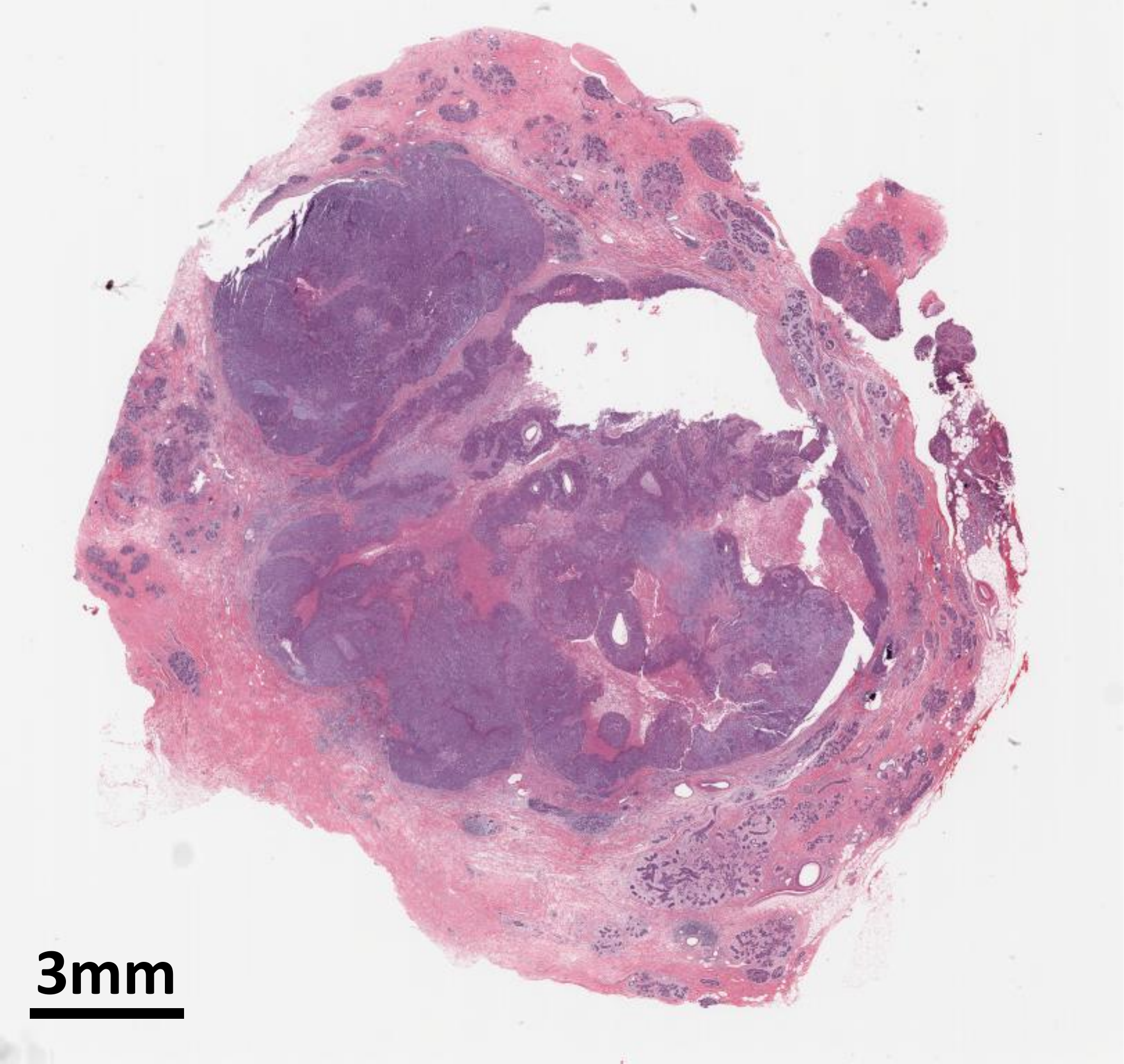,width = 0.32\textwidth}}}
\subfigure[Ground Truth]{\frame{\epsfig{figure=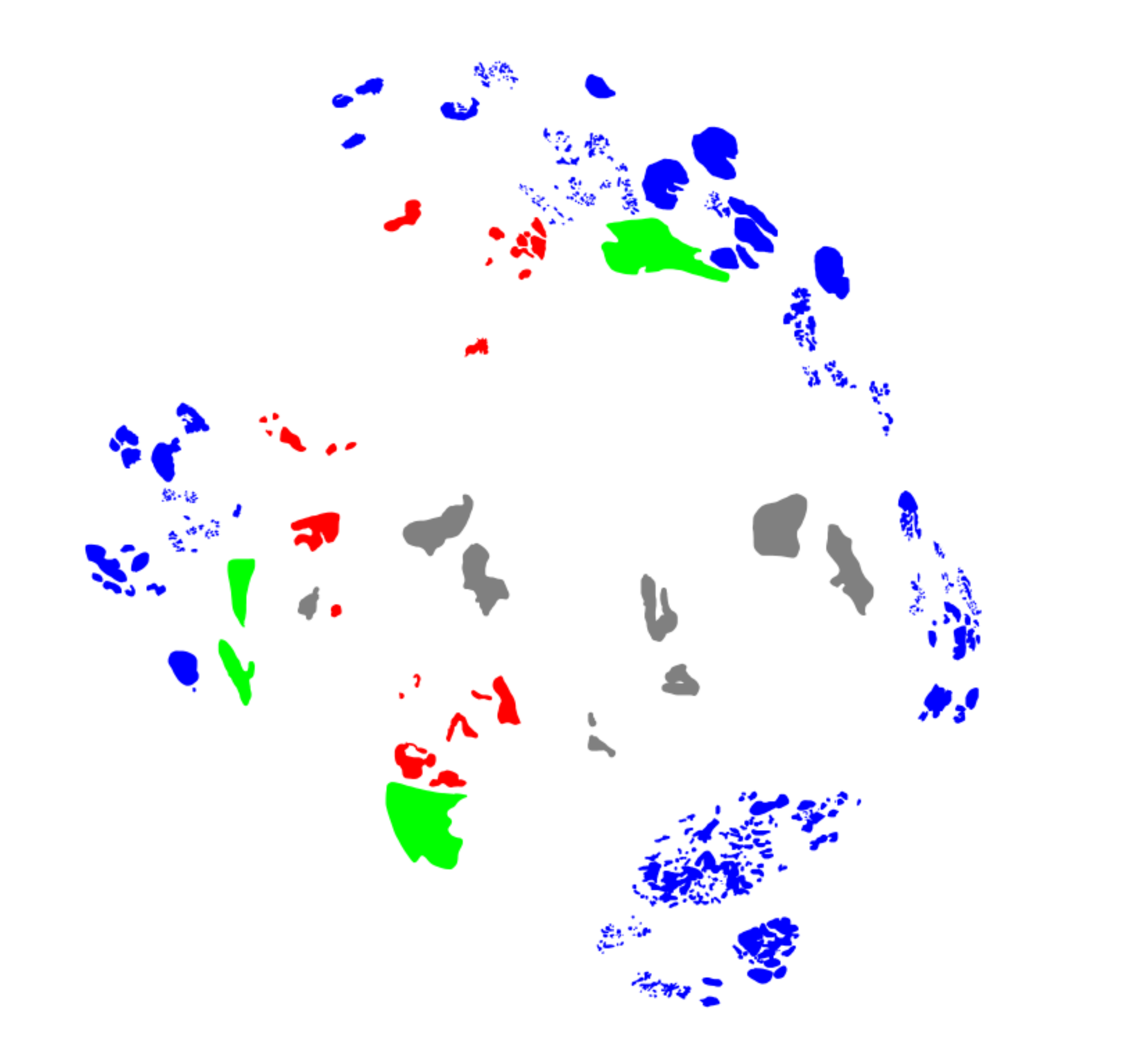,width = 0.32\textwidth}}}
\subfigure[SegNet]{\frame{\epsfig{figure=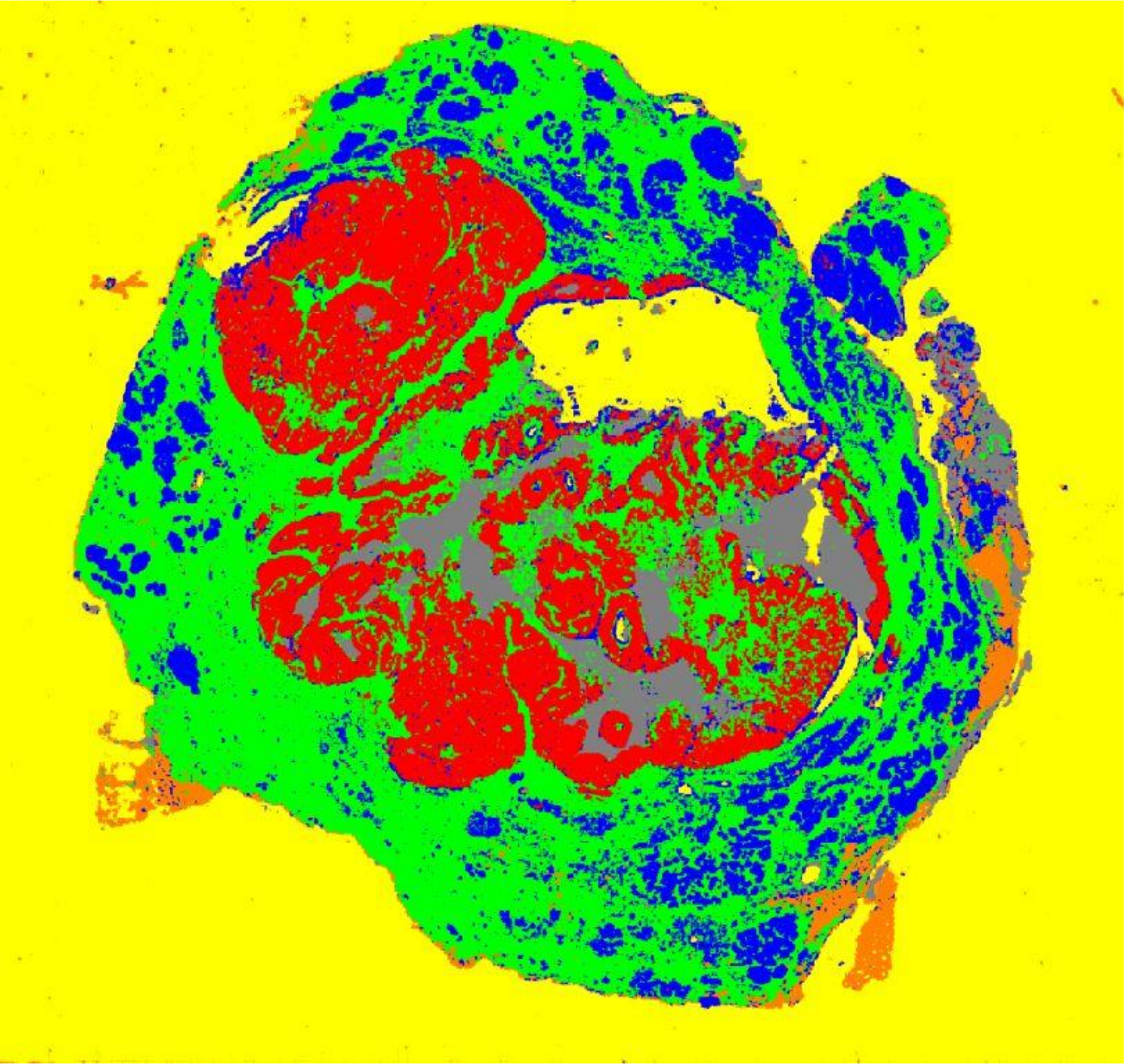,width = 0.32\textwidth}}}

\subfigure[U-Net]{\frame{\epsfig{figure=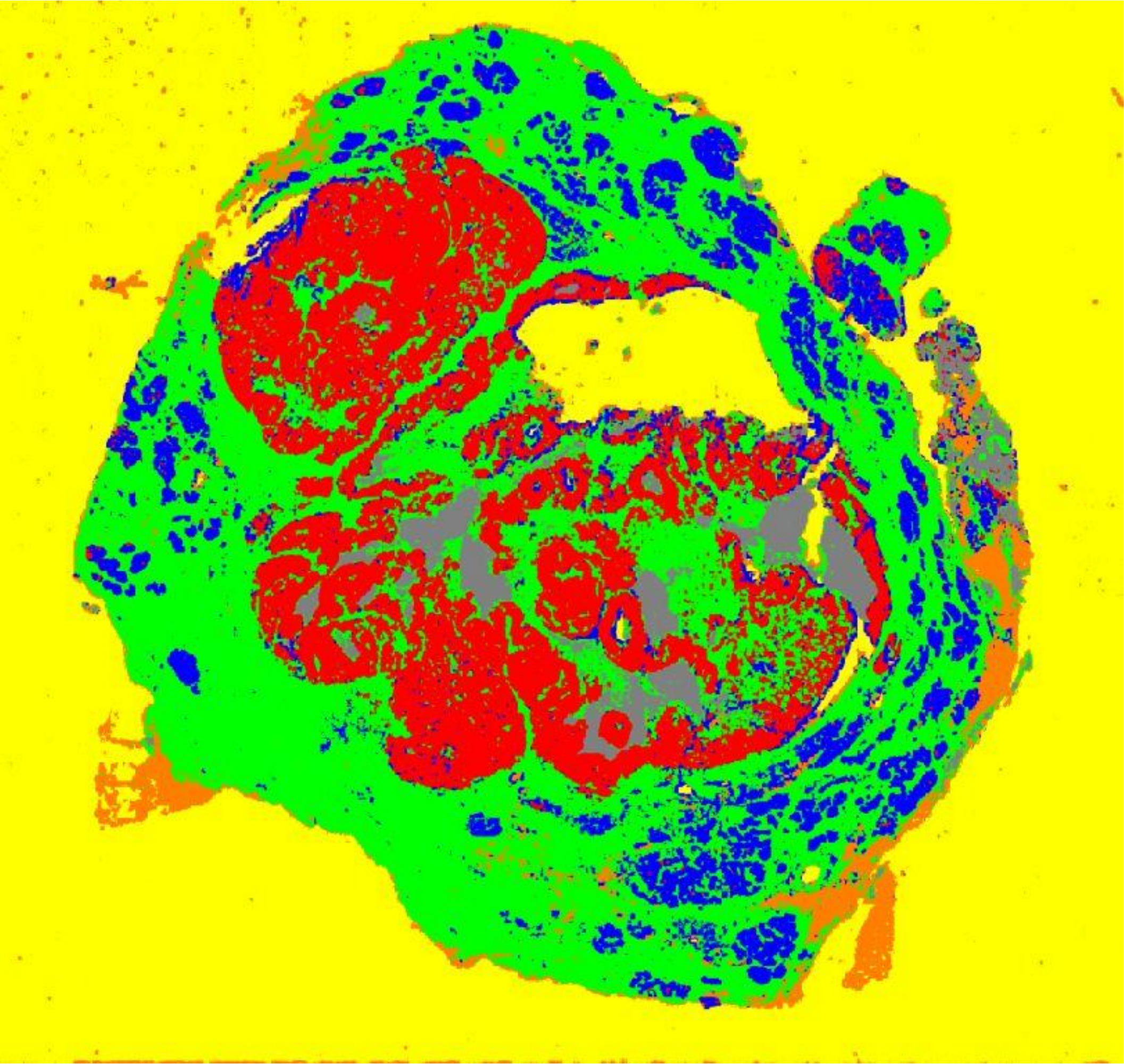,width = 0.32\textwidth}}}
\subfigure[DMMN-S2]{\frame{\epsfig{figure=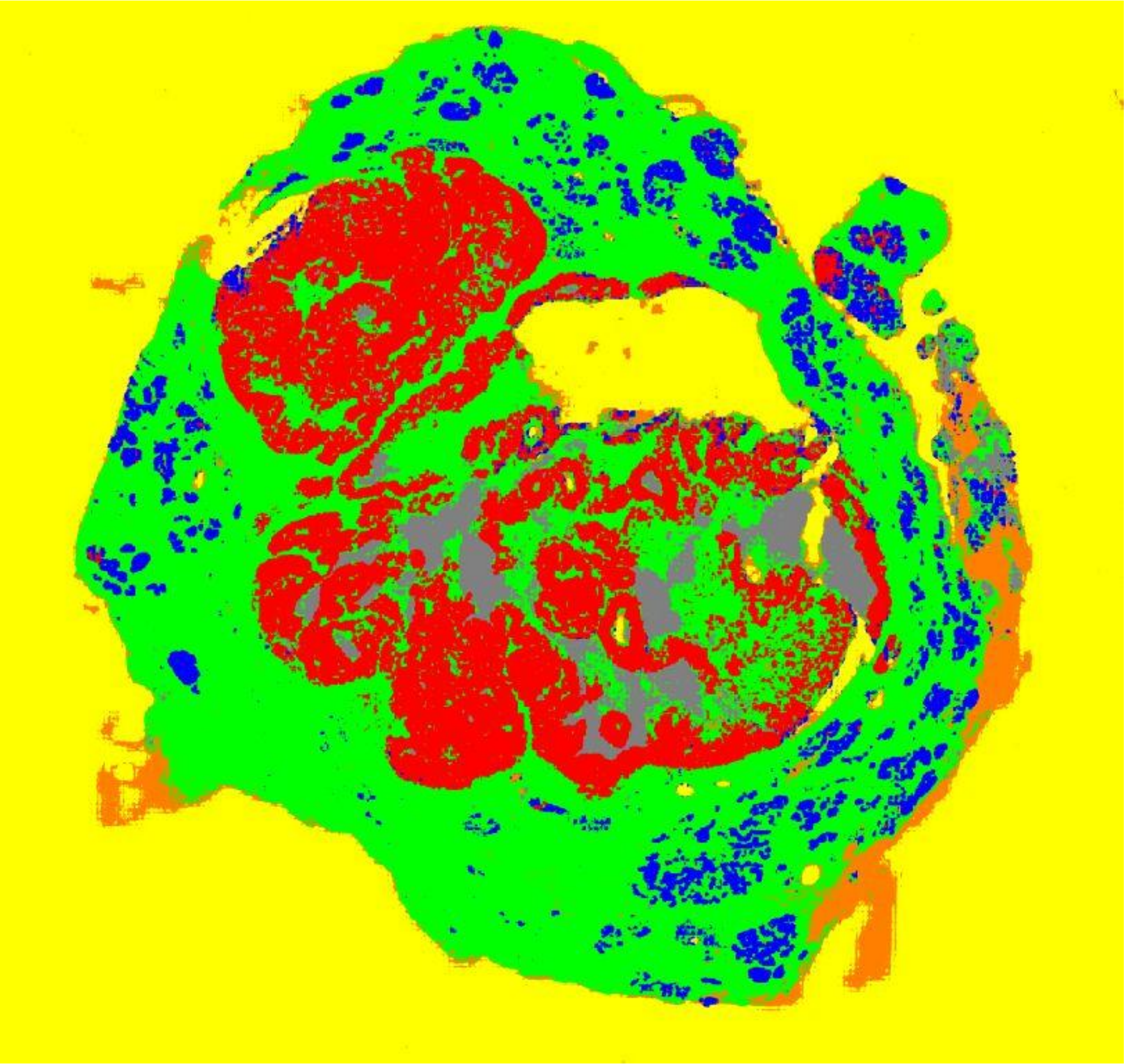,width = 0.32\textwidth}}}
\subfigure[DMMN-MS]{\frame{\epsfig{figure=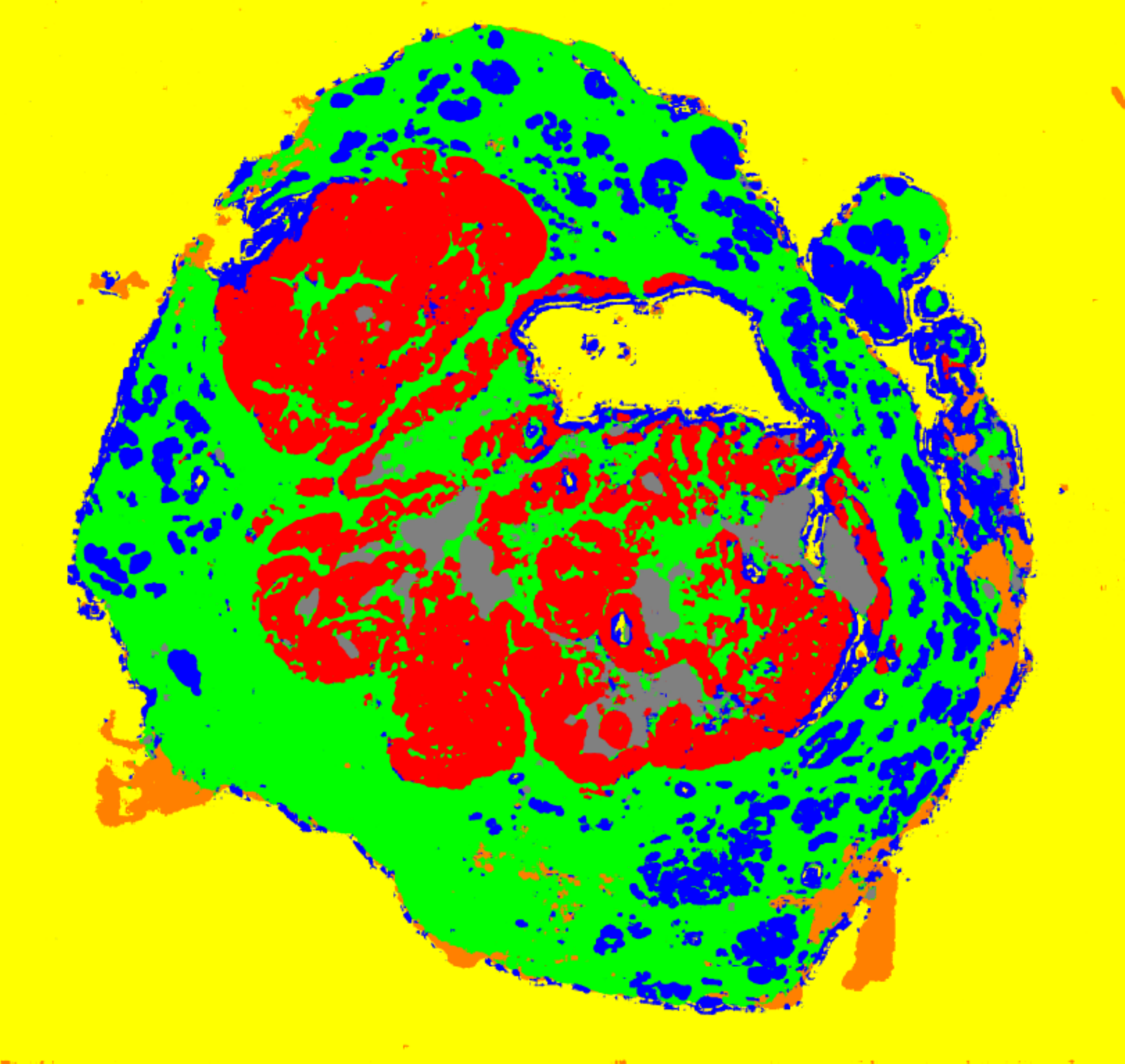,width = 0.32\textwidth}}}

\subfigure[DMMN-M2S]{\frame{\epsfig{figure=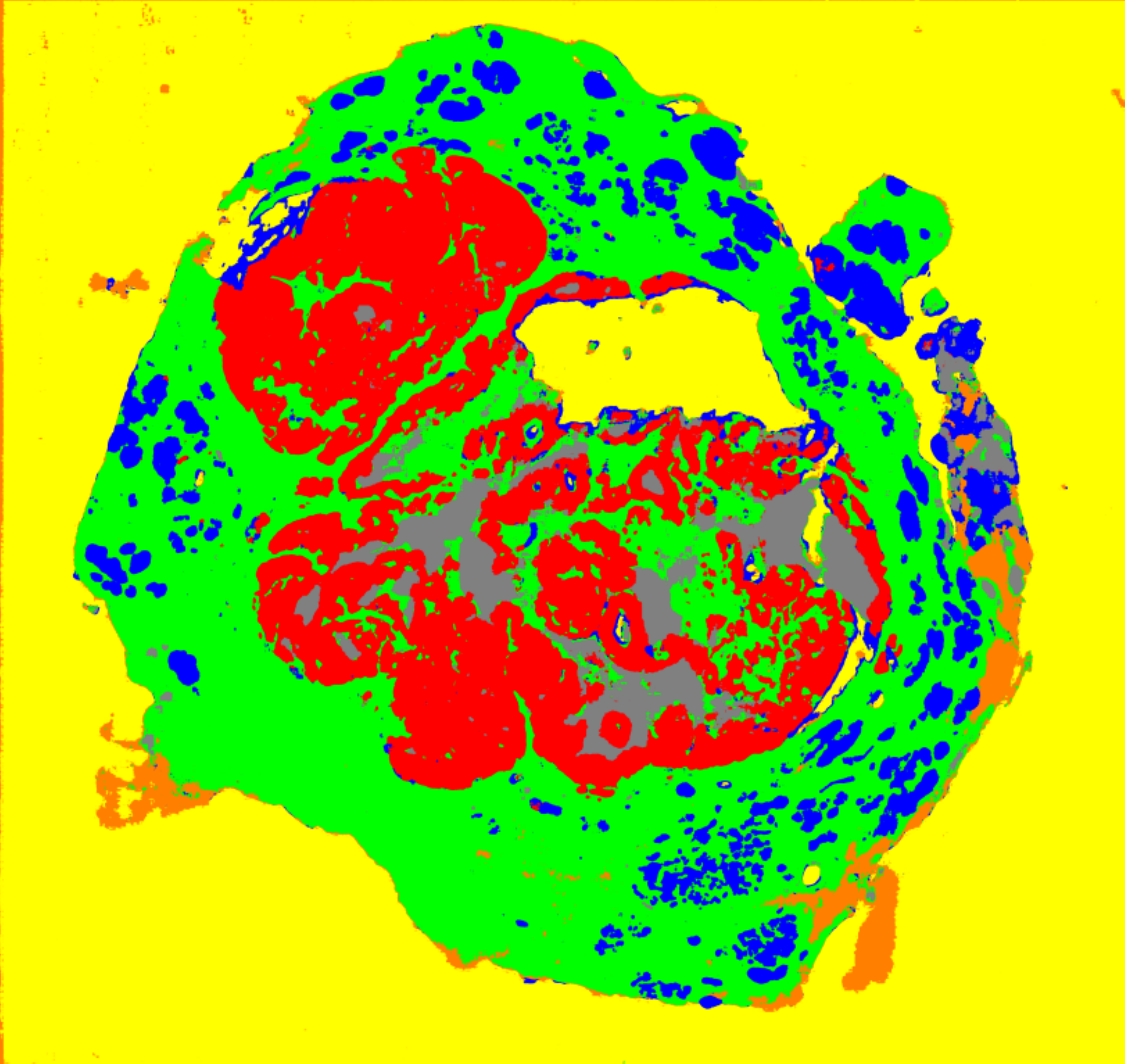,width = 0.32\textwidth}}}
\subfigure[DMMN-M3]{\frame{\epsfig{figure=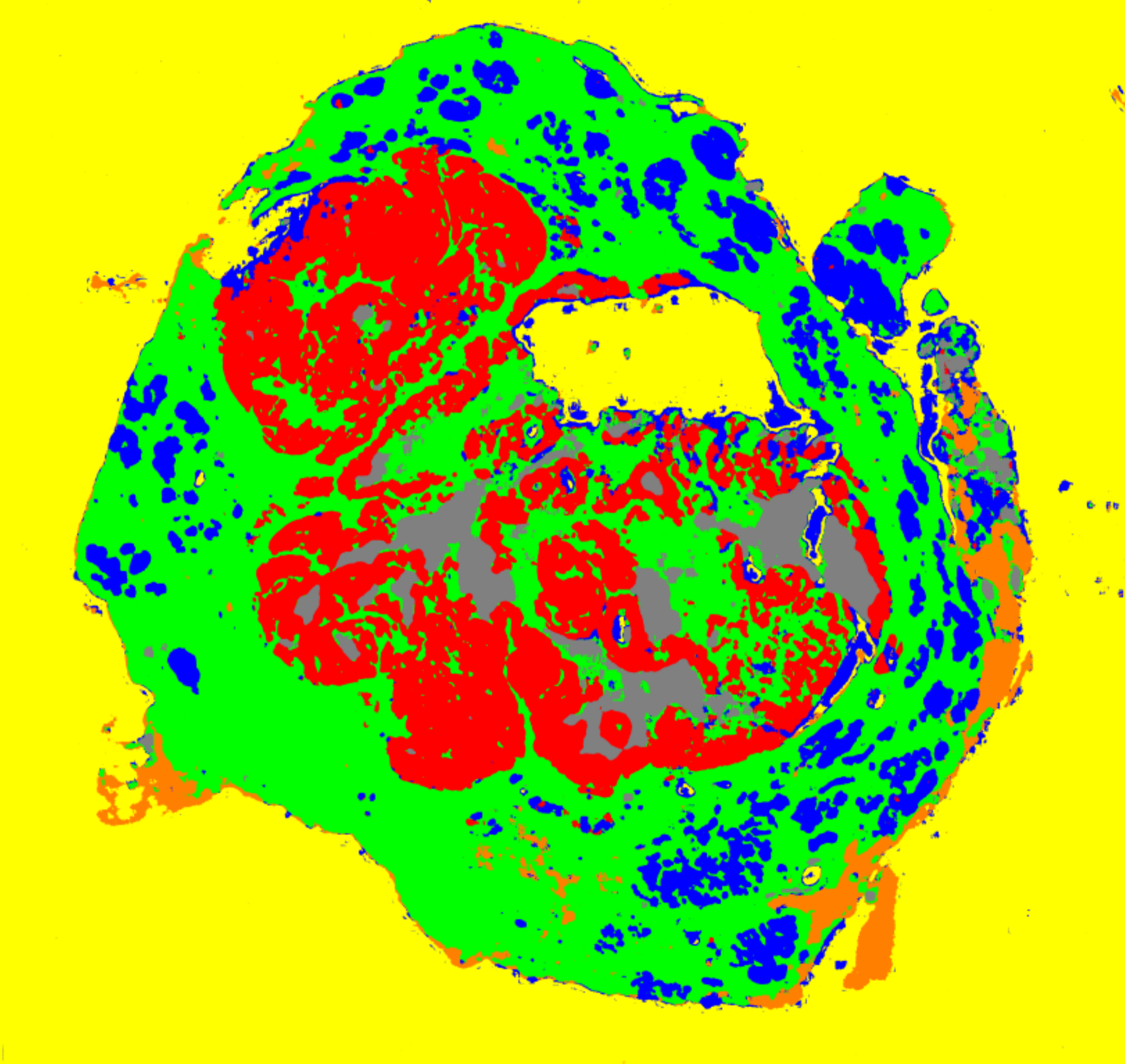,width = 0.32\textwidth}}}
\caption{Multi-class tissue segmentation predictions of a whole slide image from Dataset-I using two Deep Single-Magnification Networks (DSMNs), SegNet \cite{badrinarayanan2017} and U-Net \cite{ronneberger2015}, and four Deep Multi-Magnification Networks (DMMNs), Single-Encoder Single-Decoder (DMMN-S2), Multi-Encoder Single-Decoder (DMMN-MS), Multi-Encoder Multi-Decoder Single-Concatenation (DMMN-M2S), and our proposed Multi-Encoder Multi-Decoder Multi-Concatenation (DMMN-M3).}
\label{fig:393867_WSI}
\end{figure*}

\begin{figure*}[ht!]
\centering
\subfigure[Image]{\frame{\epsfig{figure=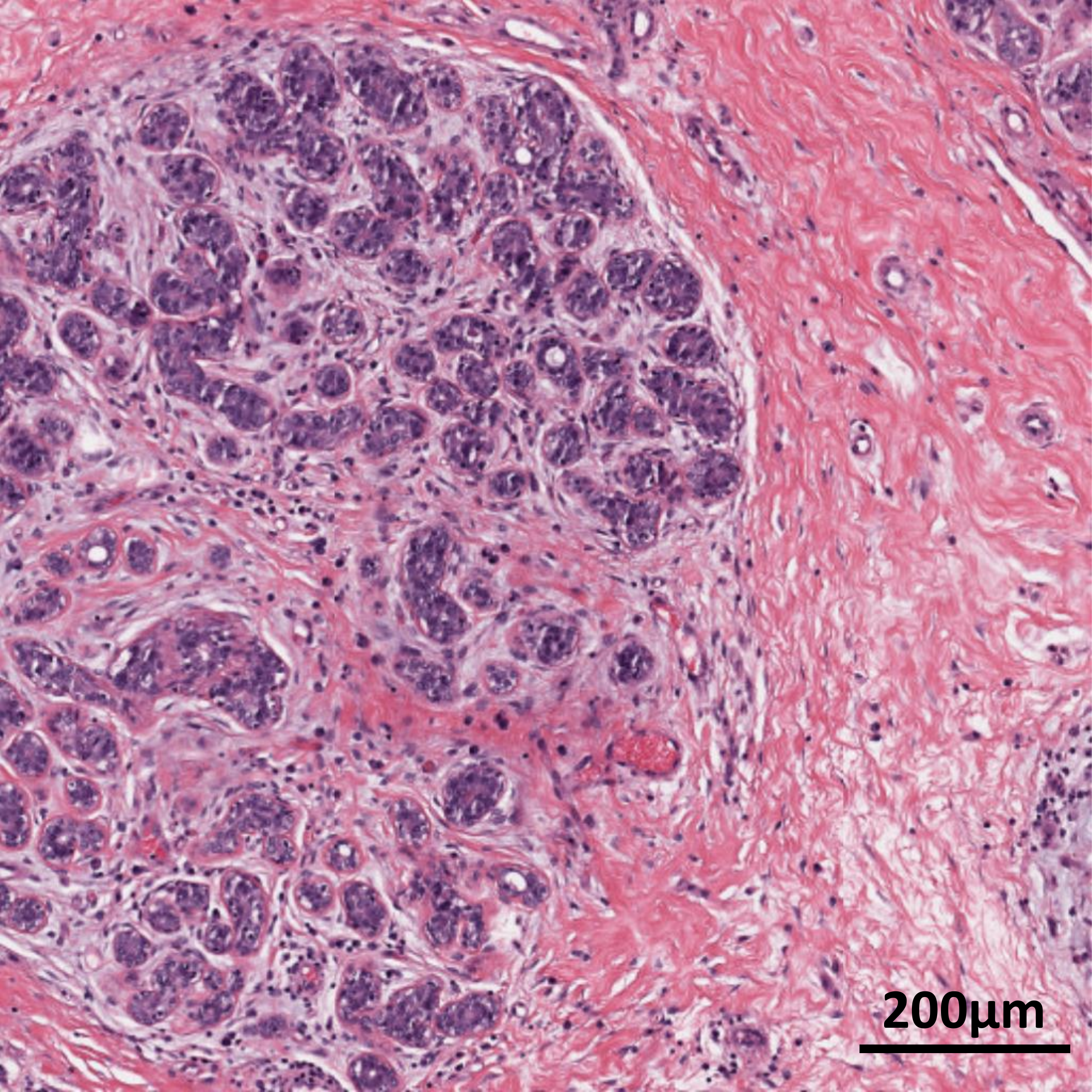,width = 0.32\textwidth}}}
\subfigure[Ground Truth]{\frame{\epsfig{figure=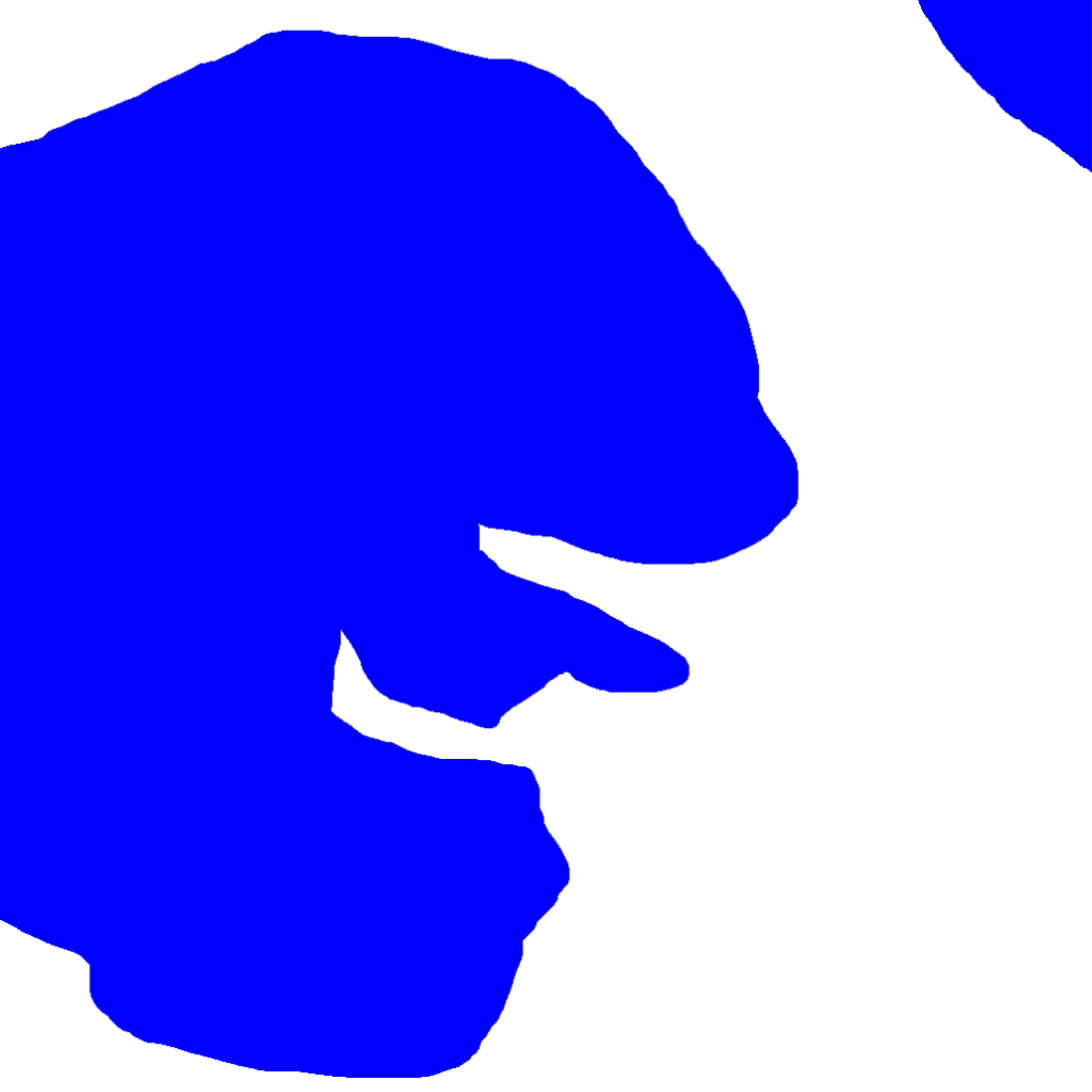,width = 0.32\textwidth}}}
\subfigure[SegNet]{\frame{\epsfig{figure=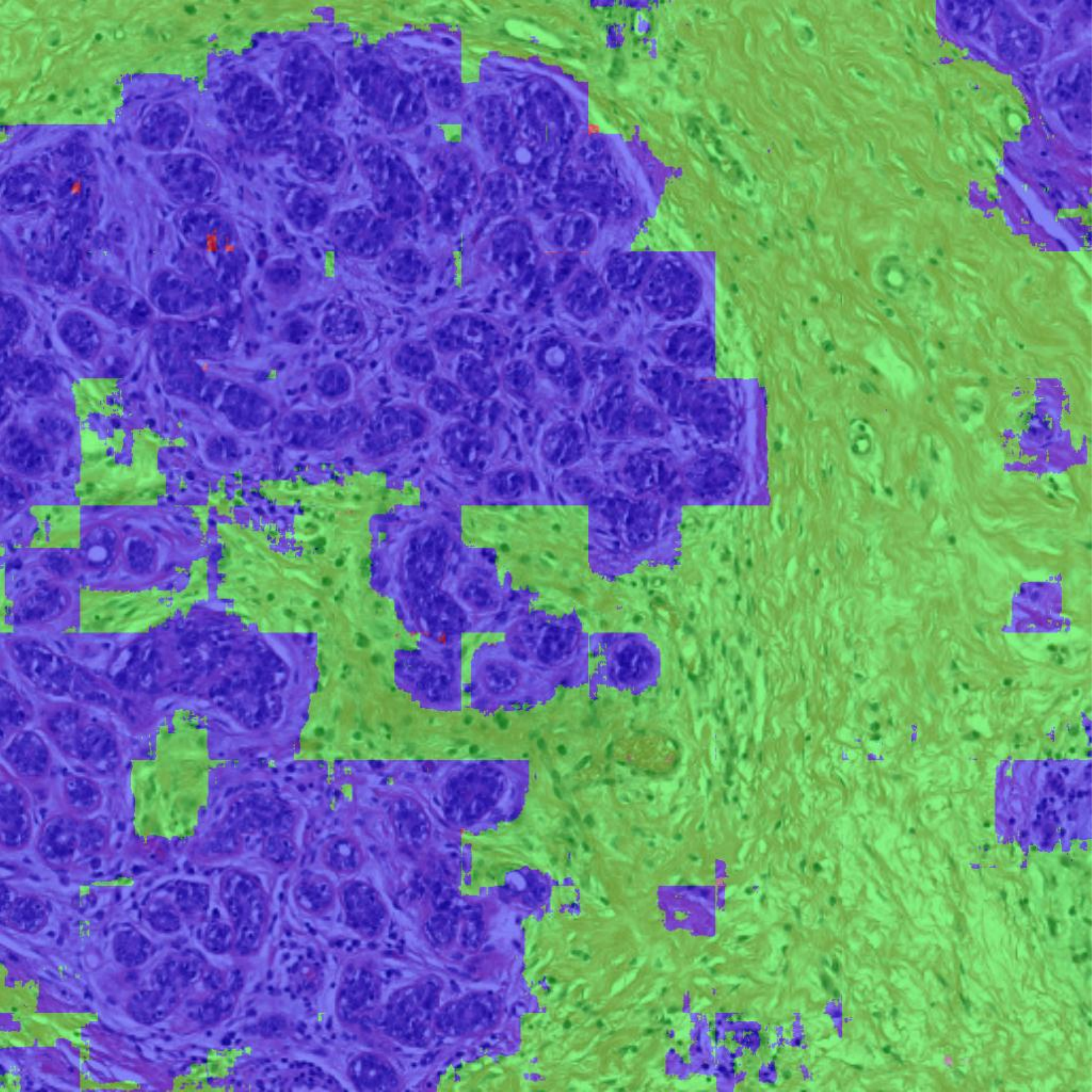,width = 0.32\textwidth}}}

\subfigure[U-Net]{\frame{\epsfig{figure=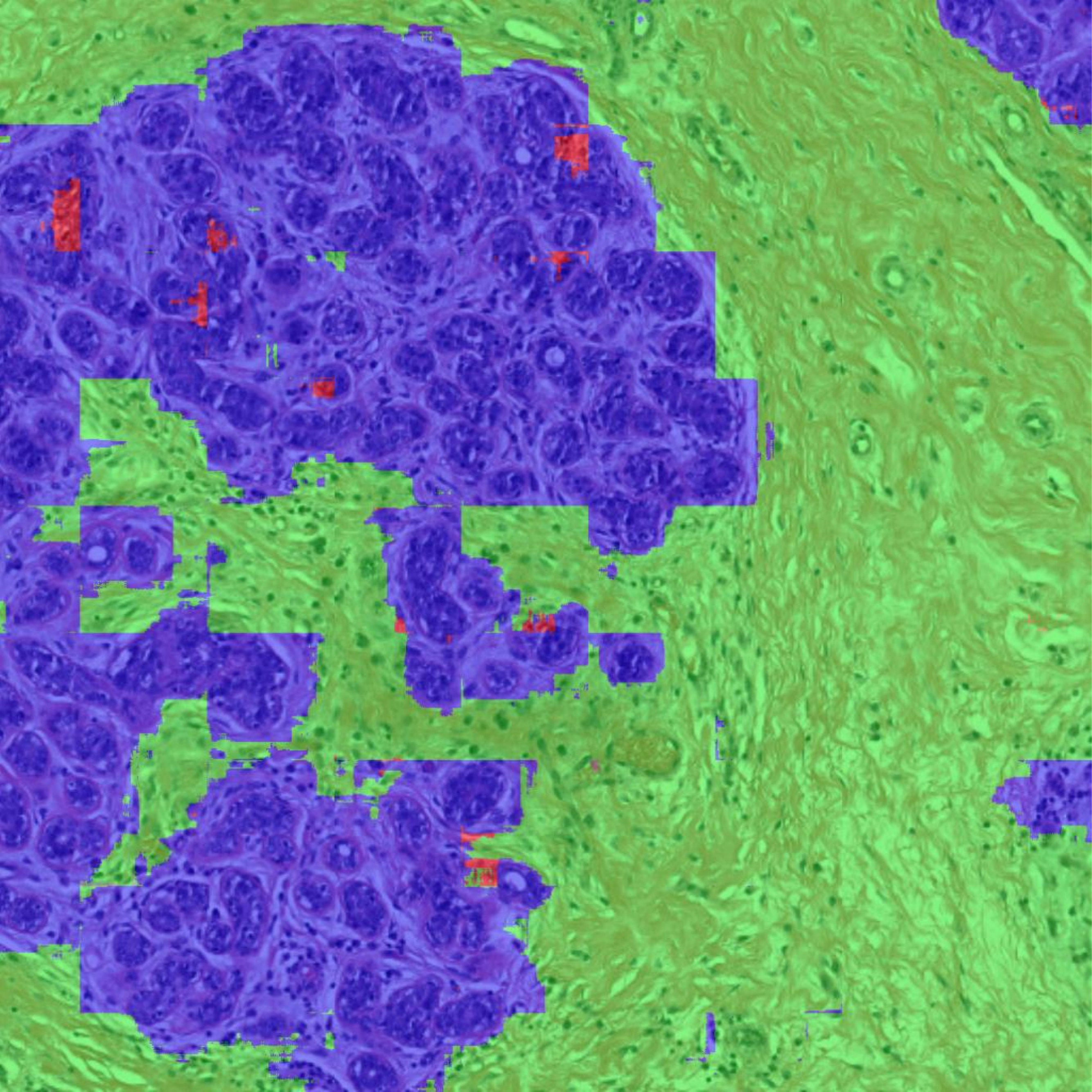,width = 0.32\textwidth}}}
\subfigure[DMMN-S2]{\frame{\epsfig{figure=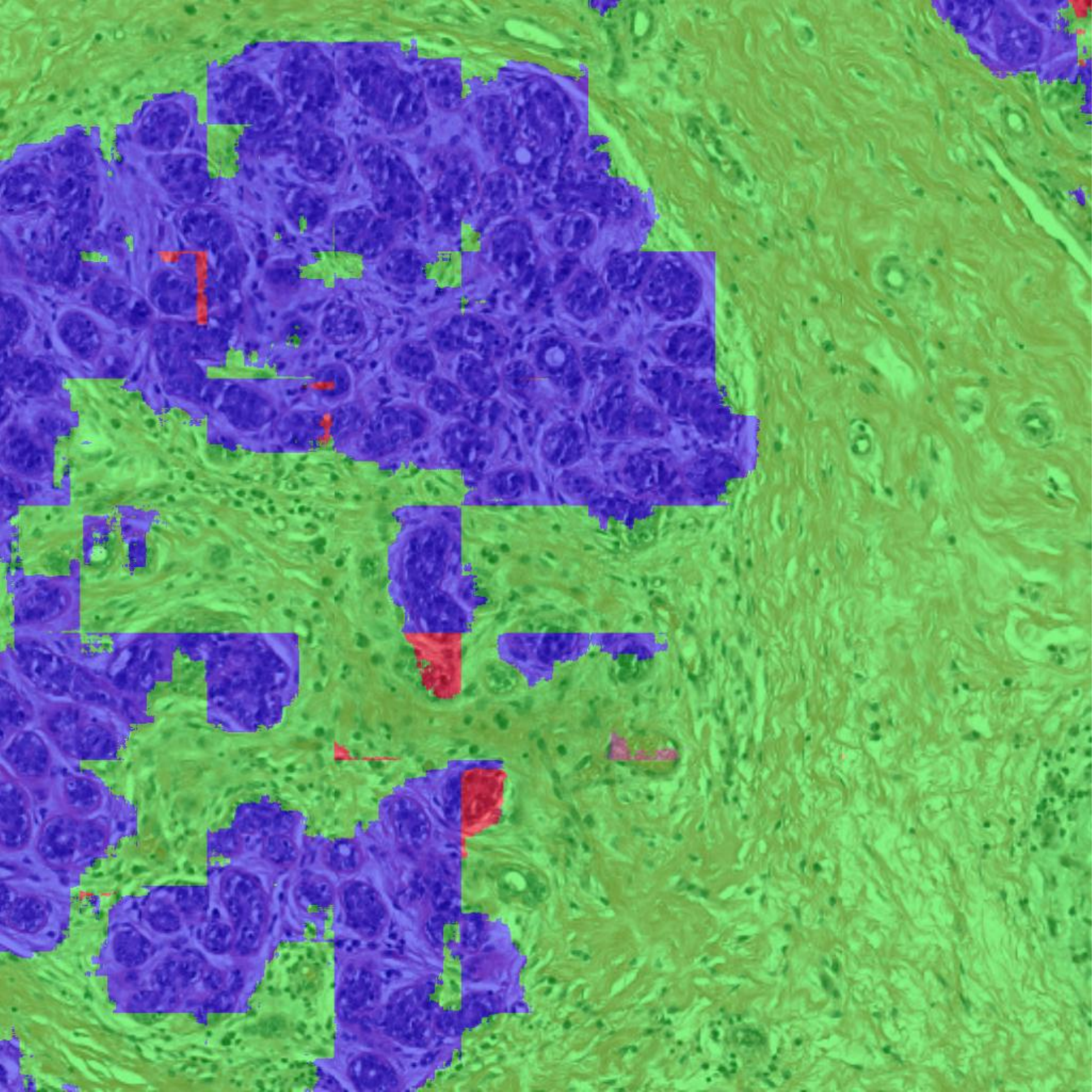,width = 0.32\textwidth}}}
\subfigure[DMMN-MS]{\frame{\epsfig{figure=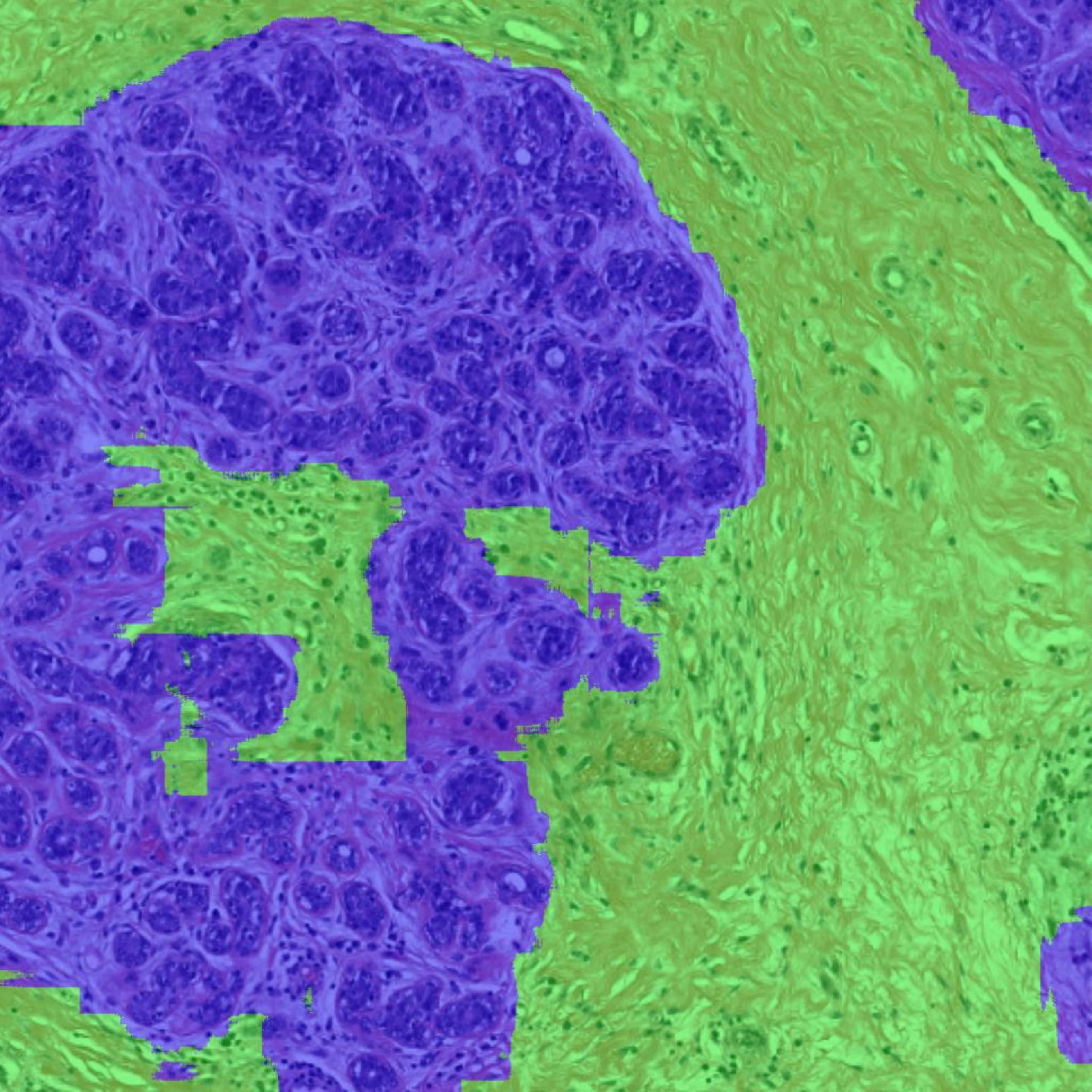,width = 0.32\textwidth}}}

\subfigure[DMMN-M2S]{\frame{\epsfig{figure=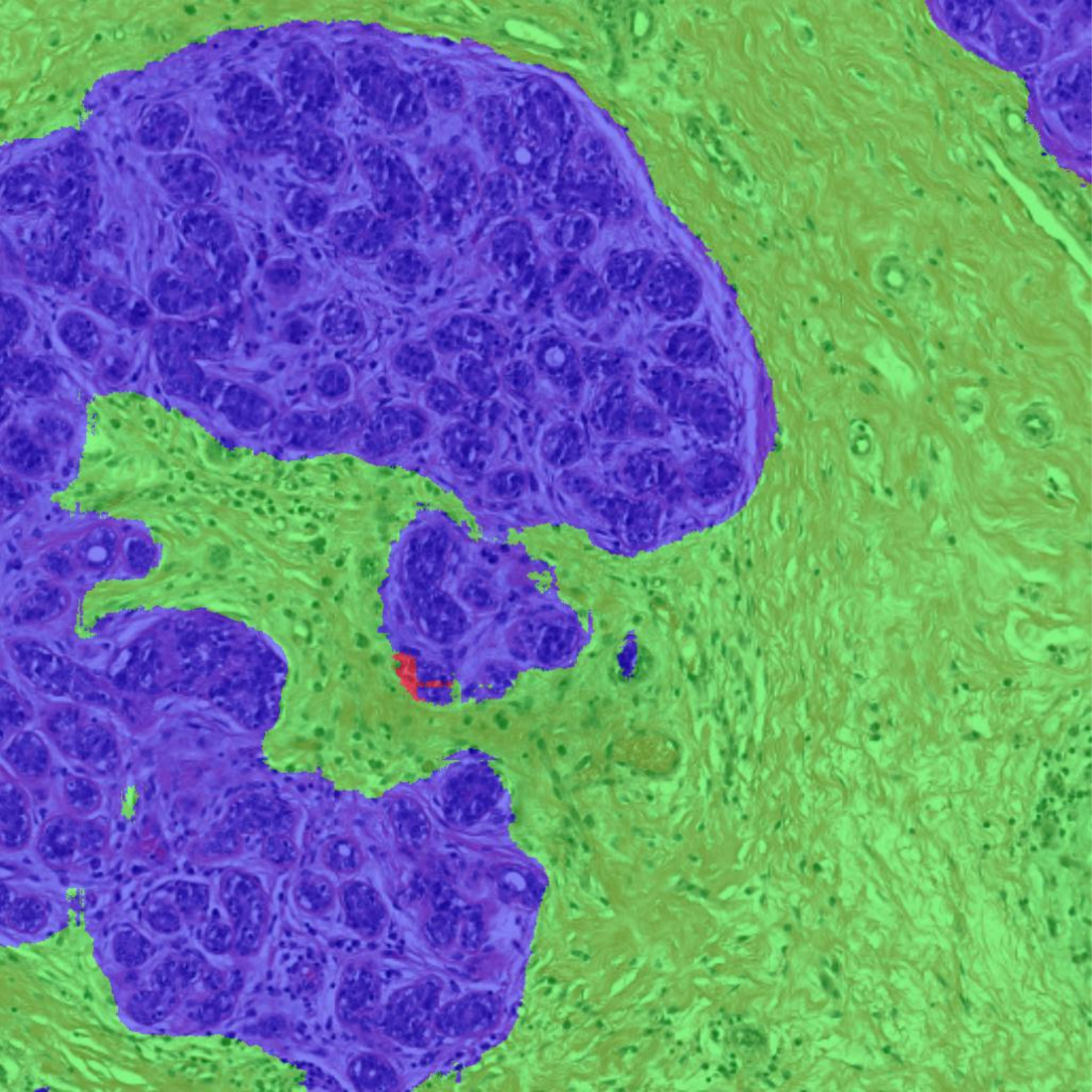,width = 0.32\textwidth}}}
\subfigure[DMMN-M3]{\frame{\epsfig{figure=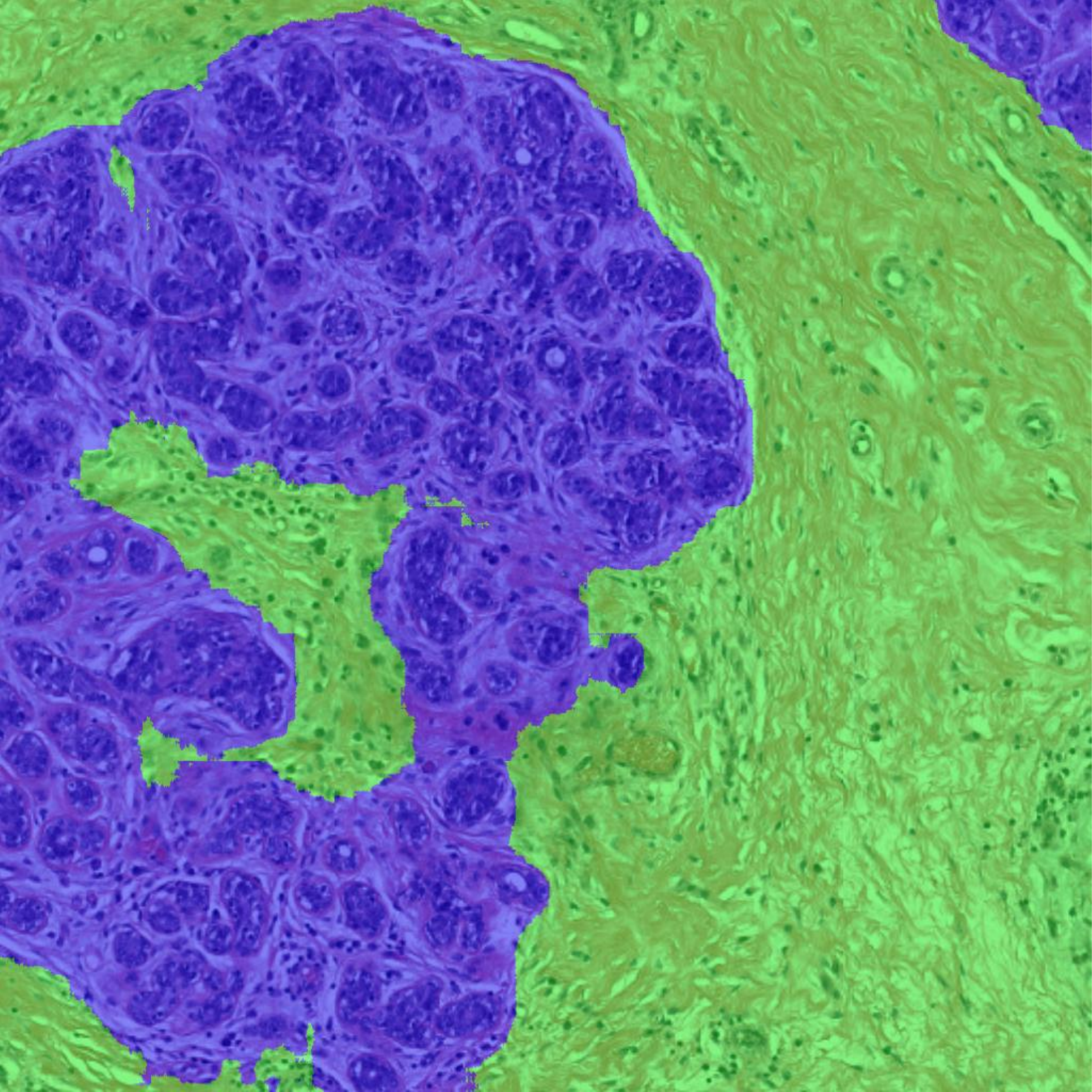,width = 0.32\textwidth}}}
\caption{Multi-class tissue segmentation predictions of benign epithelial in blue from Dataset-I using two Deep Single-Magnification Networks (DSMNs), SegNet \cite{badrinarayanan2017} and U-Net \cite{ronneberger2015}, and four Deep Multi-Magnification Networks (DMMNs), Single-Encoder Single-Decoder (DMMN-S2), Multi-Encoder Single-Decoder (DMMN-MS), Multi-Encoder Multi-Decoder Single-Concatenation (DMMN-M2S), and our proposed Multi-Encoder Multi-Decoder Multi-Concatenation (DMMN-M3).}
\label{fig:393867}
\end{figure*}

\begin{figure*}[ht!]
\centering
\subfigure[Image]{\frame{\epsfig{figure=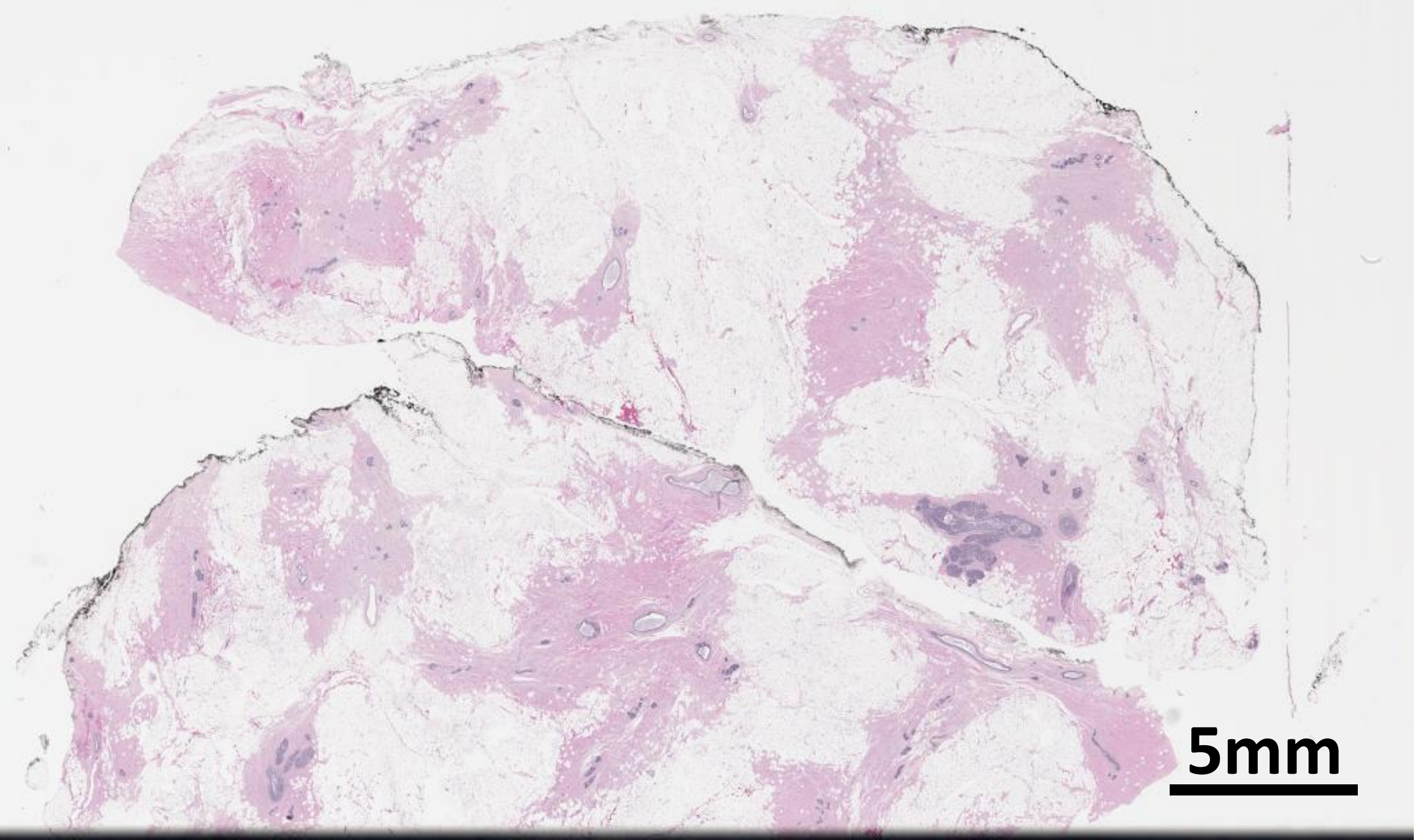,width = 0.32\textwidth}}}
\subfigure[Ground Truth]{\frame{\epsfig{figure=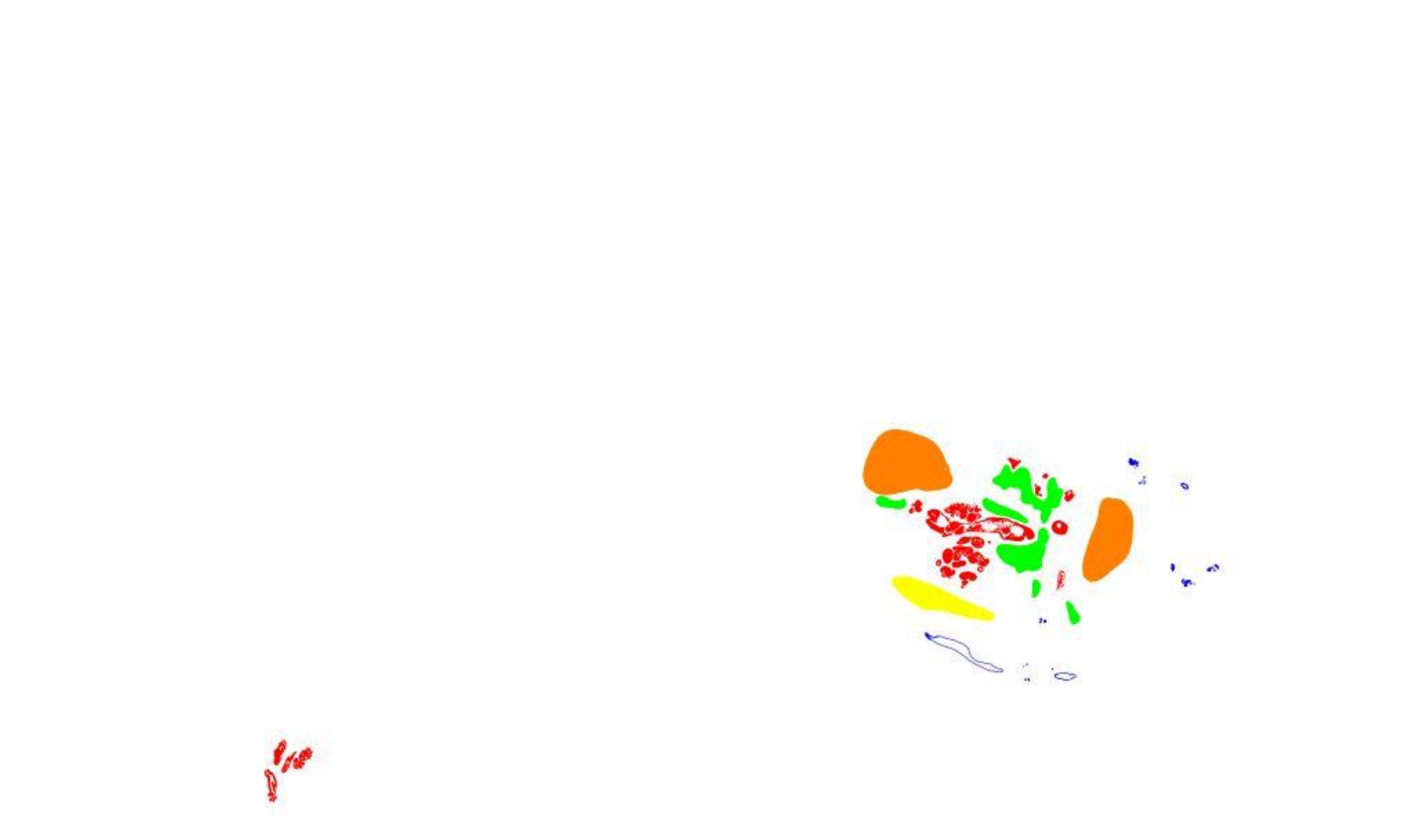,width = 0.32\textwidth}}}
\subfigure[SegNet]{\frame{\epsfig{figure=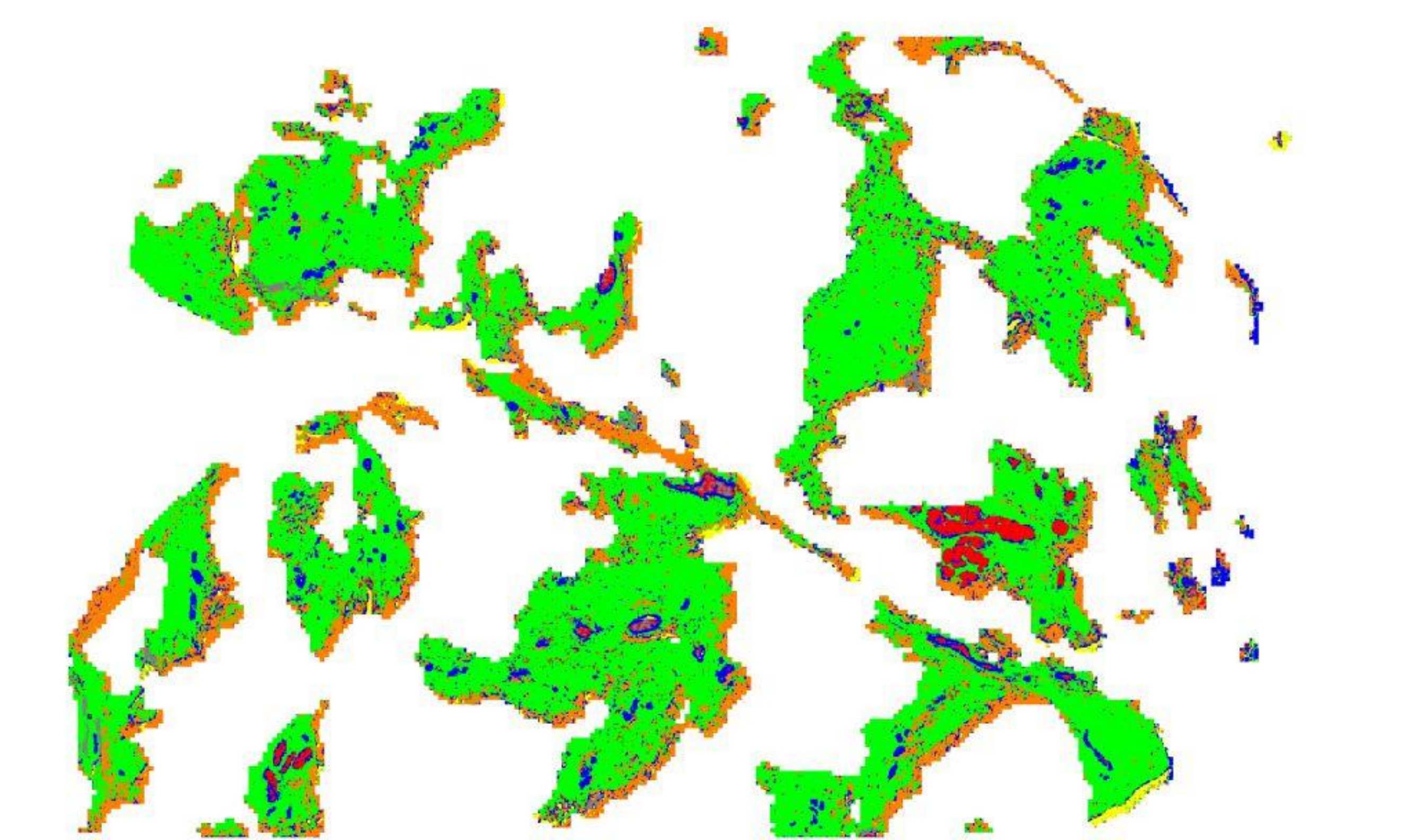,width = 0.32\textwidth}}}

\subfigure[U-Net]{\frame{\epsfig{figure=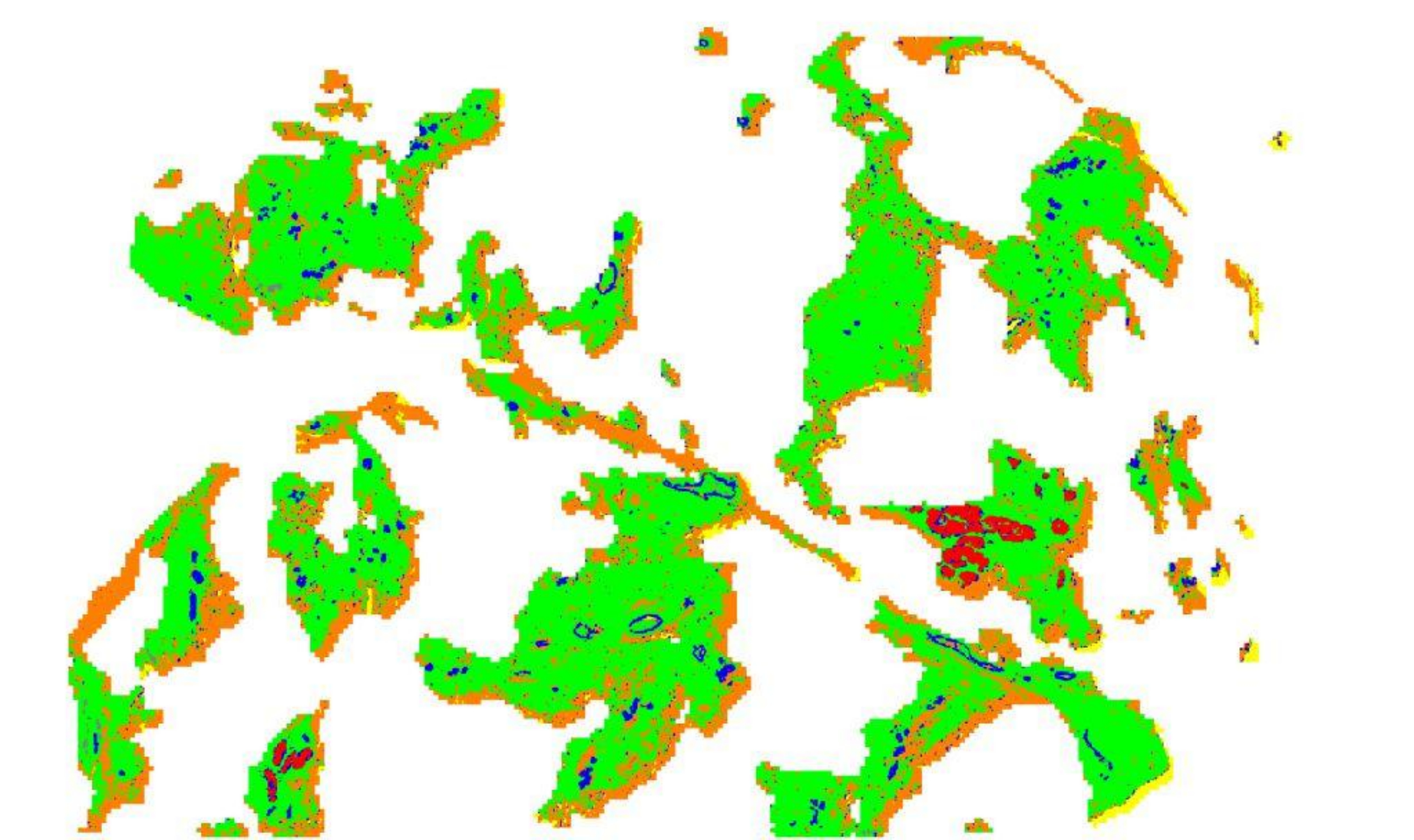,width = 0.32\textwidth}}}
\subfigure[DMMN-S2]{\frame{\epsfig{figure=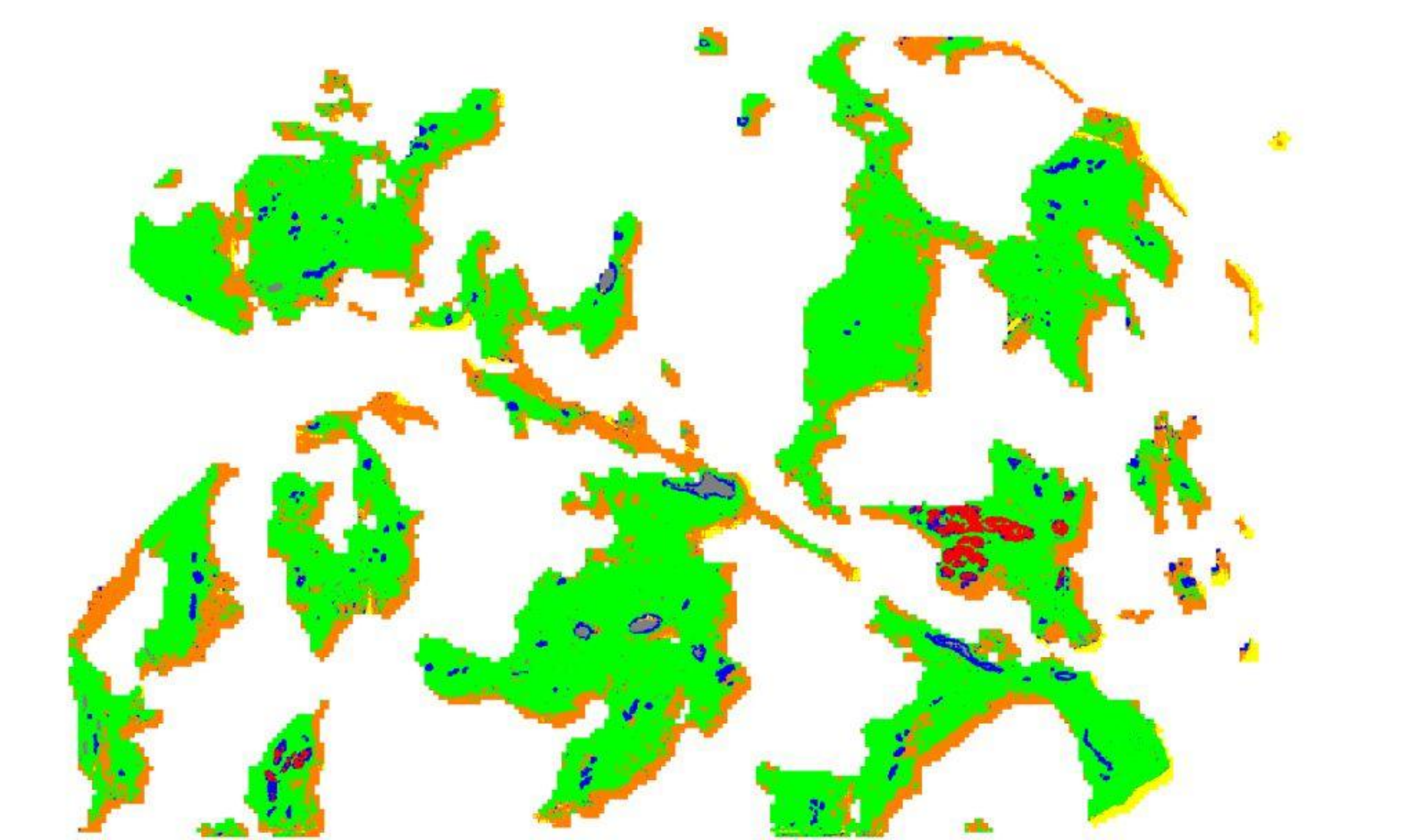,width = 0.32\textwidth}}}
\subfigure[DMMN-MS]{\frame{\epsfig{figure=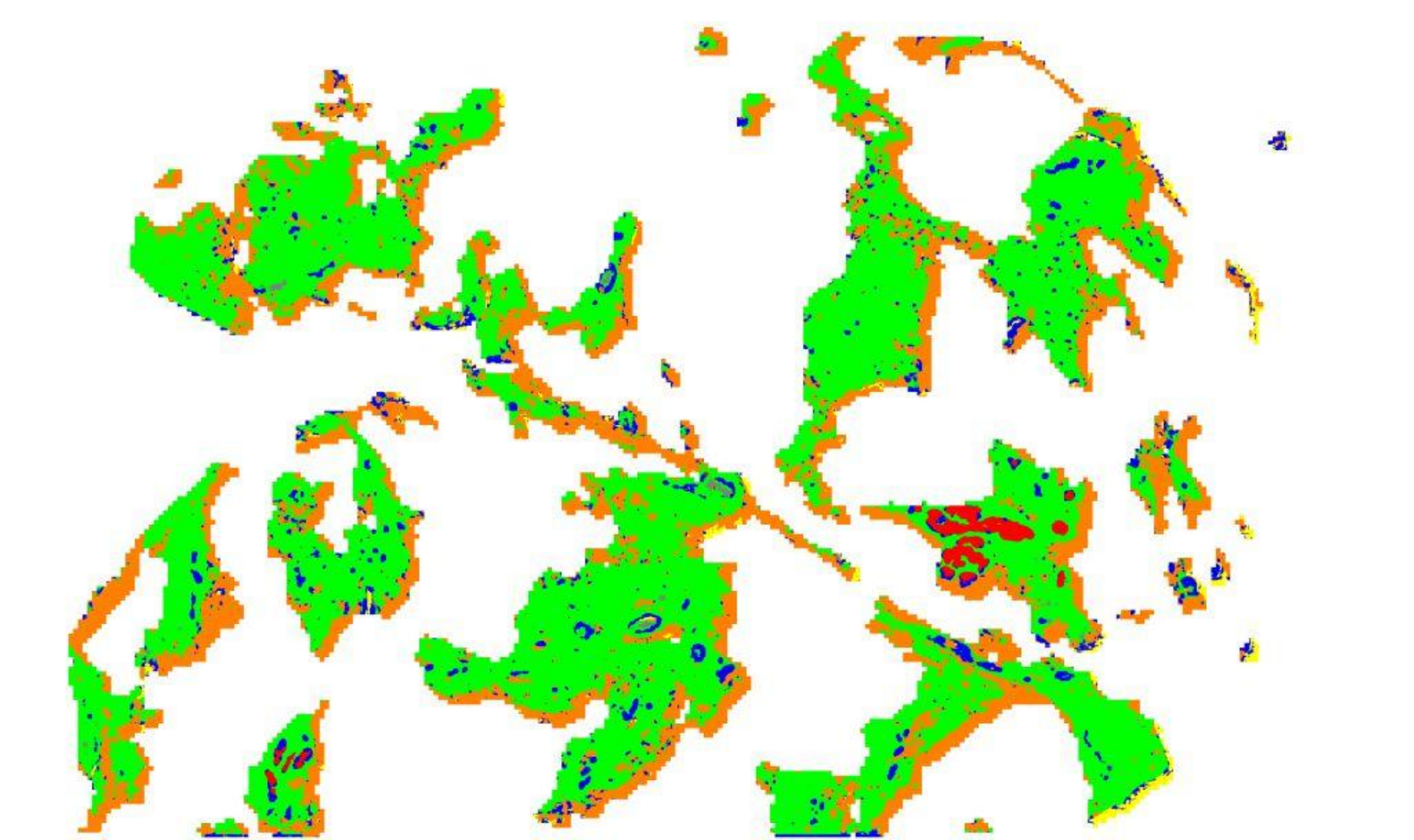,width = 0.32\textwidth}}}

\subfigure[DMMN-M2S]{\frame{\epsfig{figure=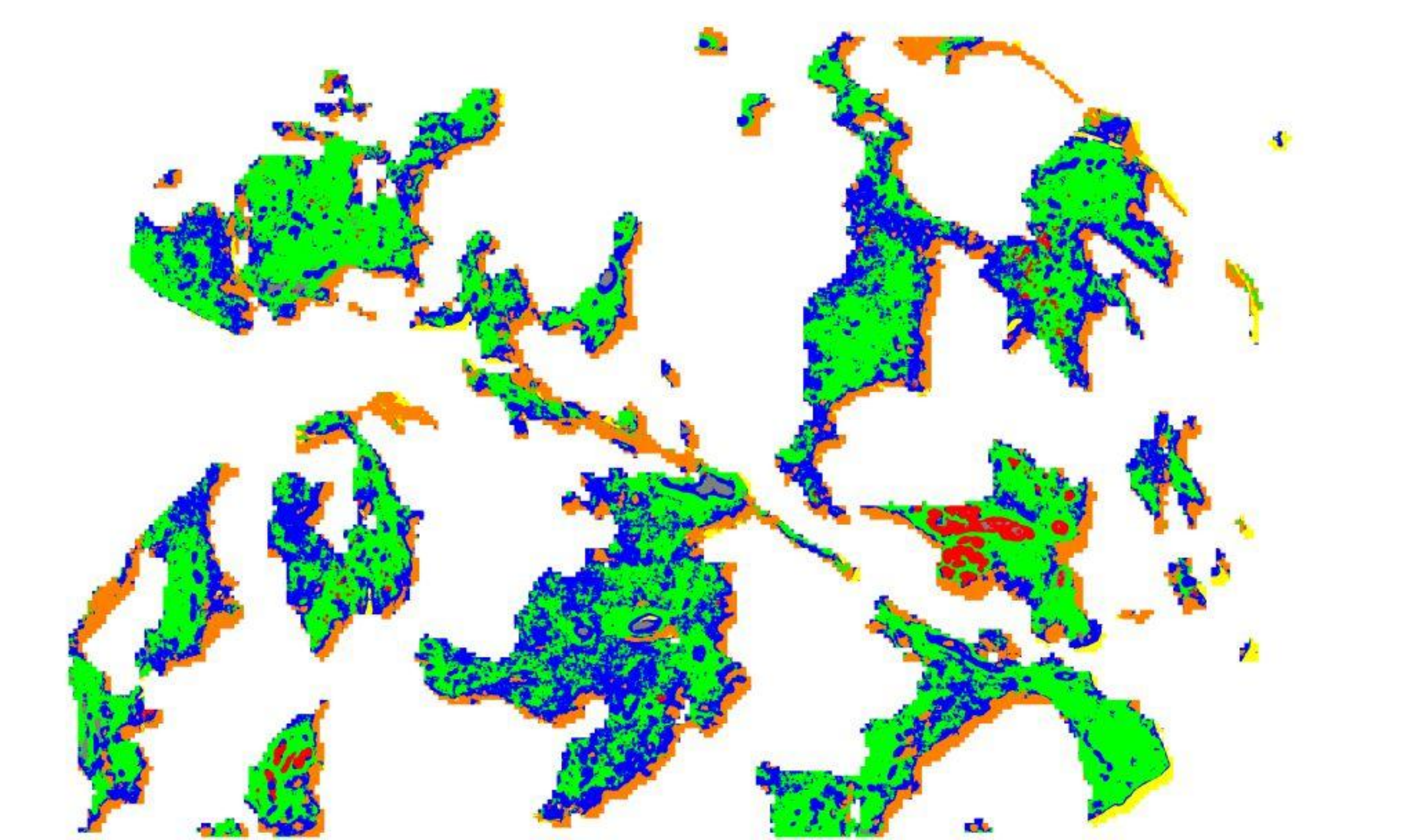,width = 0.32\textwidth}}}
\subfigure[DMMN-M3]{\frame{\epsfig{figure=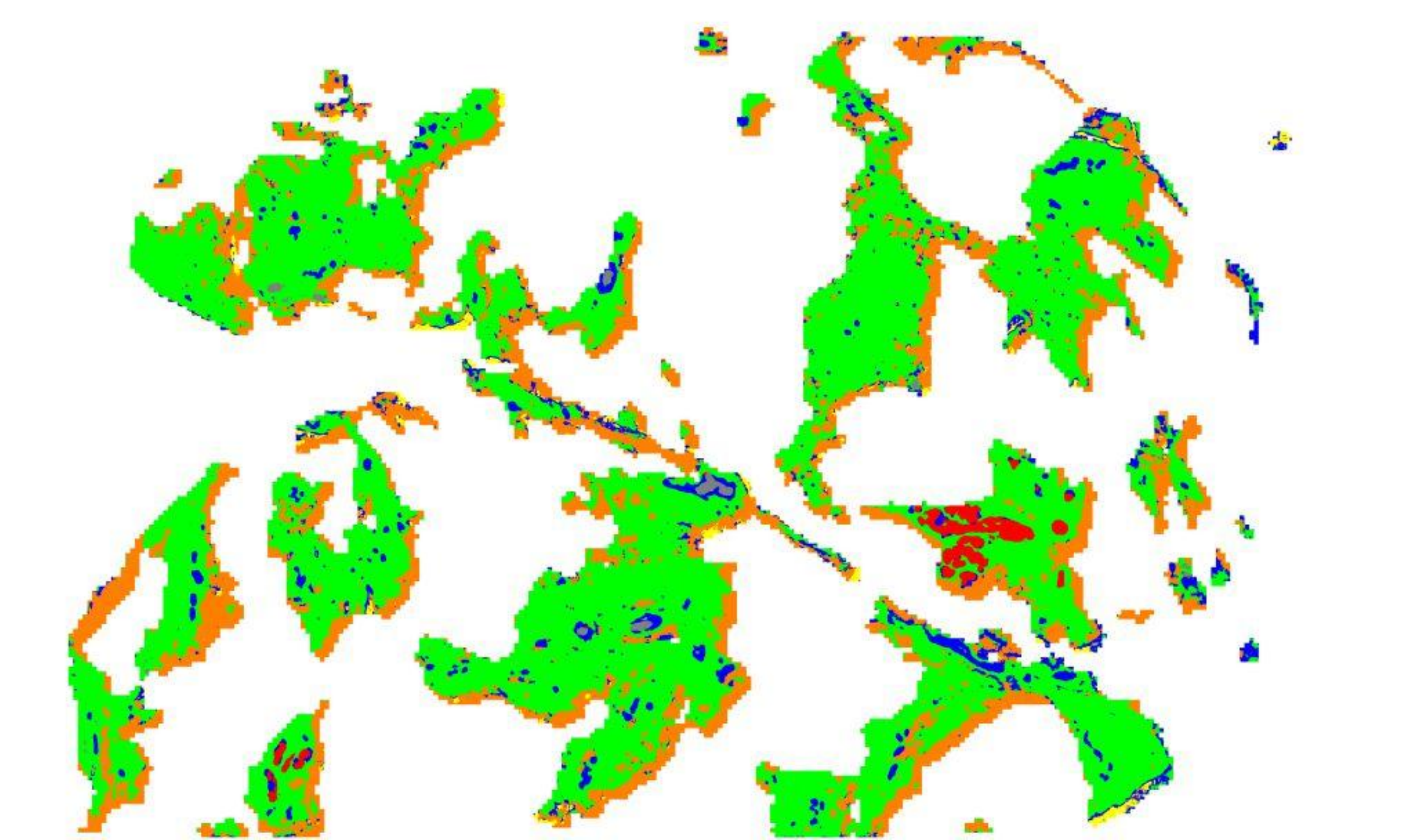,width = 0.32\textwidth}}}
\caption{Multi-class tissue segmentation predictions of a whole slide image from Dataset-II using two Deep Single-Magnification Networks (DSMNs), SegNet \cite{badrinarayanan2017} and U-Net \cite{ronneberger2015}, and four Deep Multi-Magnification Networks (DMMNs), Single-Encoder Single-Decoder (DMMN-S2), Multi-Encoder Single-Decoder (DMMN-MS), Multi-Encoder Multi-Decoder Single-Concatenation (DMMN-M2S), and our proposed Multi-Encoder Multi-Decoder Multi-Concatenation (DMMN-M3).}
\label{fig:1365648_WSI}
\end{figure*}

\begin{figure*}[ht!]
\centering
\subfigure[Image]{\frame{\epsfig{figure=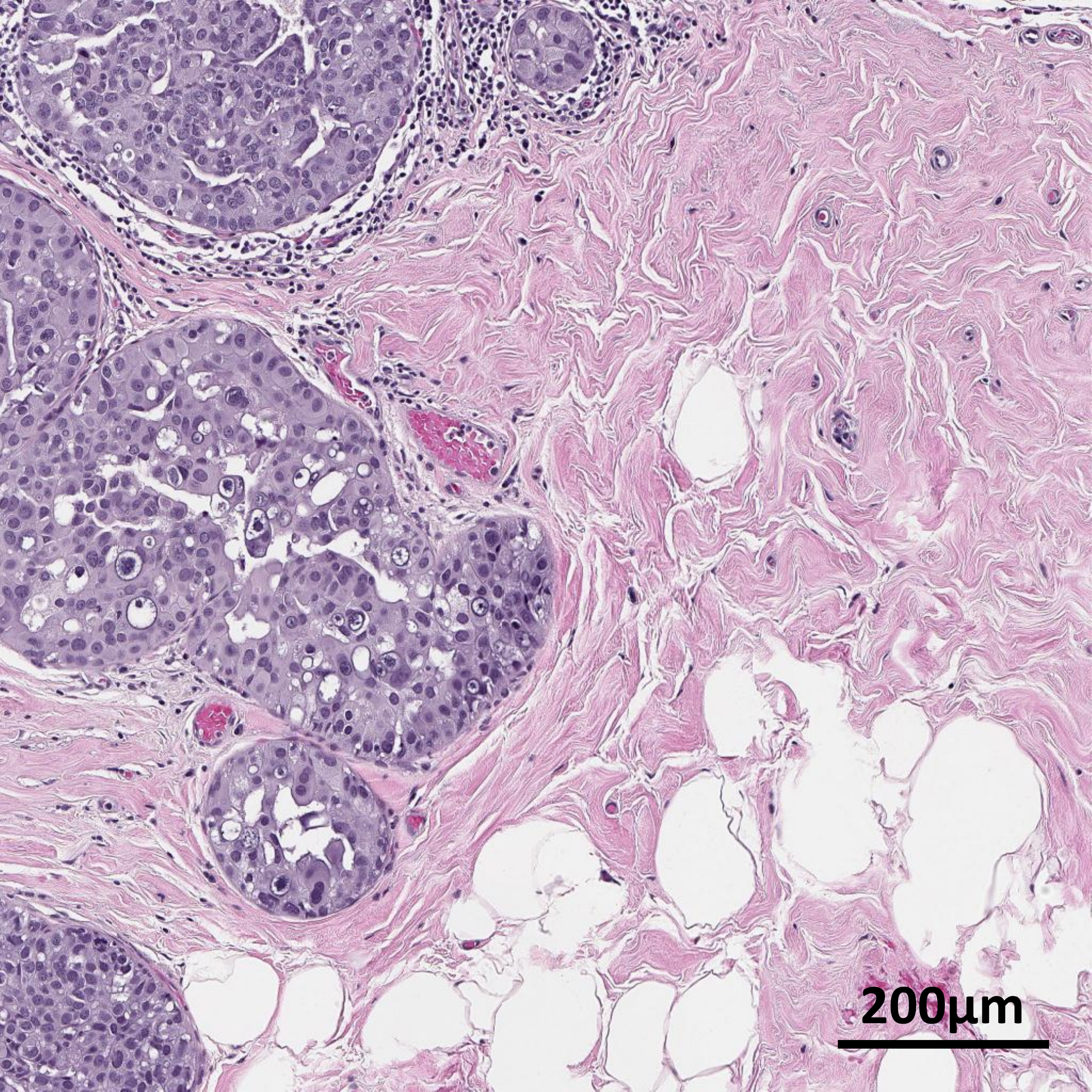,width = 0.32\textwidth}}}
\subfigure[Ground Truth]{\frame{\epsfig{figure=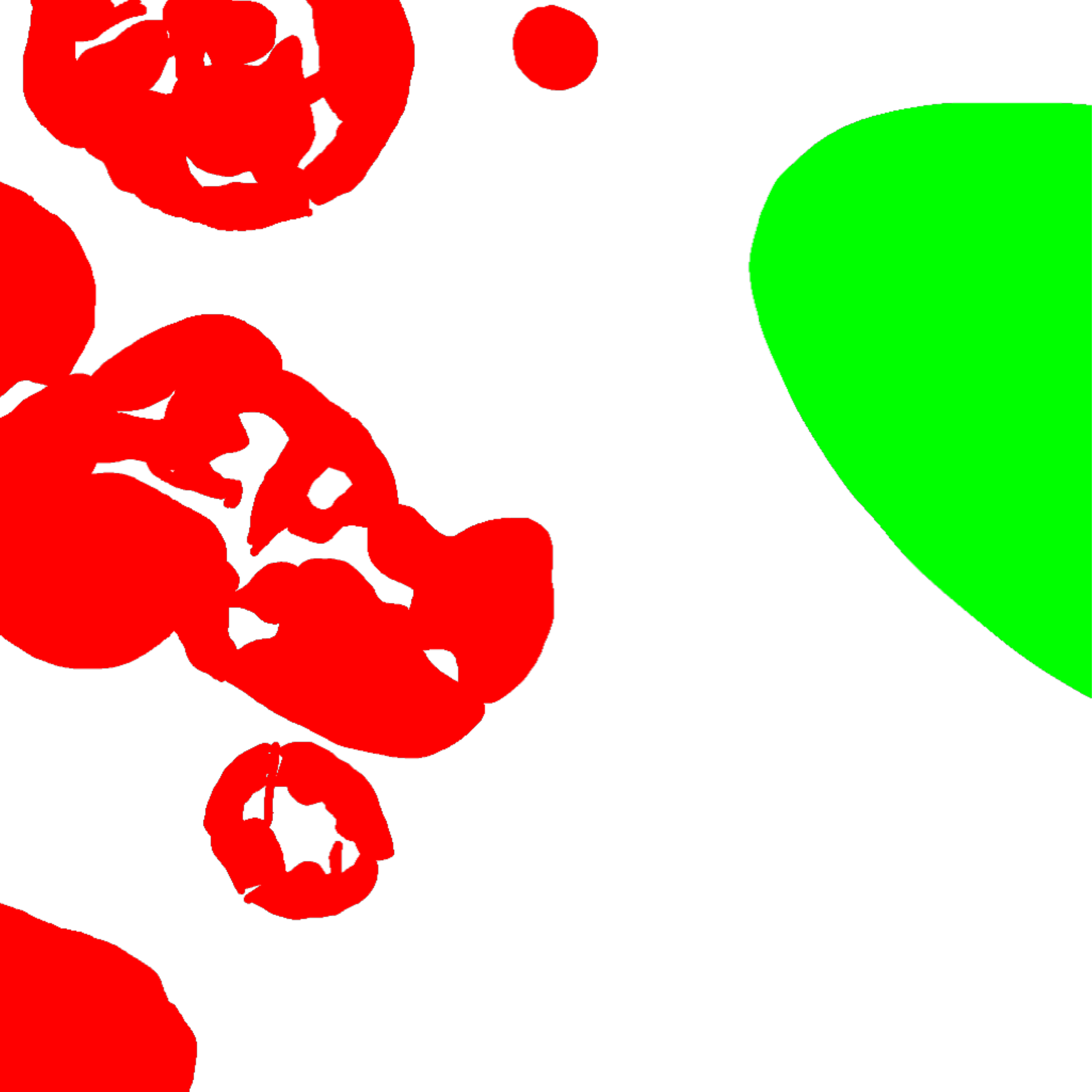,width = 0.32\textwidth}}}
\subfigure[SegNet]{\frame{\epsfig{figure=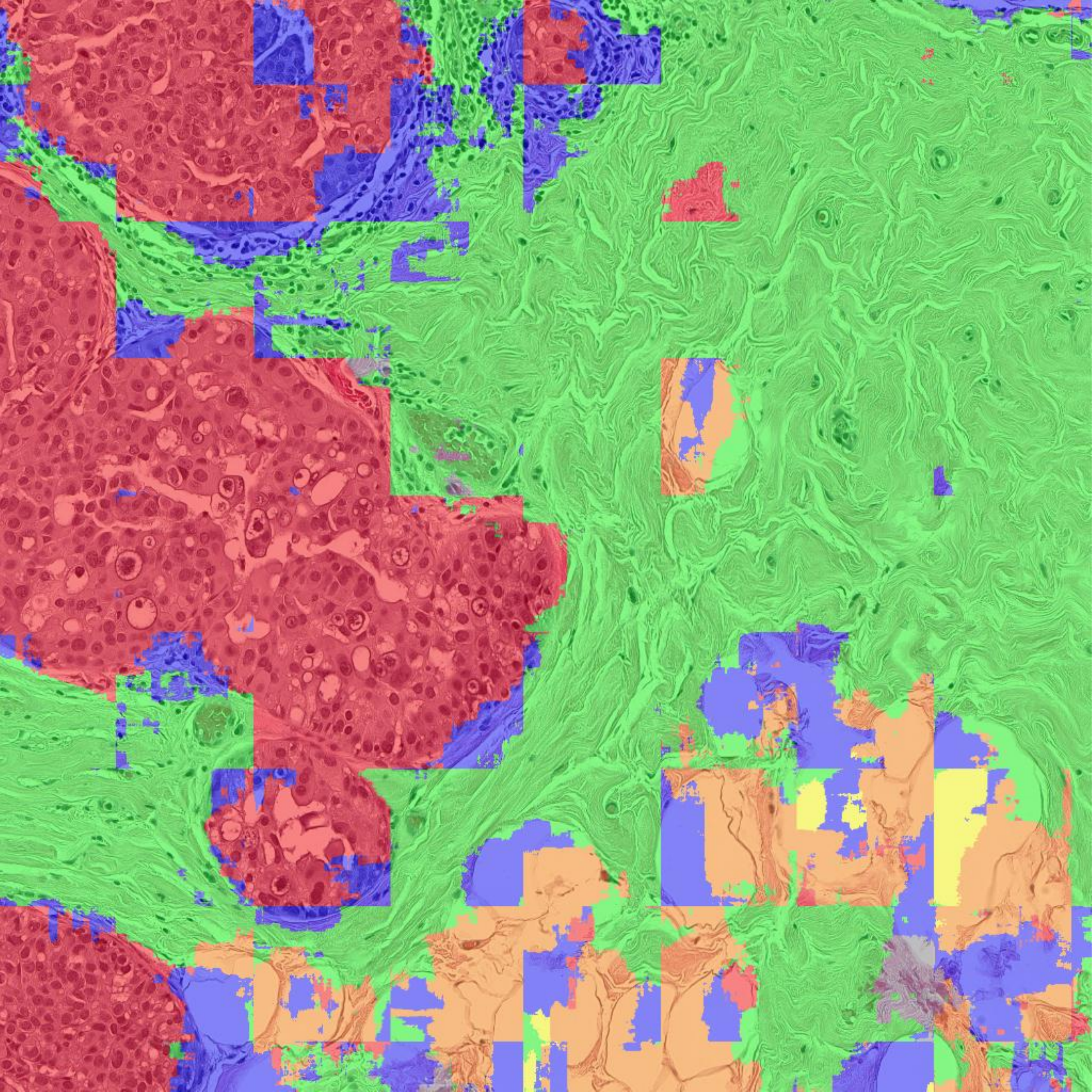,width = 0.32\textwidth}}}

\subfigure[U-Net]{\frame{\epsfig{figure=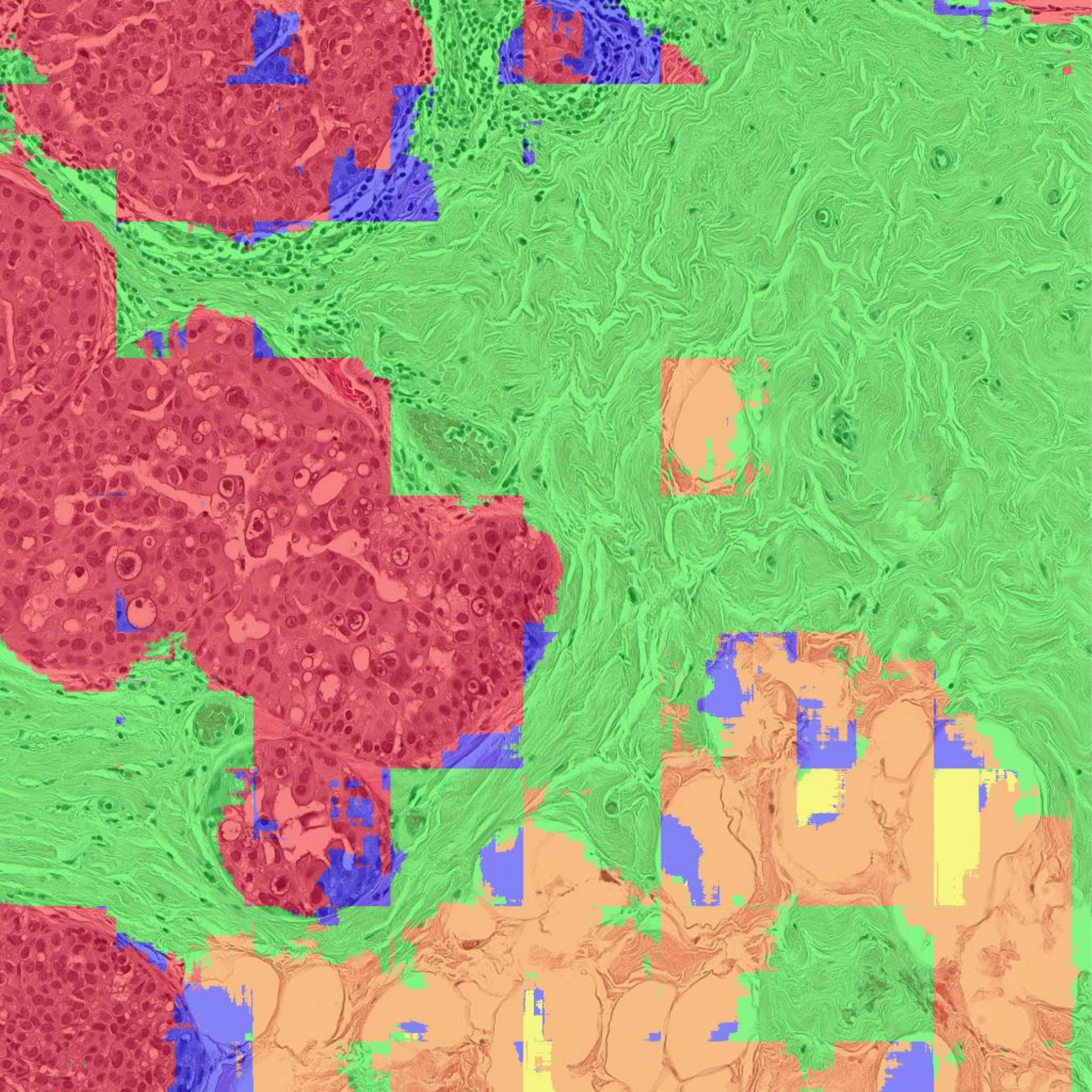,width = 0.32\textwidth}}}
\subfigure[DMMN-S2]{\frame{\epsfig{figure=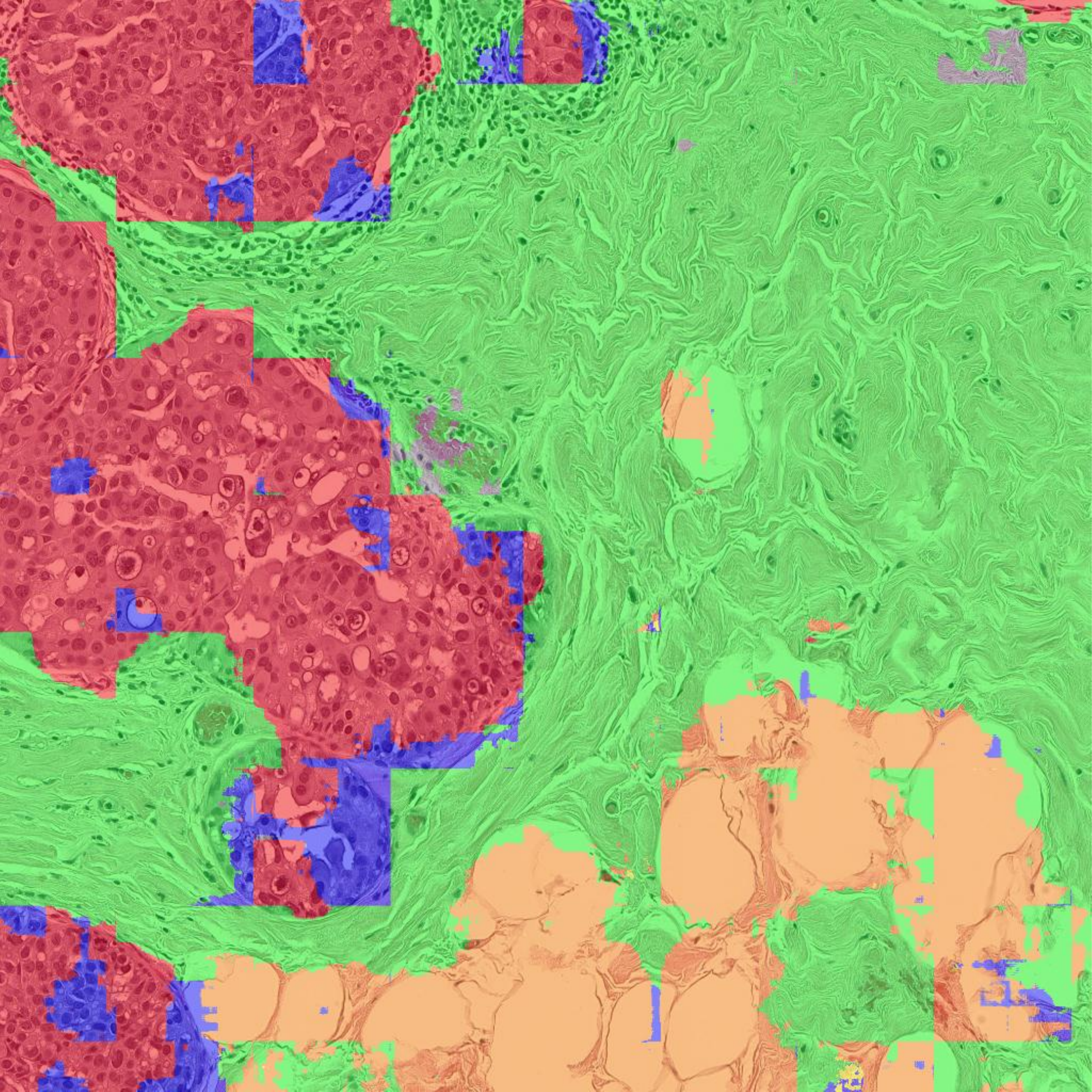,width = 0.32\textwidth}}}
\subfigure[DMMN-MS]{\frame{\epsfig{figure=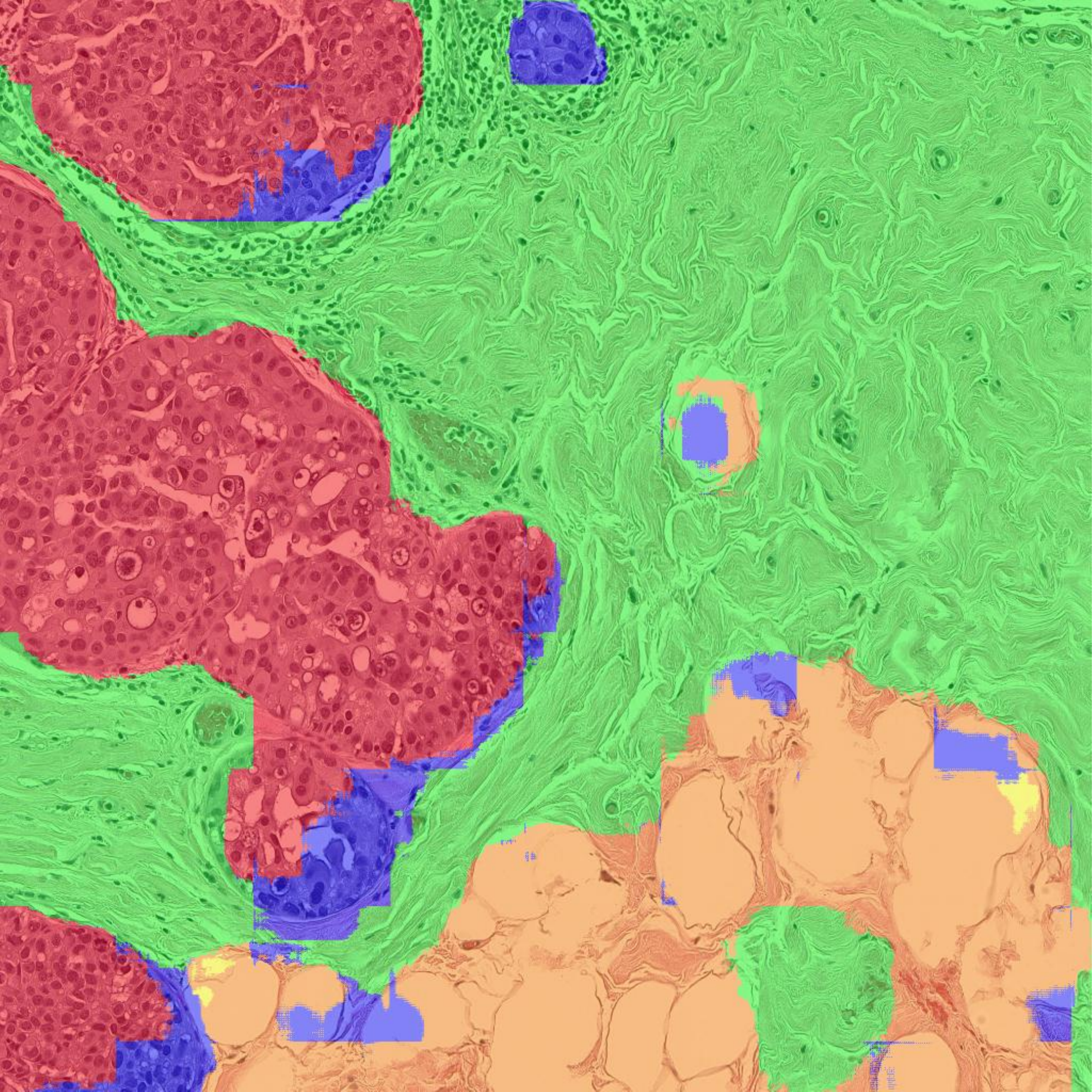,width = 0.32\textwidth}}}

\subfigure[DMMN-M2S]{\frame{\epsfig{figure=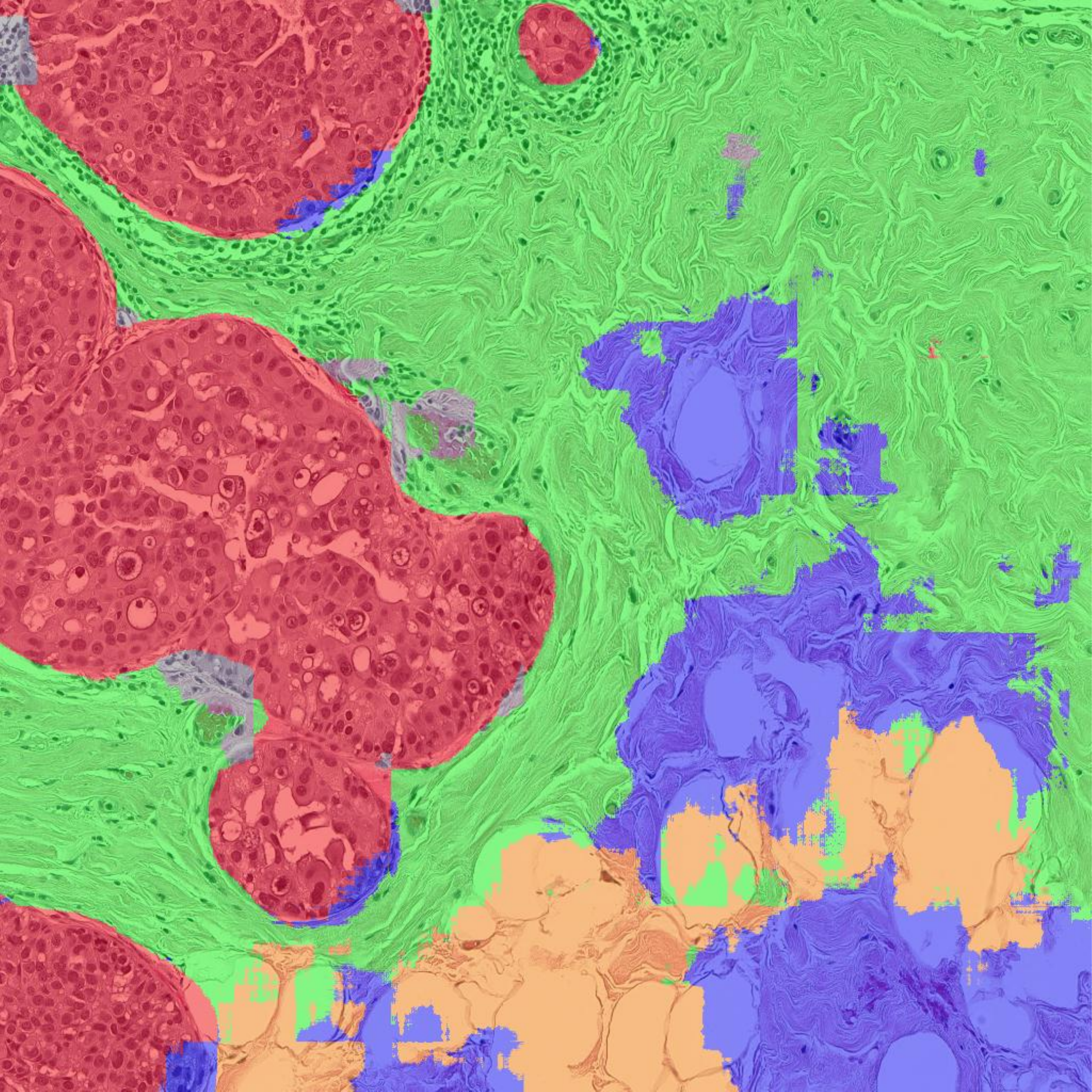,width = 0.32\textwidth}}}
\subfigure[DMMN-M3]{\frame{\epsfig{figure=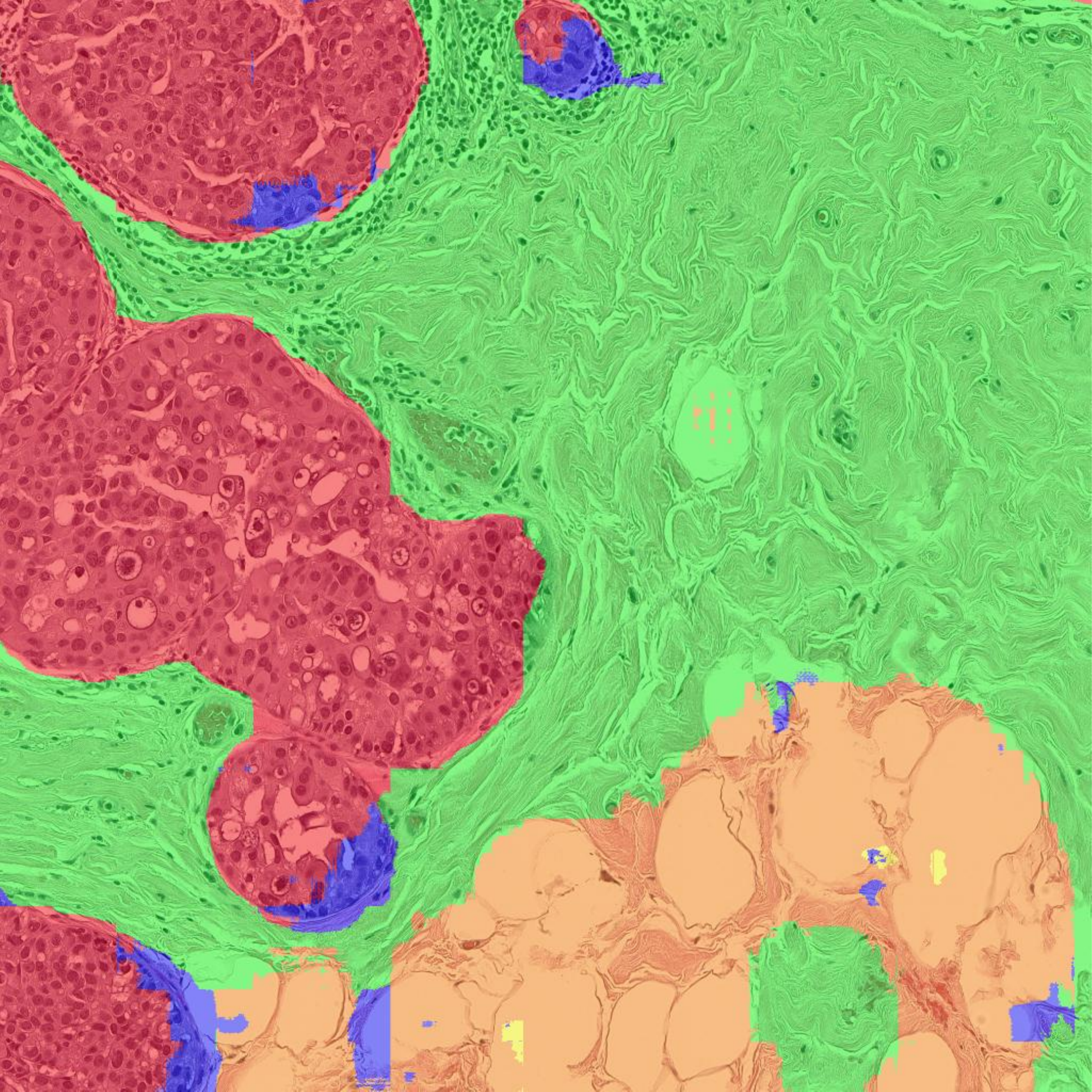,width = 0.32\textwidth}}}
\caption{Multi-class tissue segmentation predictions of ductal carcinoma in situ (DCIS) in red from Dataset-II using two Deep Single-Magnification Networks (DSMNs), SegNet \cite{badrinarayanan2017} and U-Net \cite{ronneberger2015}, and four Deep Multi-Magnification Networks (DMMNs), Single-Encoder Single-Decoder (DMMN-S2), Multi-Encoder Single-Decoder (DMMN-MS), Multi-Encoder Multi-Decoder Single-Concatenation (DMMN-M2S), and our proposed Multi-Encoder Multi-Decoder Multi-Concatenation (DMMN-M3).}
\label{fig:1365648}
\end{figure*}

We processed 55 testing images from Dataset-I and 34 testing images from Dataset-II to evaluate various models.
Figure \ref{fig:393382_WSI} depicts multi-class segmentation predictions in a WSI from Dataset-I by SegNet \cite{badrinarayanan2017} architecture, U-Net \cite{ronneberger2015} architecture, DMMN-S2 architecture, DMMN-MS architecture, DMMN-M2S architecture, and our proposed DMMN-M3 architecture, and Figure \ref{fig:393382} depicts multi-class segmentation predictions in a patch containing invasive ductal carcinoma (IDC) with size of 1024$\times$1024 pixels in 10$\times$ magnification from the WSI in Figure \ref{fig:393382_WSI}.
Similarly, Figure \ref{fig:393867_WSI} depicts multi-class segmentation predictions in a WSI from Dataset-I, Figure \ref{fig:393867} depicts multi-class segmentation predictions in a patch containing benign epithelial from the WSI in Figure \ref{fig:393867_WSI}, Figure \ref{fig:1365648_WSI} depicts multi-class segmentation predictions in a WSI from Dataset-II, and Figure \ref{fig:1365648} depicts multi-class segmentation predictions in a patch containing ductal carcinoma in situ (DCIS) from the WSI in Figure \ref{fig:1365648_WSI}.
Tissue subtypes are labeled in distinct colors such as carcinoma in red, benign epithelial in blue, background in yellow, stroma in green, necrotic in gray, and adipose in orange.
White regions in Figures \ref{fig:393382_WSI}(b), \ref{fig:393382}(b), \ref{fig:393867_WSI}(b), \ref{fig:393867}(b), \ref{fig:1365648_WSI}(b), and \ref{fig:1365648}(b) are unlabeled.
The Otsu thresholding technique \cite{otsu1979} was used to extract patches only on foreground regions of the WSIs from Dataset-II digitized from a different scanner because we observed that models are sensitive to background noise leading mis-segmentation on background regions.
White regions in Figure \ref{fig:1365648_WSI}(c-h) are removed by the Otsu technique \cite{otsu1979}.

\begin{figure*}[ht!]
\centering
\subfigure[SegNet]{\epsfig{figure=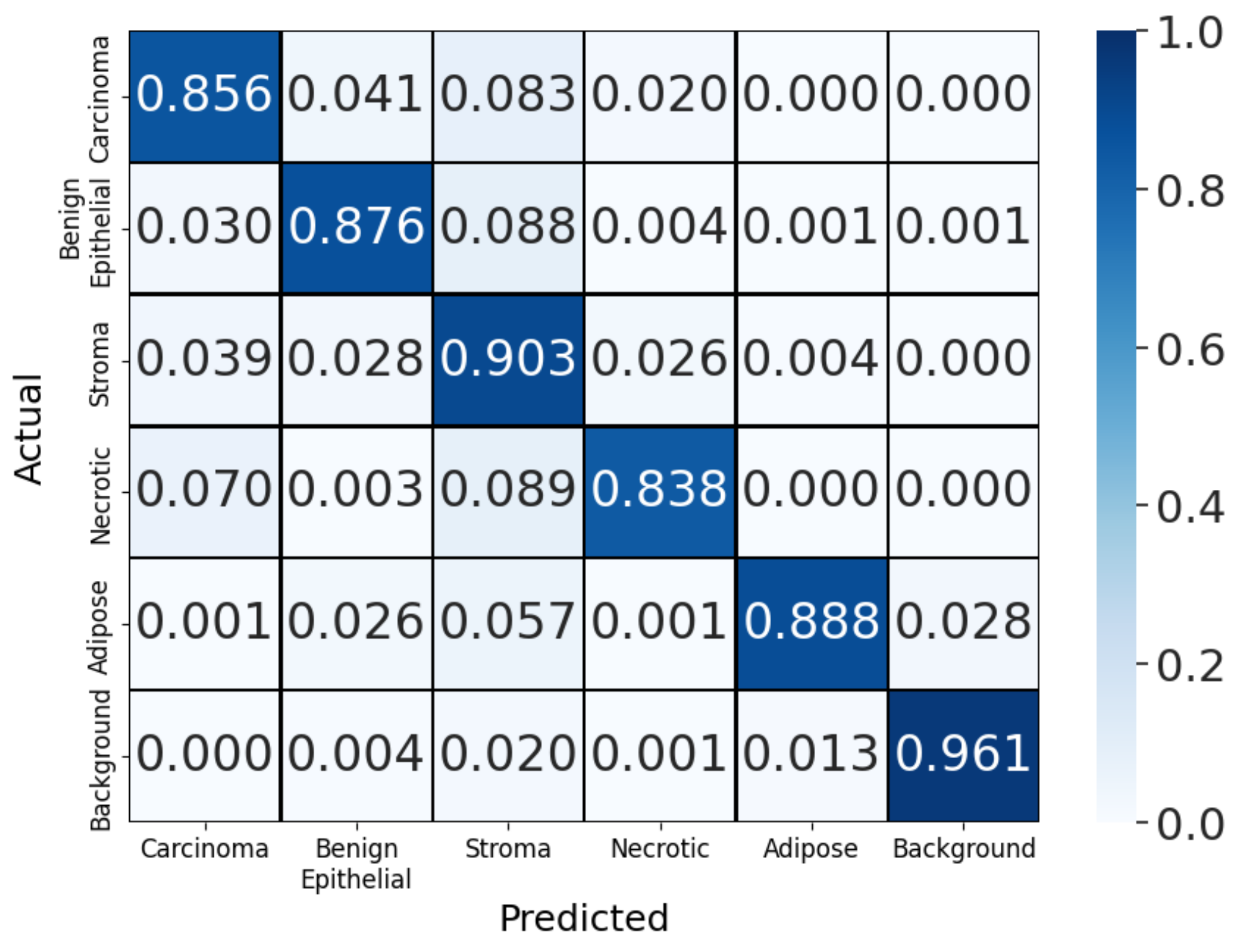,width = 0.32\textwidth}}
\subfigure[U-Net]{\epsfig{figure=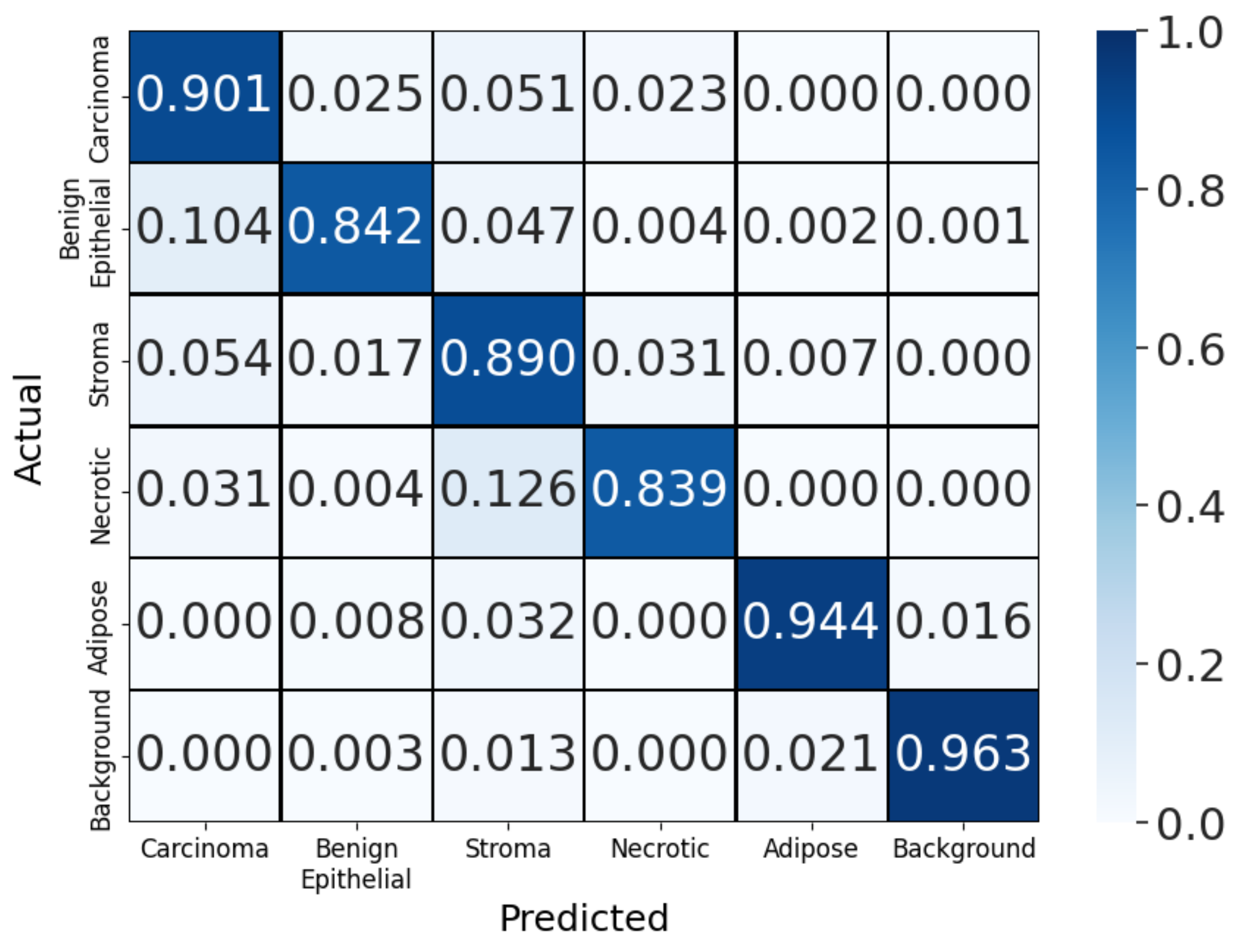,width = 0.32\textwidth}}
\subfigure[DMMN-S2]{\epsfig{figure=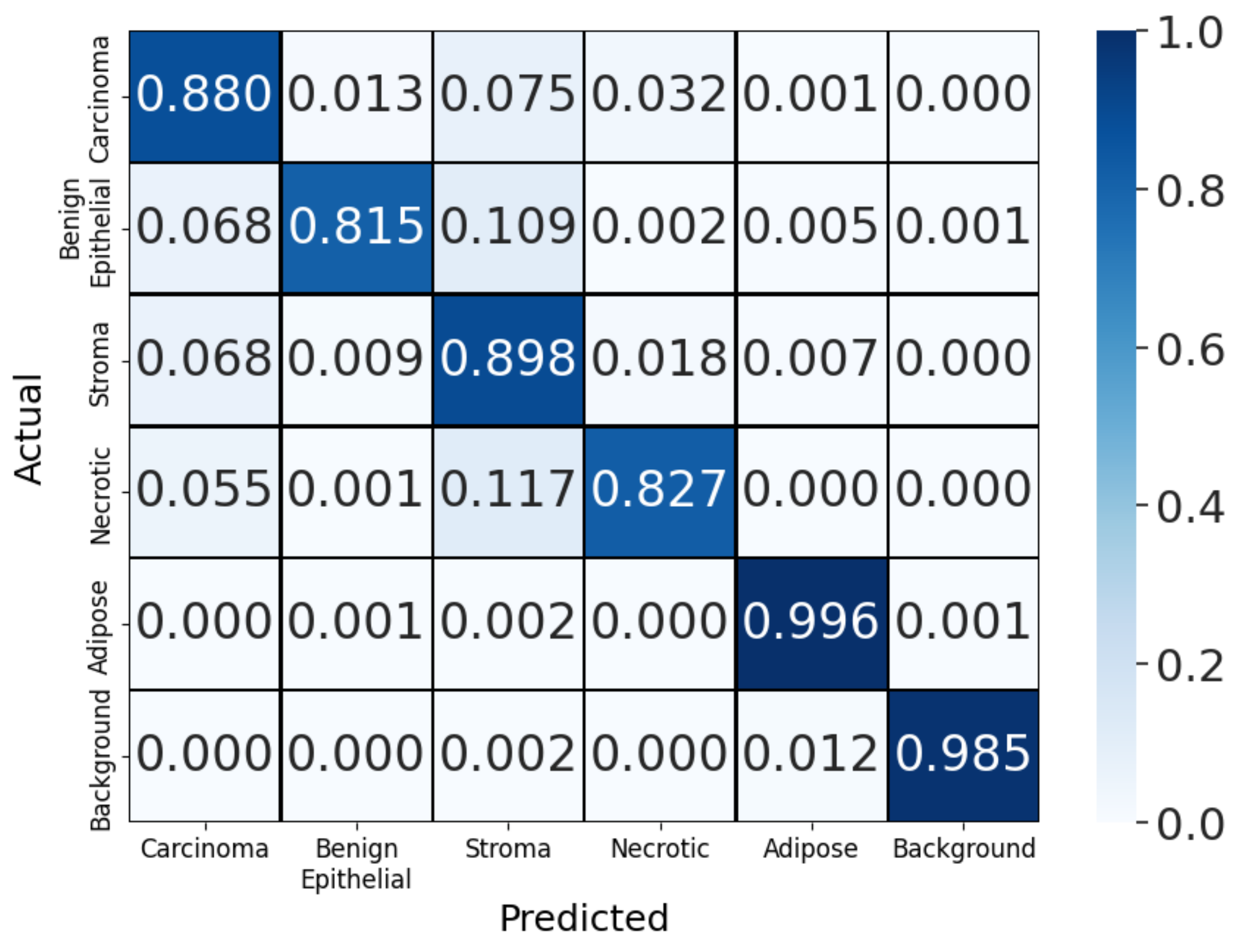,width = 0.32\textwidth}}
\subfigure[DMMN-MS]{\epsfig{figure=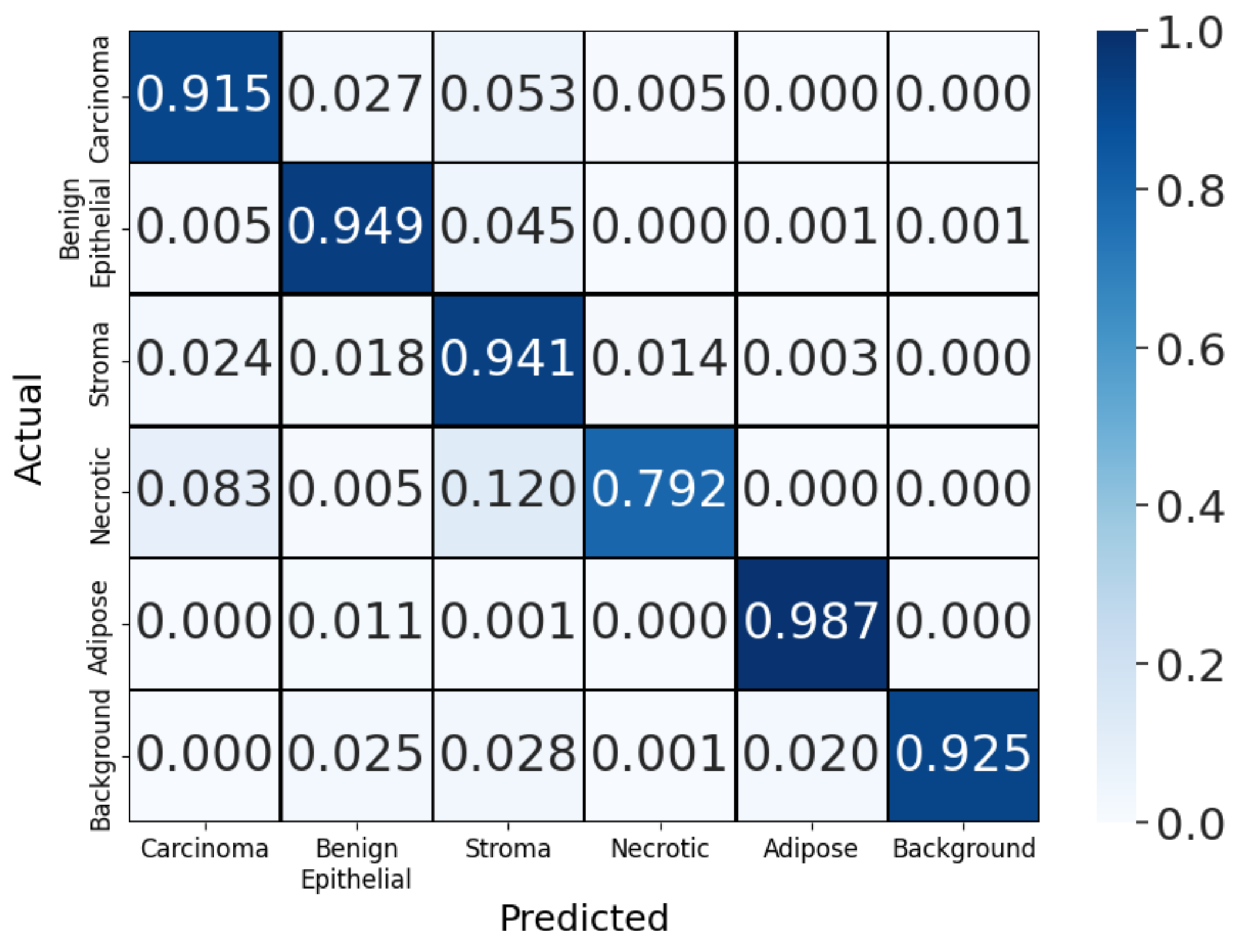,width = 0.32\textwidth}}
\subfigure[DMMN-M2S]{\epsfig{figure=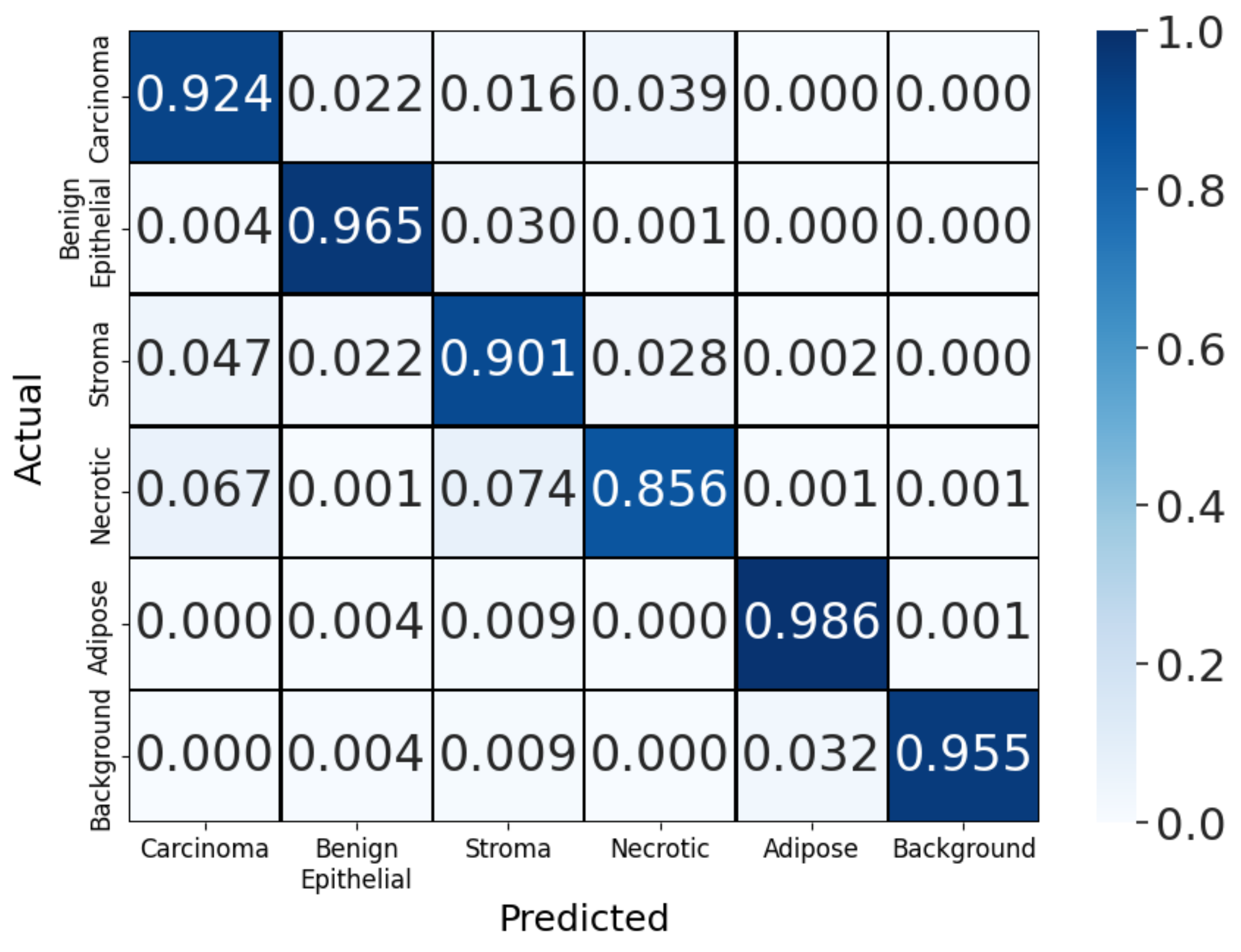,width = 0.32\textwidth}}
\subfigure[DMMN-M3]{\epsfig{figure=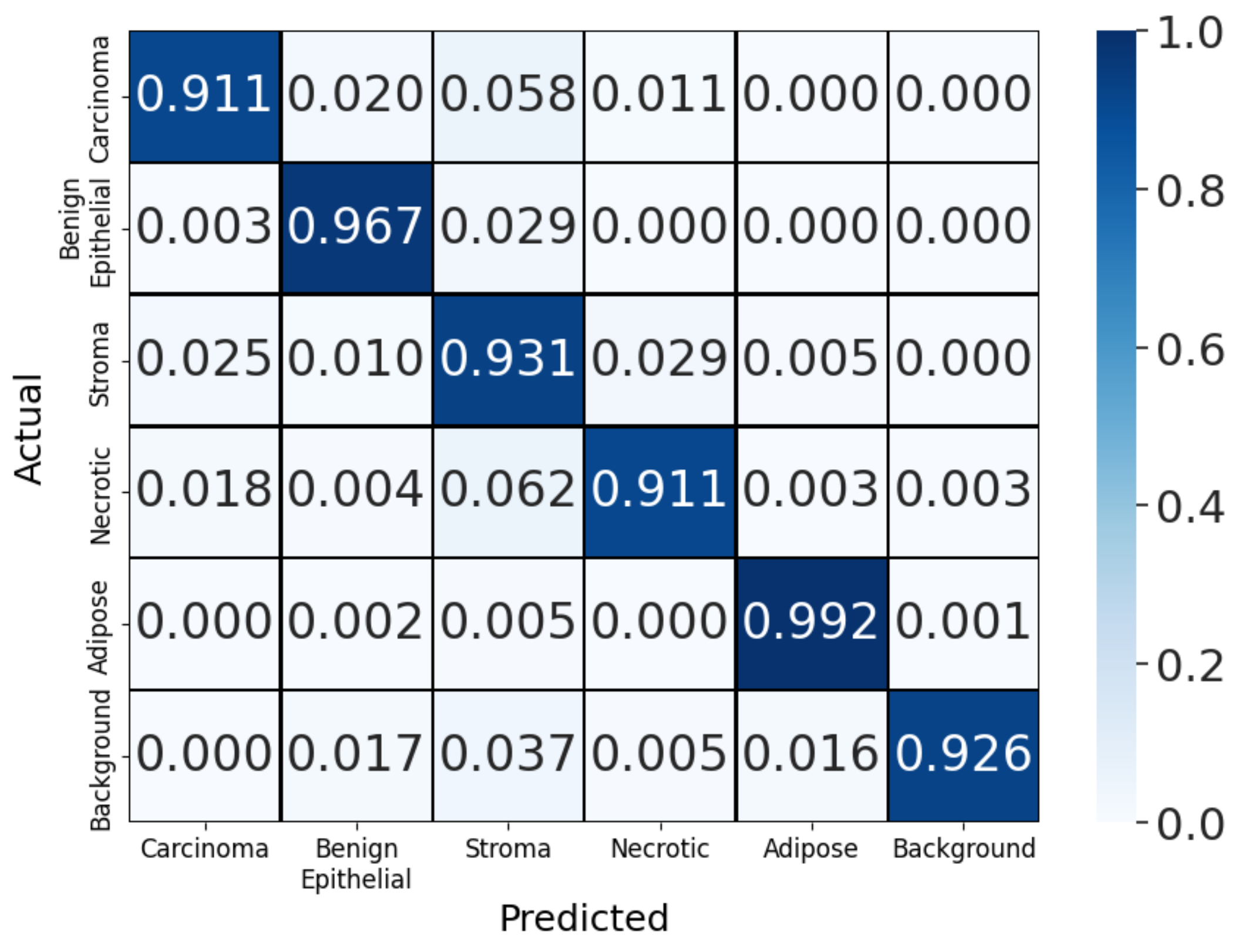,width = 0.32\textwidth}}
\caption{Confusion matrices evaluating carcinoma, benign epithelial, stroma, necrotic, adipose, and background segmentation on Dataset-I based on two Deep Single-Magnification Networks (DSMNs), SegNet \cite{badrinarayanan2017} and U-Net \cite{ronneberger2015}, and four Deep Multi-Magnification Networks (DMMNs), Single-Encoder Single-Decoder (DMMN-S2), Multi-Encoder Single-Decoder (DMMN-MS), Multi-Encoder Multi-Decoder Single-Concatenation (DMMN-M2S), and our proposed Multi-Encoder Multi-Decoder Multi-Concatenation (DMMN-M3).}
\label{fig:confusion_TNBC}
\end{figure*}

\begin{figure*}[ht!]
\centering
\subfigure[SegNet]{\epsfig{figure=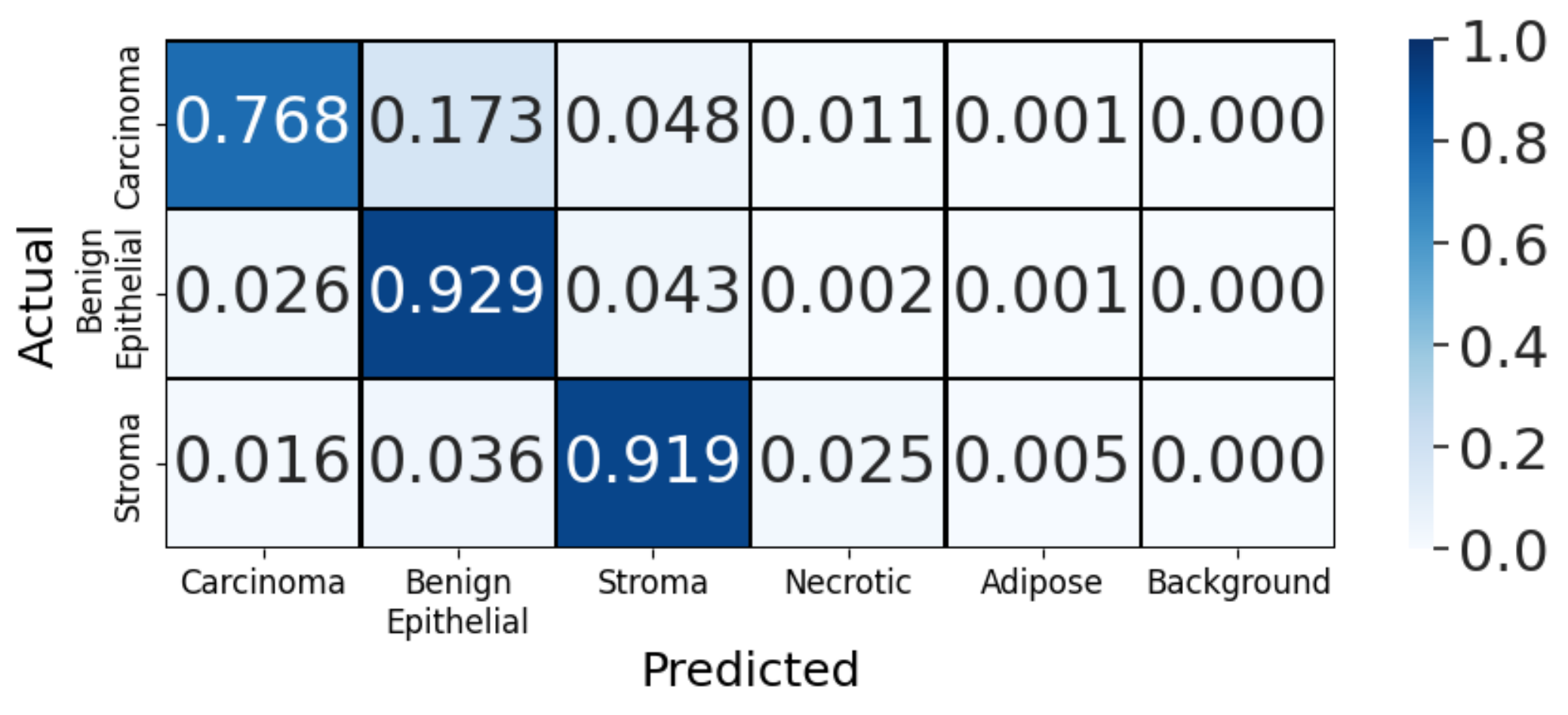,width = 0.32\textwidth}}
\subfigure[U-Net]{\epsfig{figure=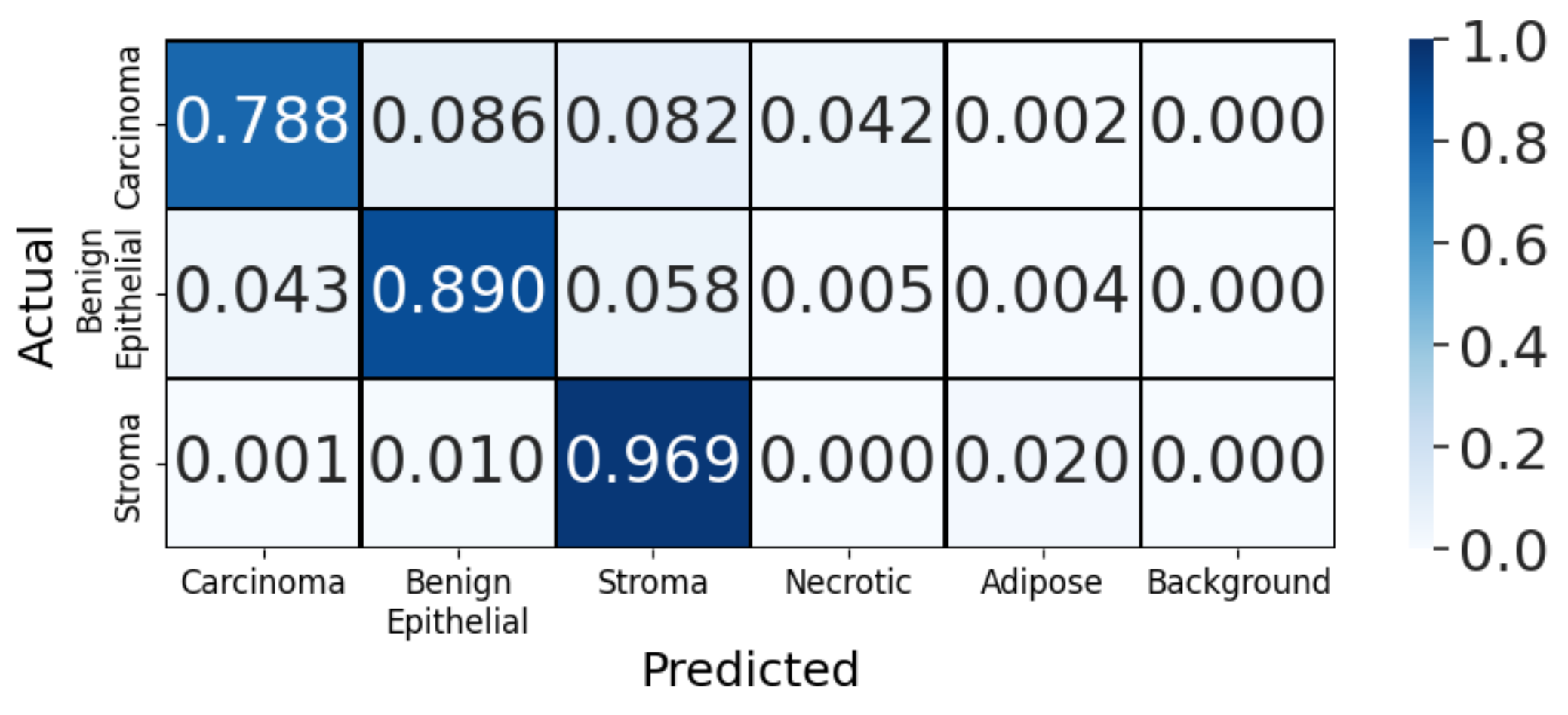,width = 0.32\textwidth}}
\subfigure[DMMN-S2]{\epsfig{figure=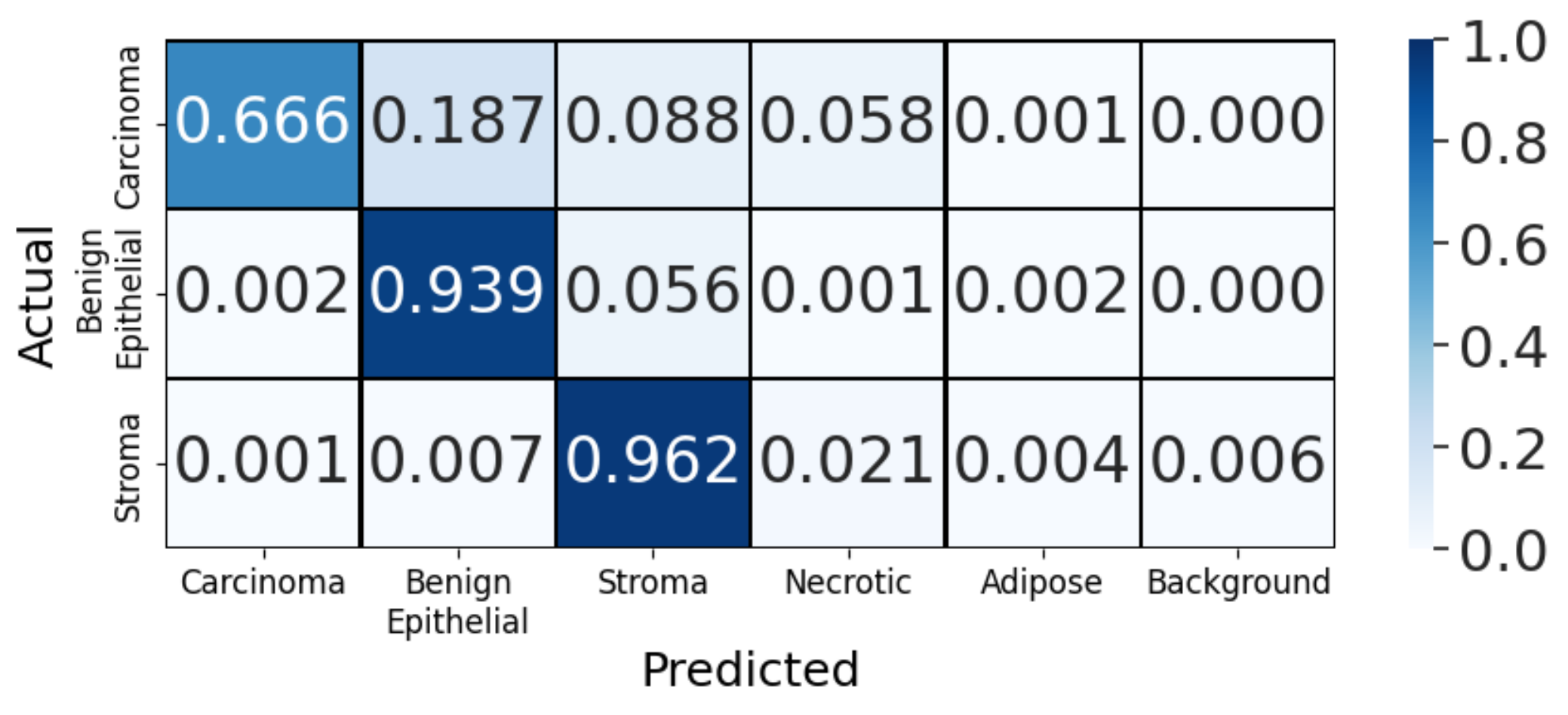,width = 0.32\textwidth}}
\subfigure[DMMN-MS]{\epsfig{figure=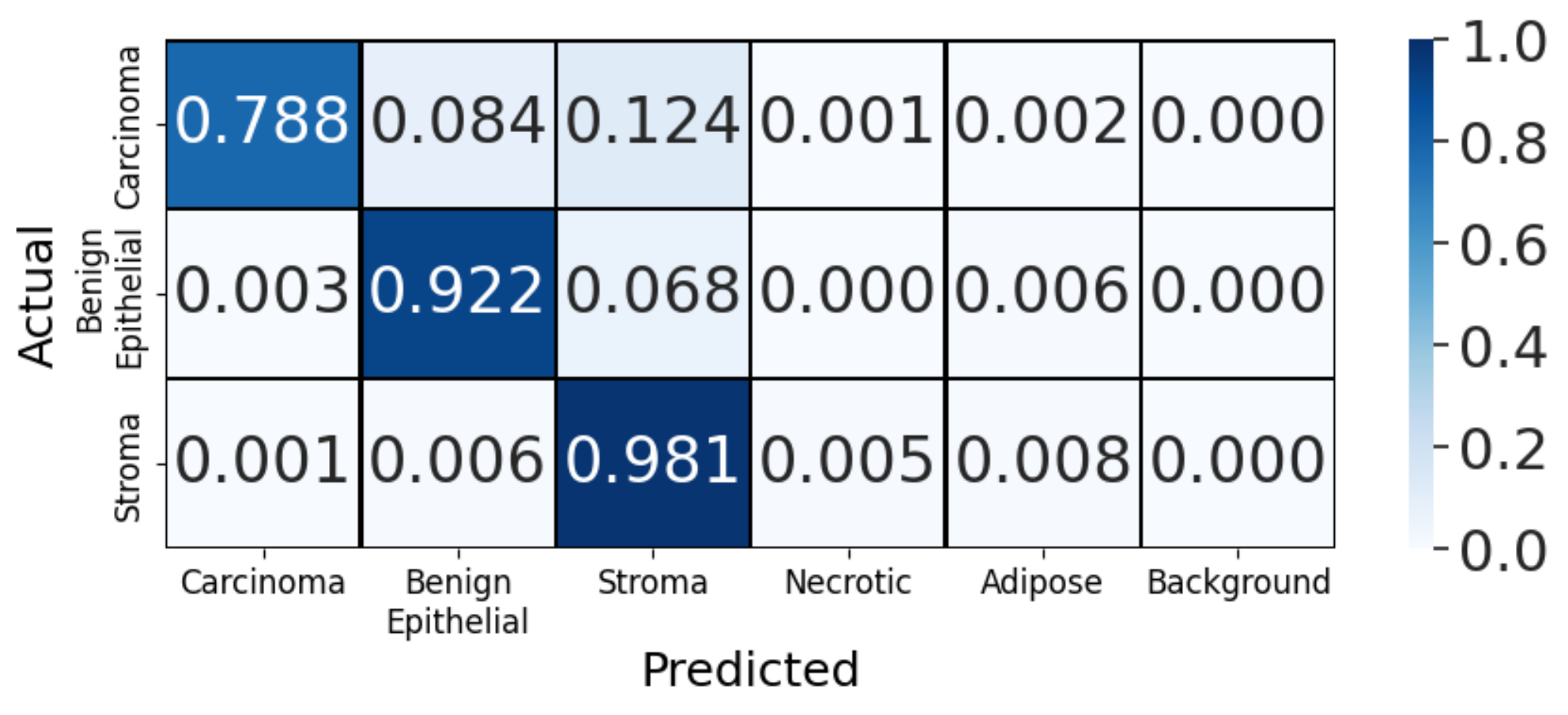,width = 0.32\textwidth}}
\subfigure[DMMN-M2S]{\epsfig{figure=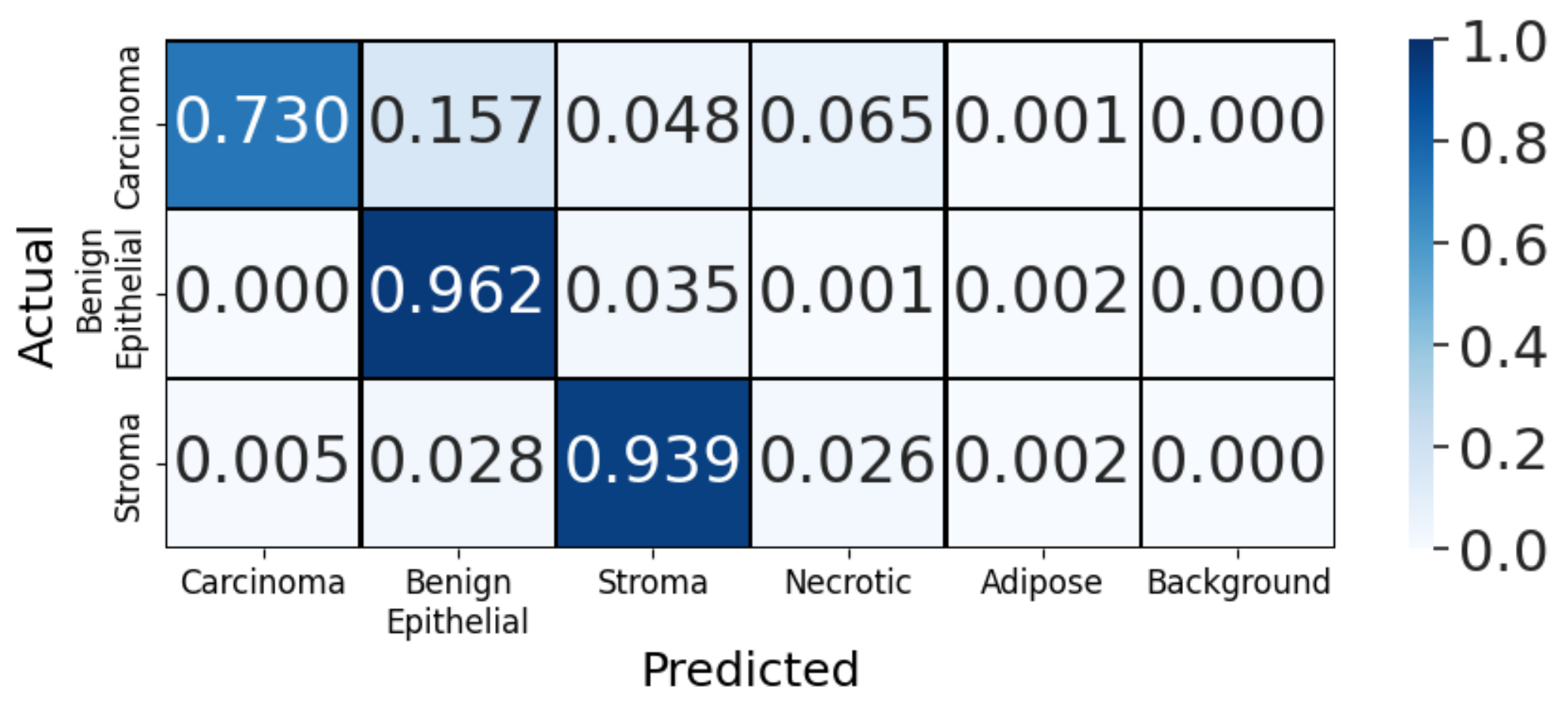,width = 0.32\textwidth}}
\subfigure[DMMN-M3]{\epsfig{figure=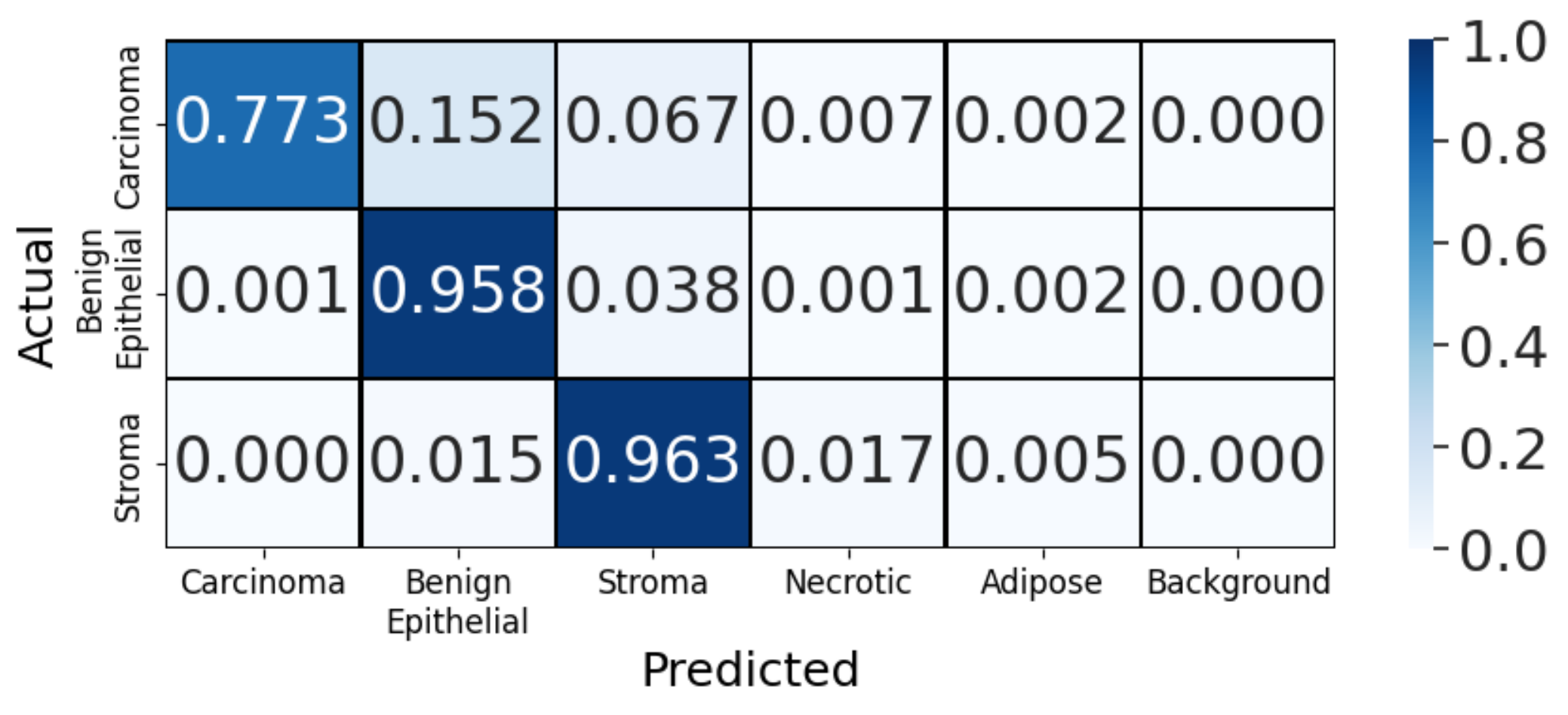,width = 0.32\textwidth}}
\caption{Confusion matrices evaluating carcinoma, benign epithelial, and stroma segmentation on Dataset-II based on two Deep Single-Magnification Networks (DSMNs), SegNet \cite{badrinarayanan2017} and U-Net \cite{ronneberger2015}, and four Deep Multi-Magnification Networks (DMMNs), Single-Encoder Single-Decoder (DMMN-S2), Multi-Encoder Single-Decoder (DMMN-MS), Multi-Encoder Multi-Decoder Single-Concatenation (DMMN-M2S), and our proposed Multi-Encoder Multi-Decoder Multi-Concatenation (DMMN-M3). Necrotic, adipose, and background are excluded from the confusion matrices on Dataset-II due to the lack of pixels being evaluated.}
\label{fig:confusion_margin}
\end{figure*}

We evaluated our predictions numerically using intersection-over-union (IOU), recall, and precision which are defined as:
\begin{equation}
    IOU = \frac{N_{TP}}{N_{TP} + N_{FP} + N_{FN}}
\end{equation}
\begin{equation}
    Recall = \frac{N_{TP}}{N_{TP} + N_{FN}}
\end{equation}
\begin{equation}
    Precision = \frac{N_{TP}}{N_{TP} + N_{FP}}
\end{equation}
where $N_{TP}$, $N_{FP}$, and $N_{FN}$ are the number of pixels for true-positive, false-positive, and false-negative, respectively.
Tables \ref{tab:TNBC} and \ref{tab:margins} shows mean IOU (mIOU), mean recall (mRecall), and mean precision (mPrecision) on Dataset-I and Dataset-II, respectively, where mIOU is used as our main evaluation metric to select the best performing model.
Figures \ref{fig:confusion_TNBC} and \ref{fig:confusion_margin} show confusion matrices from the models on Dataset-I and Dataset-II, respectively.
Note that necrotic, adipose, and background were excluded for evaluating Dataset-II because (1) Dataset-II does not contain large necrotic regions and (2) most of adipose and background regions were not segmented due to the Otsu technique \cite{otsu1979}.

\begin{table}[ht]
\centering
{
\caption{Mean IOU, Recall, and Precision on Dataset-I}
\begin{tabular}{| c | c | c | c |}
	\hline
	Model & mIOU & mRecall & mPrecision\\
	\hline
	SegNet \cite{badrinarayanan2017} & 0.766 & 0.887 & 0.850\\
    \hline
    U-Net \cite{ronneberger2015} & 0.803 & 0.896 & 0.879\\
    \hline
    DMMN-S2 & 0.833 & 0.900 & 0.910\\
    \hline
    DMMN-MS & 0.836 & 0.918 & 0.906\\
    \hline
    DMMN-M2S & 0.848 & 0.931 & 0.904\\
    \hline
    DMMN-M3 & \textbf{0.870} & \textbf{0.939} & \textbf{0.922}\\
    \hline
\end{tabular}
\label{tab:TNBC}
}
\end{table}

\begin{table}[ht]
\centering
{
\caption{Mean IOU, Recall, and Precision on Dataset-II}
\begin{tabular}{| c | c | c | c |}
	\hline
	Model & mIOU & mRecall & mPrecision\\
	\hline
	SegNet \cite{badrinarayanan2017} & 0.682 & 0.872 & 0.784\\
    \hline
    U-Net \cite{ronneberger2015} & \textbf{0.726} & 0.882 & \textbf{0.819}\\
    \hline
    DMMN-S2 & 0.639 & 0.855 & 0.764\\
    \hline
    DMMN-MS & 0.720 & 0.897 & 0.806\\
    \hline
    DMMN-M2S & 0.693 & 0.877 & 0.801\\
    \hline
    DMMN-M3 & 0.706 & \textbf{0.898} & 0.795\\
    \hline
\end{tabular}
\label{tab:margins}
}
\end{table}

Based on our visual and numerical evaluations on Dataset-I, both DSMNs had blocky boundaries between subtypes, shown in Figures \ref{fig:393382}(c,d) and \ref{fig:393867}(c,d) due to their narrow field-of-view.
DMMN-S2 also had blocky boundaries between subtypes, shown in Figures \ref{fig:393382}(e) and \ref{fig:393867}(e), because patches from multiple magnifications are concatenated early in the model so various features from multiple magnifications could not be fully extracted.
These blockly predictions led to low mIOU, low mRecall, and low mPrecision in Table \ref{tab:TNBC}.
DMMN-MS and DMMN-M2S had smoother boundaries between subtypes, but they did not have consistent predictions throughout subtypes.
For example, DMMN-MS and DMMN-M2S cannot predict necrotic successfully according to Figure \ref{fig:confusion_TNBC}(d,e).
Our proposed DMMN-M3 has shown accurate predictions throughout all subtypes, shown in Figure \ref{fig:confusion_TNBC}(f), leading to the best mIOU, mRecall, and mPrecision in Table \ref{tab:TNBC}.

Our models were trained on Dataset-I and we kept aside images in Dataset-II, annotated by a different pathologist, for our testing set.
We still observed blocky boundaries on predictions done by SegNet, U-Net, and DMMN-S2 on Dataset-II, shown in Figure \ref{fig:1365648}(c,d,e).
We noticed predictions by DMMN-M2S were not successful where large regions are falsely segmented as benign epithelial in Figures \ref{fig:1365648_WSI}(g) and \ref{fig:1365648}(g).
Figure \ref{fig:1365648}(h) demonstrates that our proposed DMMN-M3 segments subtypes with smoother and clearer boundaries.

\begin{figure*}[ht!]
\centering
\subfigure[Image]{\frame{\epsfig{figure=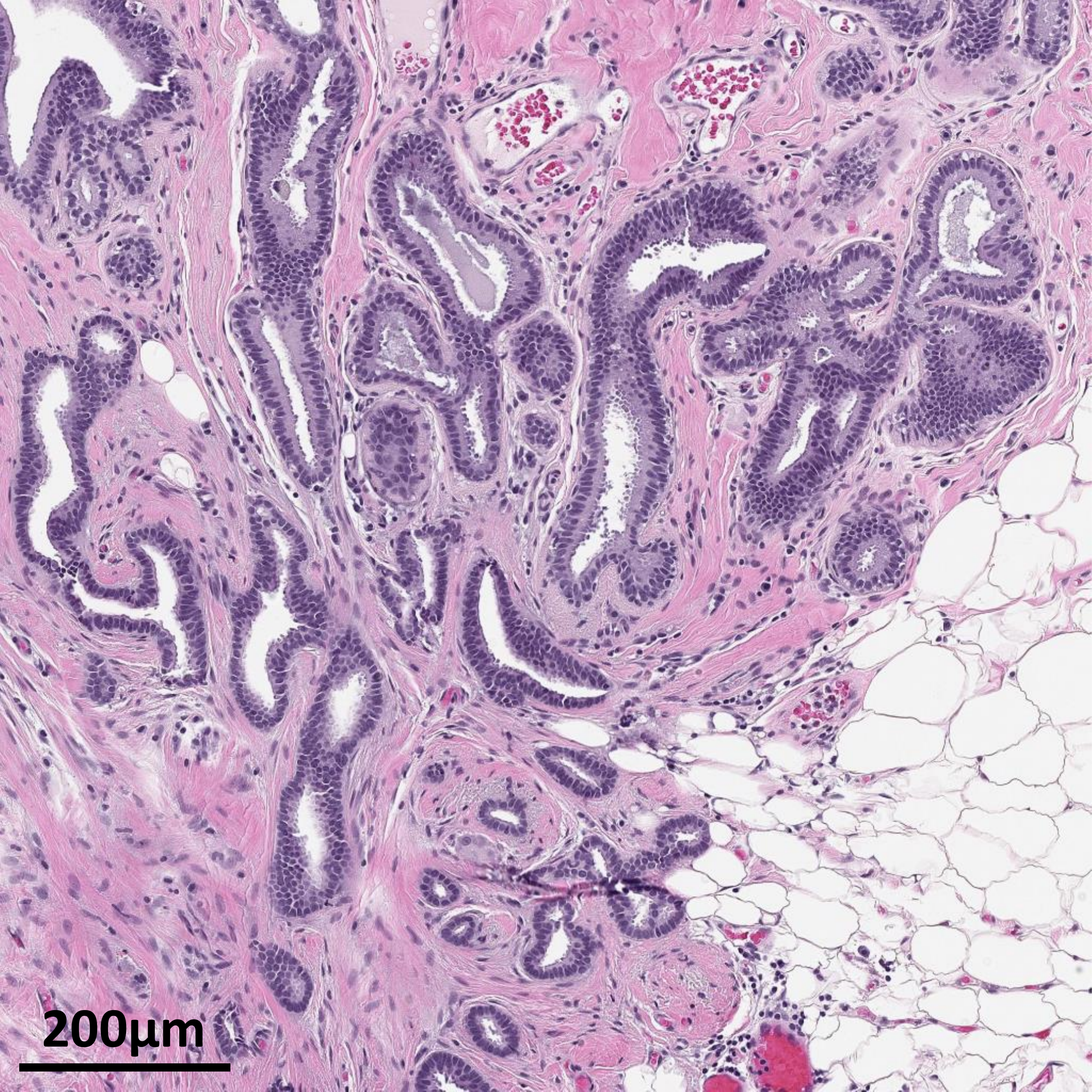,width = 0.32\textwidth}}}
\subfigure[Ground Truth]{\frame{\epsfig{figure=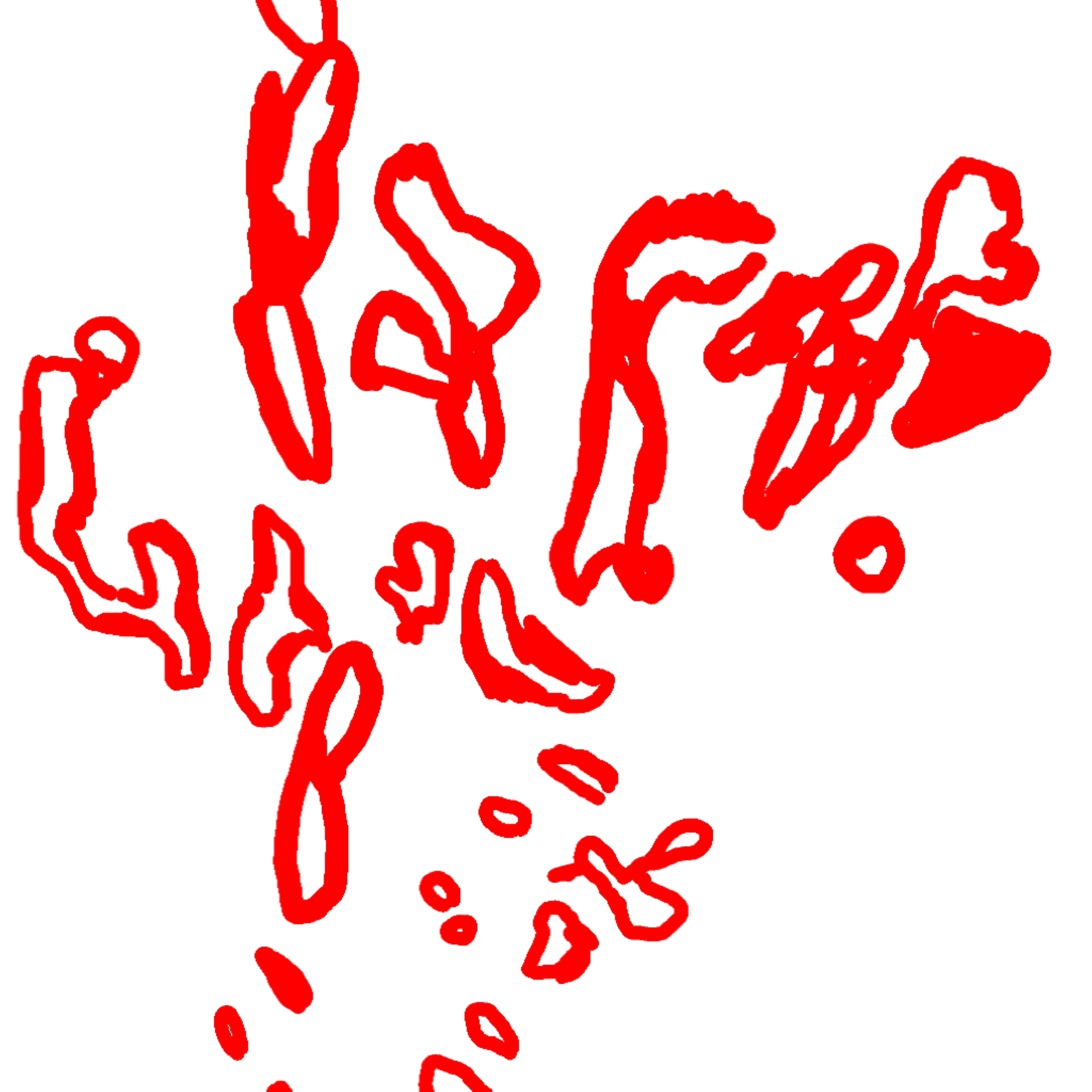,width = 0.32\textwidth}}}
\subfigure[Segmentation]{\frame{\epsfig{figure=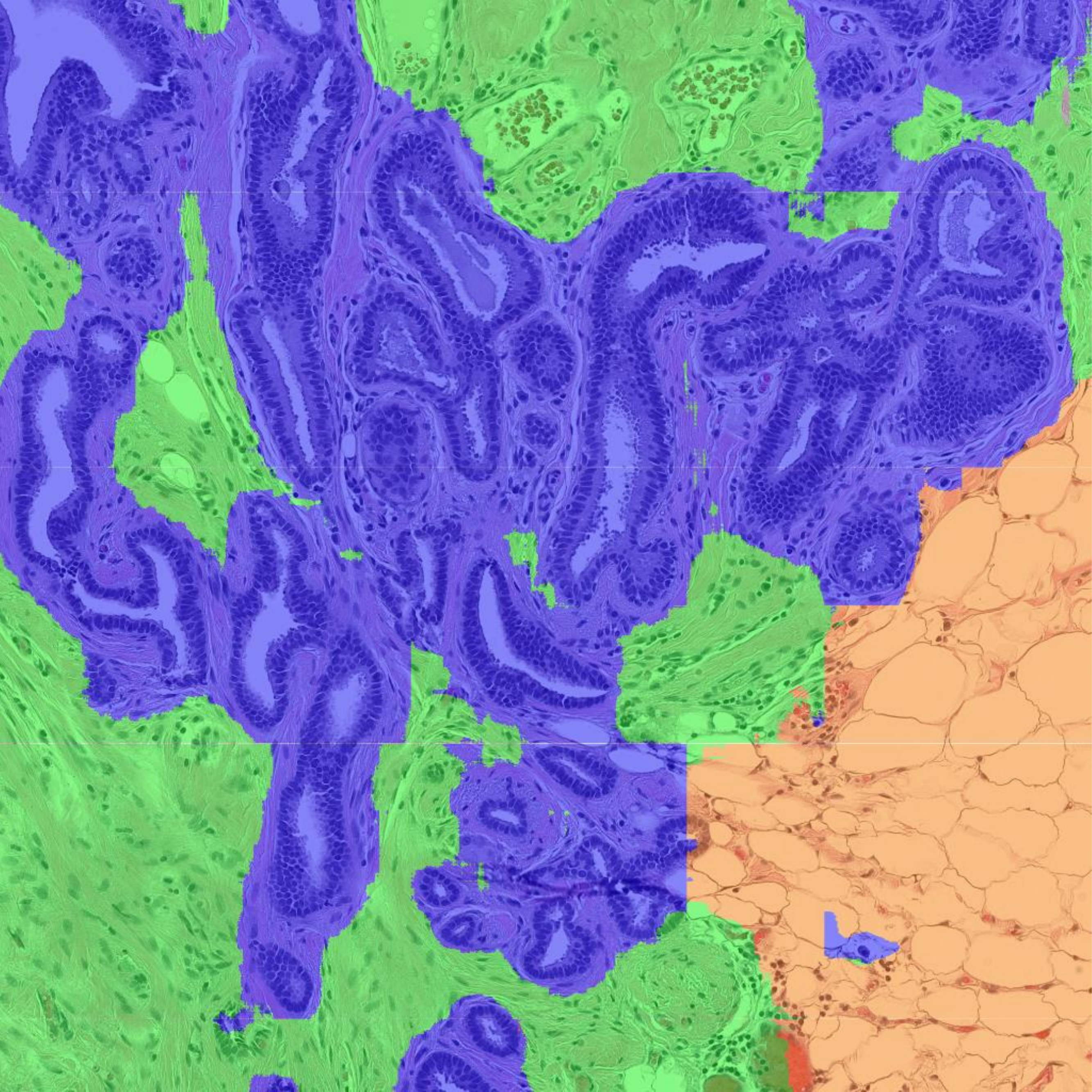,width = 0.32\textwidth}}}

\subfigure[Image]{\frame{\epsfig{figure=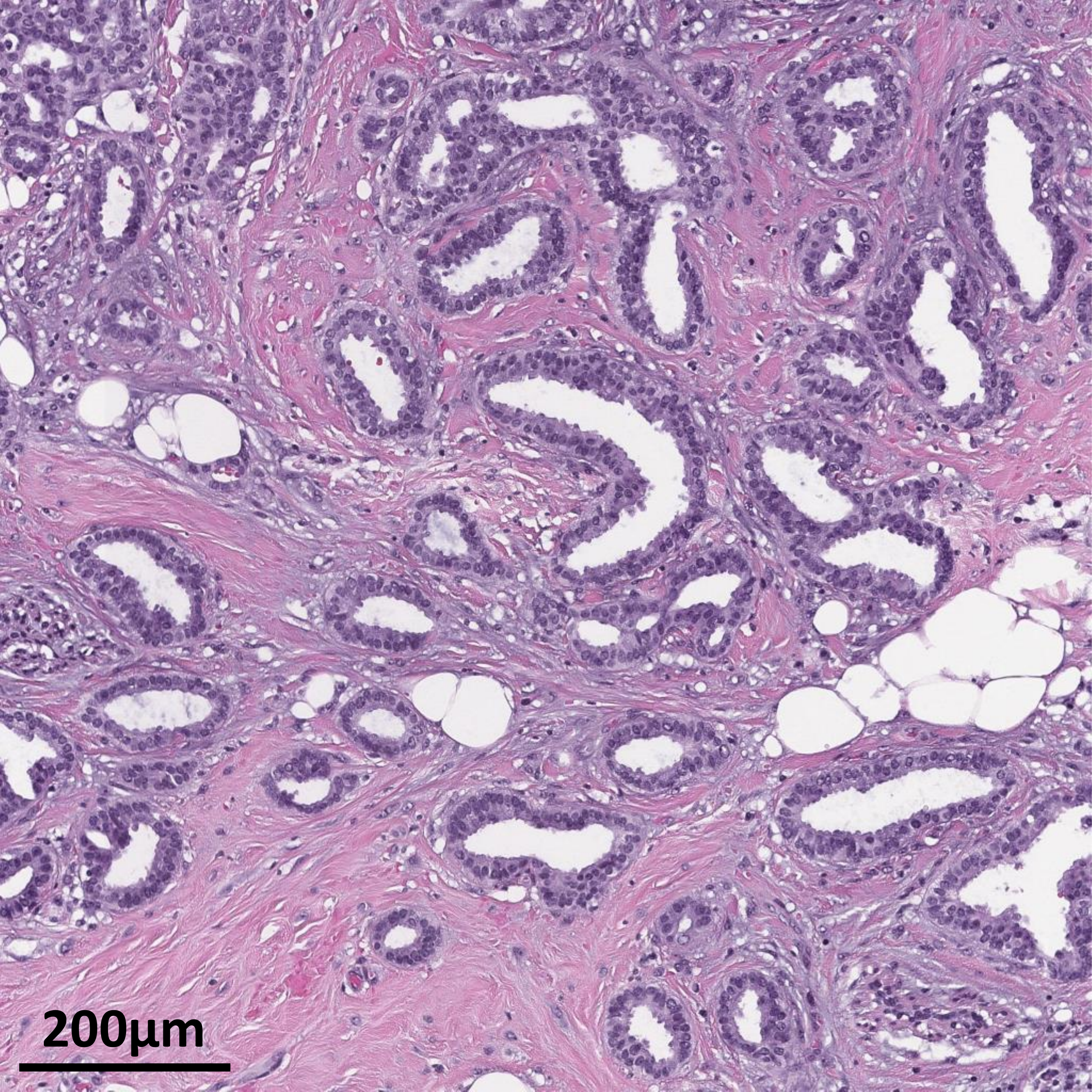,width = 0.32\textwidth}}}
\subfigure[Ground Truth]{\frame{\epsfig{figure=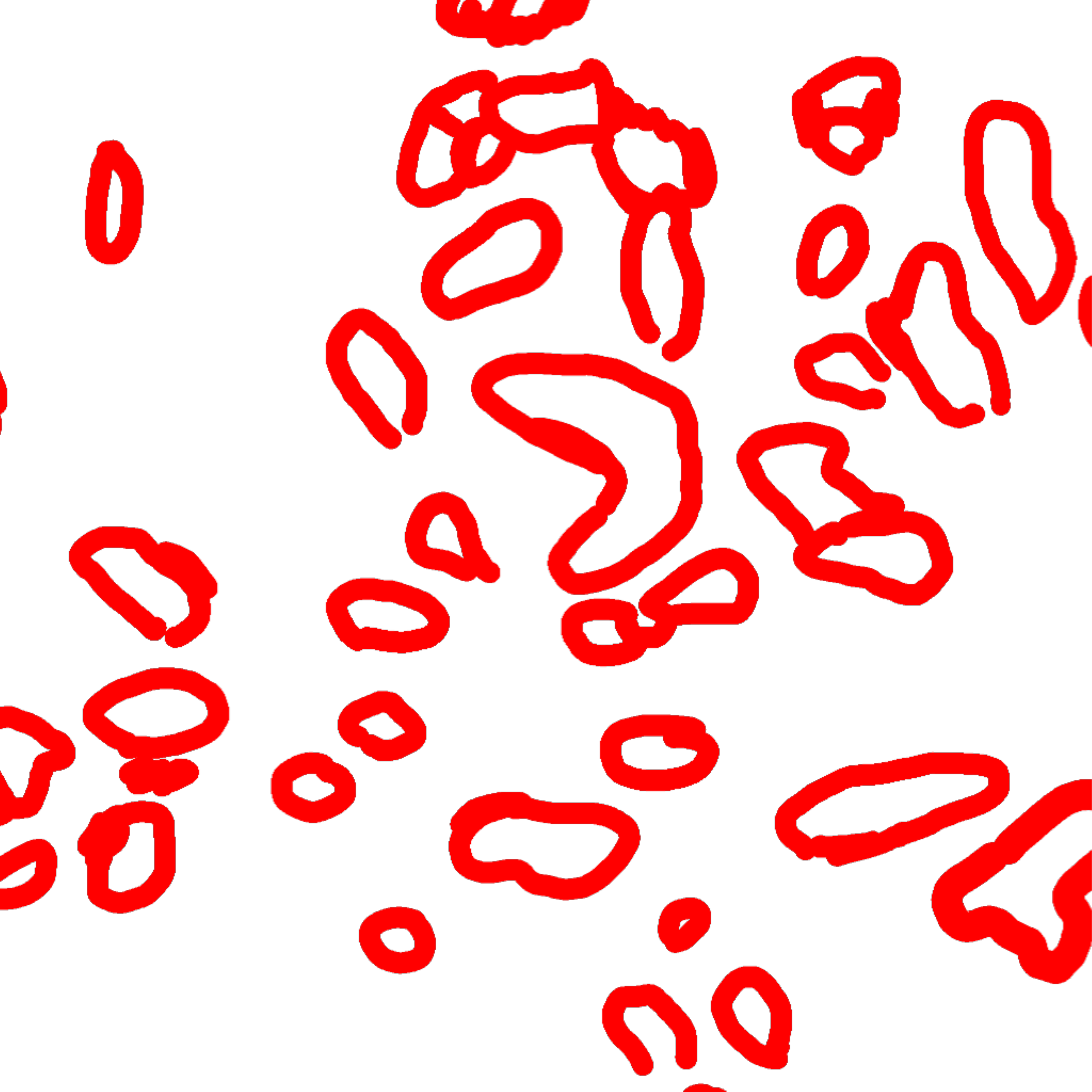,width = 0.32\textwidth}}}
\subfigure[Segmentation]{\frame{\epsfig{figure=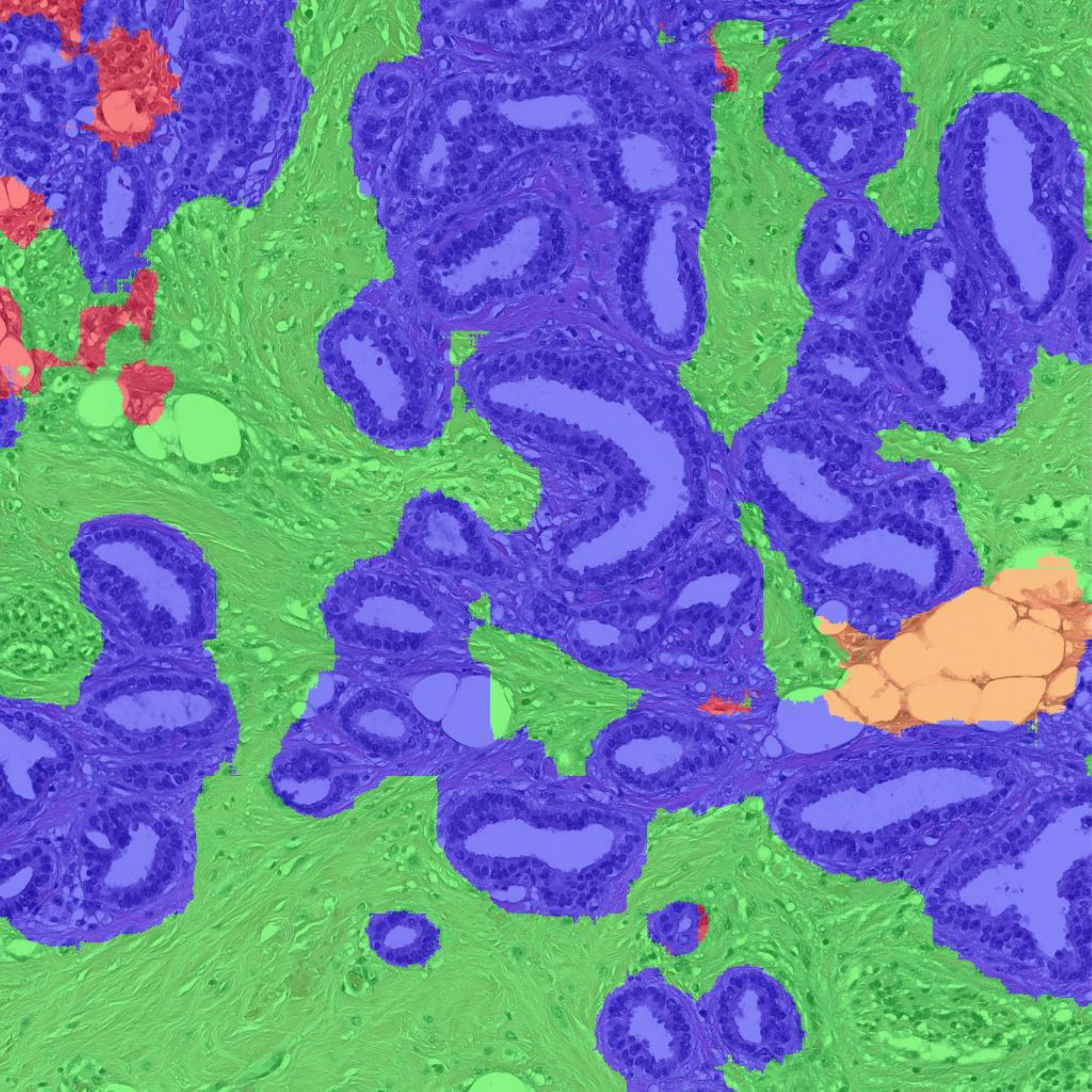,width = 0.32\textwidth}}}
\caption{Multi-class tissue segmentation predictions of well-differentiated carcinomas in red from Dataset-II using our proposed Multi-Encoder Multi-Decoder Multi-Concatenation (DMMN-M3).}
\label{fig:well-differentiated}
\end{figure*}

According to Figure \ref{fig:confusion_margin}(f), DMMN-M3 segmented many carcinoma pixels as benign epithelial on Dataset-II causing low mIOU in Table \ref{tab:margins}.
We observed that well-differentiated carcinomas were segmented as benign epithelial by DMMN-M3, shown in Figure \ref{fig:well-differentiated}.
Well-differentiated carcinomas, known to be morphologically similar to benign cells, were not presented in Dataset-I composed of high grade Triple-Negative Breast Cancer (TNBC).
DMMN-M3 trained by Dataset-I alone would not be able to learn morphological features of well-differentiated carcinomas, causing inaccurate segmentation on them in Dataset-II.
We numerically evaluated the 6 models on 29 WSIs in Dataset-II by excluding 5 images with well-differentiated carcinomas, and Table \ref{tab:margins_no_wd} shows DMMN-M3 outperforms other methods for higher histologic grades based on mIOU, mRecall, and mPrecision.
Although our current DMMN-M3 model may be challenged to segment well-differentiated carcinomas due to their absence in the training set, our proposed model, with additional training annotations of well-differentiated carcinomas, can be used to successfully segment breast whole slide images to assist pathologists for breast cancer assessments.

\begin{table}[ht]
\centering
{
\caption{Mean IOU, Recall, and Precision on a subset of Dataset-II excluding whole slide images with well-differentiated carcinomas}
\begin{tabular}{| c | c | c | c |}
	\hline
	Model & mIOU & mRecall & mPrecision\\
	\hline
	SegNet \cite{badrinarayanan2017} & 0.717 & 0.887 & 0.806\\
    \hline
    U-Net \cite{ronneberger2015} & 0.757 & 0.892 & 0.845\\
    \hline
    DMMN-S2 & 0.670 & 0.870 & 0.780\\
    \hline
    DMMN-MS & 0.759 & 0.910 & 0.836\\
    \hline
    DMMN-M2S & 0.758 & 0.901 & 0.846\\
    \hline
    DMMN-M3 & \textbf{0.782} & \textbf{0.923} & \textbf{0.847}\\
    \hline
\end{tabular}
\label{tab:margins_no_wd}
}
\end{table}

\section{Conclusion}
\label{sec:conclusion}
We described a Deep Multi-Magnification Network (DMMN) for an accurate multi-class tissue segmentation of whole slide images.
Our model is trained by partially-annotated images to reduce time and effort for annotators.
Although the annotation was partially done, our model was able to learn not only spatial characteristics within a class but also spatial relationship between classes.
Our DMMN architecture looks at all 20$\times$, 10$\times$, and 5$\times$ magnifications to have a wider field-of-view to make more accurate predictions based on feature maps from multiple magnifications.
We were able to improve previous DMMNs by transferring intermediate feature maps from decoders in 10$\times$ and 5$\times$ to a decoder in 20$\times$ to enrich feature maps.
Our implementation achieved outstanding segmentation performance on breast datasets that can be used to decide patients' future treatment.
One main challenge we encountered is that our model may not successfully segment well-differentiated carcinomas presented in breast images because well-differentiated carcinomas were not included in training annotation.
We also observed that our model can be sensitive to background noises potentially leading to mis-segmentation on background regions if whole slide images are digitized by other scanners.
In the future, we plan to develop a more accurate DMMN model where various cancer structures and background noise patterns are included during training.

\section{Acknowledgments}
This work was supported by the Warren Alpert Foundation Center for Digital and Computational Pathology at Memorial Sloan Kettering Cancer Center and the NIH/NCI Cancer Center Support Grant P30 CA008748.

\section{Conflict of interest}
T.J.F. is the Chief Scientific Officer, co-founders and equity holders of Paige.AI. 
M.G.H. is a consultant for Paige.AI and on the medical advisory board of PathPresenter. 
D.J.H. and T.J.F. have intellectual property interests relevant to the work that is the subject of this paper. 
MSK has financial interests in Paige.AI. and intellectual property interests relevant to the work that is the subject of this paper.

\bibliography{mybibfile}

\end{document}